\documentclass{new_tlp}

\makeatletter
\let\O@argtabularcr\@argtabularcr
\def\O@xtabularcr{\@ifnextchar[\O@argtabularcr{\ifnum 0=`{\fi}\cr}}
\let\O@tabacol\@tabacol
\let\O@tabclassiv\@tabclassiv
\let\O@tabclassz\@tabclassz
\let\O@tabarray\@tabarray
\def\author@tabular{\authorsize\def\@halignto{}\@authortable}
\let\endauthor@tabular=\endtabular
\def\author@tabcrone{{\ifnum0=`}\fi\O@xtabularcr\affilsize\itshape
 \let\\=\author@tabcrtwo\ignorespaces}
\def\author@tabcrtwo{{\ifnum0=`}\fi\O@xtabularcr[-3\p@]\affilsize\itshape
 \let\\=\author@tabcrtwo\ignorespaces}
\def\@authortable{\leavevmode \hbox \bgroup $\let\@acol\O@tabacol
 \let\@classz\O@tabclassz \let\@classiv\O@tabclassiv
 \let\\=\author@tabcrone \ignorespaces \O@tabarray}
\makeatother
 
\usepackage[utf8]{inputenc}
\usepackage{cancel}

\usepackage{caption}

\usepackage{enumitem}
\usepackage{graphicx}
\graphicspath{{.}}

\usepackage{amsmath}
\usepackage{wrapfig}
\usepackage{framed}
\usepackage{color}
\definecolor{shadecolor}{RGB}{224,238,238}
\usepackage{mwe} %
\usepackage{floatpag}
\usepackage{varwidth}
\usepackage{booktabs}

\usepackage[section]{placeins}

\usepackage{ifthen}

\usepackage[section]{placeins}

\newcommand{\secbeg}{\vspace*{-3mm}}
\newcommand{\secend}{\vspace*{-3mm}}
\renewcommand{\secbeg}{}
\renewcommand{\secend}{}

\newlength{\figwidth}
\newcommand{\ignore}[1]{}

\newcommand{\tuple}[3]{\ensuremath{\tuplek{\langle #1,\allowbreak #2\rangle}{#3}\xspace}}
\newcommand{\tuplek}[2]{\ensuremath{#1 \mapsto \allowbreak #2}\xspace}
\newcommand{\tupleu}[2]{\ensuremath{#1 \mapsto \allowbreak #2}\xspace}
\newcommand{\apair}[2]{\langle #1,#2 \rangle}
\newcommand{\arc}[5]{\ensuremath{\apair{#1}{#2}%
    \rightarrow_{#3}
    \allowbreak\apair{#4}{#5}}\xspace}
\newcommand{\arck}[2]{\ensuremath{#1 \rightarrow \allowbreak #2}\xspace}

\newcommand{\imports}[1]{\ensuremath{\mathsf{imports}}(#1)\xspace}
\newcommand{\exports}[1]{\ensuremath{\mathsf{exports}}(#1)\xspace}

\newcommand{\diff}[1]{\ensuremath{\Delta_{\mathit{#1}}}\xspace}

\newcommand{\DD}{\mbox{$D_\alpha$}}
\newcommand{\CD}{\mbox{$D$}}
\newcommand{\QQ}{\mbox{\ensuremath{\Q_\alpha}\xspace}}

\newcommand{\pdesc}{A}
\newcommand{\qdesc}{B}
\newcommand{\phead}{A}
\newcommand{\pbody}{A}

\newcommand{\CP}{\mbox{\ensuremath{\lambda^c}\xspace}}
\newcommand{\AP}{\mbox{\ensuremath{\lambda^s}\xspace}}
\newcommand{\PP}{\mbox{\ensuremath{\lambda^p}\xspace}}
\newcommand{\RP}{\mbox{\ensuremath{\lambda^r}}} %
\newcommand{\CDP}[2]{\apair{#1}{#2}\xspace}
\newcommand{\ACP}{\CDP{A}{\CP}}

\newcommand{\fulledge}[6]{\ensuremath{#1 \rightarrow_{#4,#5} #6}\xspace}

\newcommand{\cls}{\ensuremath{\mathit{Cls}}\xspace}
\newcommand{\assign}{\ensuremath{:=}\ \xspace}

\newcommand{\A}{\mbox{$\cal A$}}
\newcommand{\Q}{\mbox{$Q$}}

\usepackage{fancybox}
\usepackage{latexsym}
\usepackage{verbatim}
\usepackage{xcolor}
\usepackage{amssymb}
\usepackage{algorithm}
\usepackage[noend]{algpseudocode}

\newcommand{\alglinetitle}[1]{\textbf{\color{black}{#1}}}
\newcommand{\algline}[2]{\alglinetitle{#1} #2}

\usepackage[T1]{fontenc}
\usepackage{latexsym}
\usepackage{stmaryrd}
\usepackage{xspace}
\usepackage{multirow}

\usepackage{longtable}
\usepackage{tikz}
   \usetikzlibrary{positioning} %
   \usetikzlibrary{backgrounds}
   \usetikzlibrary{matrix}

\newcommand{\tikzagconf}{%
  \scriptsize
  \usetikzlibrary{arrows}
  \usetikzlibrary{shapes}
  \tikzstyle{every node}=[draw=black,minimum height=28pt,text width=65pt,thick,ellipse,align=center,node distance=1cm]
  \tikzstyle{every edge}=[thick,draw=black]
    }
    
\usepackage{lineno}

\usepackage{listings}
\usepackage{textcomp}

\definecolor{ciaoframe}     {rgb}{  0,    0,  0.3}
\definecolor{ciaostring}    {rgb}{0.6, 0.46, 0.33}
\definecolor{ciaooperators} {rgb}{0.1, 0.15,  0.6}
\definecolor{ciaokeywords}  {rgb}{0.1, 0.15,  0.6}
\definecolor{ciaoassertions}{rgb}{0.1, 0.15,  0.6}
\definecolor{ciaotrust}     {RGB}{200, 130,     0}
\definecolor{ciaocheck}     {rgb}{0.1, 0.2,   0.8}
\definecolor{ciaochecked}   {rgb}{0.2, 0.34,  0.1}
\definecolor{ciaotrue}      {rgb}{0.2, 0.34,  0.1}
\definecolor{ciaofalse}     {rgb}{0.6,  0.0, 0.09}
\definecolor{ciaoprops}     {rgb}{0.1,  0.2,  0.8}
\definecolor{ciaocomment}   {rgb}{0.5,  0.5,  0.5}

\newcommand{\prettylstciao}[0]{
\lstset{language=Prolog,
  frameround=fttt,
  frame=ltrb,
  rulecolor=\color{ciaoframe},
  numbers=left,numberstyle=\tiny,stepnumber=1,numbersep=8pt,
  tabsize=4,
  breaklines=true,breakatwhitespace=true,
  basicstyle=\scriptsize\ttfamily, %
  showlines=true,
  showspaces=false,
  showtabs=false,
  escapechar=@,
  escapeinside={~~},
  commentstyle=\color{ciaocomment},
  stringstyle=\color{ciaostring},
  showstringspaces=false,
  deletekeywords={true}, %
  keywordstyle={\color{ciaooperators}\bfseries}, %
  classoffset=1, %
        otherkeywords={>,<,>=,=<,.,;,-,!,=,*,\&,+,:-,[,],|,->,:,:=,\#},
        keywordstyle={\color{ciaokeywords}\bfseries},
  classoffset=2,
       morekeywords={module,use_module,dynamic,export,import,impl_defined},
       keywordstyle={\color{ciaokeywords}\bfseries},
       morekeywords={pred,prop,calls,success,comp},
       keywordstyle={\color{ciaoassertions}\bfseries},
  classoffset=4,
       morekeywords={trust,trust_default,entry},
       keywordstyle={\color{ciaotrust}\bfseries},
  classoffset=5,
       morekeywords={check},
       keywordstyle={\color{ciaocheck}\bfseries},
  classoffset=6,
       morekeywords={checked},
       keywordstyle={\color{ciaochecked}\bfseries},
  classoffset=7,
       morekeywords={true},
       keywordstyle={\color{ciaotrue}\bfseries},
  classoffset=8,
       morekeywords={false},
       keywordstyle={\color{ciaofalse}\bfseries},
  classoffset=9,
       morekeywords={even,nat,int,flt,atm,term,num,var,list,ground,mshare,
                    rsize,cardinality,not_fails,exp,cost,costb,steps_ub,steps_lb,
                    size_ub,size_lb,covered,mut_exclusive,head_cost,literal_cost,
                    is_det,length,terminates,steps_o,resource,socket,seff,string
       },
       keywordstyle={\color{ciaoprops}\bfseries},
  classoffset=0, %
}}

\newcommand{\gfigpath}{.}

\newenvironment{example-box}{\begin{example}\rm}{$\Box$\end{example}}

\newif\iflongfigs
\longfigstrue %

\newif\iflongaccfigs

\newtheorem{theorem}{{\bf Theorem}}
\newtheorem{lemma}{{\bf Lemma}}
\newtheorem{example}{{\bf Example}}

\newtheorem{definition}{{\bf Definition}}

\usepackage{soul}
\usepackage{siunitx}

\newcommand{\figendsp}{\vspace*{-3mm}}

\usepackage{makecell}

\newcommand{\apar}{\ensuremath{b}}

\newcommand{\graph}{\ensuremath{G = (V,E)}}

\newcommand{\aproject}{{\tt Aproj}}
\newcommand{\acall}{{\tt Acall}}
\newcommand{\aextend}{{\tt Aextend}}
\newcommand{\alub}{{\tt Ageneralize}}
\newcommand{\update}{{\sf upd}}

\renewcommand{\mod}[1]{\ensuremath{\mathsf{mod}}(#1)\xspace}
\renewcommand{\implies} {\mbox{$\Rightarrow$}}

\usepackage{calrsfs}
\DeclareMathAlphabet{\pazocal}{OMS}{zplm}{m}{n}
\newcommand{\loc}{\ensuremath{\mathcal{L}}}
\newcommand{\glo}{\ensuremath{\pazocal{G}}}

\usepackage{thmtools, thm-restate}

\newcommand{\GAG}{\ensuremath{\glo}\xspace}

\newcommand{\LAG}{\ensuremath{\loc}\xspace}

\newcommand{\banalyze}{\textsc{IncAnalyze}\xspace}

\newcommand{\lfp}{\textit{lfp}\xspace}

\newcommand{\manalyze}[1]{\textsc{ModAnalyze}\ensuremath{#1}\xspace}
\newcommand{\mianalyze}[1]{\textsc{ModIncAnalyze}\ensuremath{#1}\xspace}
\newcommand{\lanalyze}[1]{\textsc{LocIncAnalyze}\ensuremath{#1}\xspace}

\newcommand{\andtree}{{\sc and} tree\xspace}
\newcommand{\andtrees}{{\sc and} trees\xspace}
\newcommand*{\andnode}[1][L]{\ensuremath{\langle {#1},\theta^c,\theta^s
    \rangle}}
\newcommand{\csemantics}{\ensuremath{\llbracket P \rrbracket_{\Q}}}

\newcommand{\callingcontext}{\ensuremath{{\sf calling\_context}}}
\newcommand{\answers}{\ensuremath{{\sf answers}}}

\newcommand{\calls}{\ensuremath{\mathit{Calls}}}
\newcommand{\AGDD}{\ensuremath{\mathit{AG}}}

\title[Incremental and Modular Context-sensitive Analysis]{Incremental and
  Modular Context-sensitive Analysis$^{
    \thanks{%
      Research partially funded by
      MINECO
      MICINN PID2019-108528RB-C21 \emph{ProCode} project, FPU grant
      16/04811, and the Madrid M141047003 \emph{N-GREENS} and
      P2018/TCS-4339 \emph{BLOQUES-CM} programs. We are also grateful
      to the anonymous reviewers, editors, and to Ignacio Fábregas for
      their comments.}}$ }

 \author[Isabel Garcia-Contreras, Jos\'{e} F.~Morales,
  and Manuel V.~Hermenegildo]{\hspace*{-10mm}
       ISABEL GARCIA-CONTRERAS$^{1,2}$
    ~JOS\'{E} F.~MORALES$^{1}$ 
    ~MANUEL V.~HERMENEGILDO$^{1,2}$
\ \\
\ \\
  \hspace*{-8mm}$^1$IMDEA Software Institute \\
  \hspace*{-8mm}$^2$Universidad Polit\'{e}cnica de Madrid (UPM)\\
  \hspace*{-8mm}\email{\{isabel.garcia, josef.morales,
     manuel.hermenegildo\}@imdea.org} \vspace*{1mm} \\
}

\usepackage{hyperref}

\begin{document}
\label{firstpage}
\maketitle

\begin{abstract}
  
  \medskip
  Context-sensitive global analysis of large code bases can be expensive,
  which can make its use impractical 
  during software development.
  However, there are many situations in which modifications are small and
  isolated within a few components, and it is desirable to reuse as much as
  possible previous analysis results.
  This has been achieved to date
  through incremental global analysis fixpoint algorithms 
  that achieve cost reductions at fine 
  levels of granularity, such as changes
  in program lines. However, these fine-grained techniques are not directly
  applicable to modular programs, nor are they designed to take advantage of
  modular structures.
  This paper describes, implements, and evaluates an algorithm that performs
  efficient context-sensitive 
  analysis incrementally on modular partitions of programs.
  The experimental results show that the proposed modular algorithm
  shows significant improvements,
  in both time and memory consumption, 
  when compared
  to existing non-modular, fine-grain incremental analysis
  techniques. 
  Furthermore, thanks to the proposed inter-modular propagation of
  analysis information, our algorithm also outperforms traditional
  modular analysis even when analyzing from scratch.

\end{abstract}


\secbeg
\section{Introduction and motivation}
\secend
\label{sec:intro}
Large, real-life programs typically have a complex structure combining a number
of modules with system libraries. 
Context-sensitive global analysis of such large code bases can be
expensive, and this can be specially problematic in interactive uses
of analyzers.
An example is
detecting and reporting bugs
as the program is being edited, by running the analysis
in the background 
at small intervals, e.g., each time a set of changes is made, when a
file is saved, or when a commit is made in the version control system.
Other such scenarios include
reanalyzing after performing source-to-source transformations and/or
optimizations,
or updating analysis results after dynamic program modifications
(reanalysis at run time).
In these scenarios, triggering a complete reanalysis for each change set is
often too costly for larger programs.
However, 
a key observation is that very often changes in the program are small
and isolated inside a small number of components. 
Ideally this characteristic can be taken advantage of to reduce the
cost of re-analysis in two ways: reusing as much information as
possible from previous analyses, and avoiding the maintenance of
analysis information for unaffected components.

In the field of abstract interpretation, there have been proposals to
deal with the following two cases: a) context-sensitive incremental
fixpoint algorithms~\cite{%
  incanal-iclp95,%
  inc-fixp-sas,%
  clpr-anal,%
  incanal-toplas,%
  albertcpr12,%
  DBLP:conf/icse/ArztB14,%
  DBLP:conf/kbse/SzaboEV16%
}, which reuse information but still need to work with the program as
a whole (incremental but \emph{monolithic} analyzers)%
; and b) \emph{modular} algorithms, aimed at reducing the memory consumption or
working set size~\cite{%
  modular-anal-lopstr,%
  CousotModular02,%
  mod-an-lopstrbook,%
  modbenchmarks-lopstr05,%
  ccfmmr09,clousot-2010%
}, which work on a module at a time but
do not support changes in the program. %
Surprisingly, the combination of both techniques has not been explored
to date. 
The monolithic incremental analyzers are not directly applicable in
the modular setting due to two issues:
first, these
analyzers do not deal with code that is partially available, i.e.,
they have no provisions to make assumptions about code that is
external. Even though one could see builtin operations of the language
as external calls, as they are obviously not defined in the module,
the semantics of these are typically ``hardwired'' in the analyzer as
\emph{transfer functions}.
This leads to the second issue: even though the monolithic analyzers
can make assumptions using this mechanism, these algorithms are not
prepared to deal in a correct and precise way with updates to these
assumptions.

In order to bridge this gap, using a monolithic incremental analysis
algorithm as a starting point, we develop a modular, incremental
analyzer capable of performing fine-grain incremental analysis across
modular program partitions.
Our algorithm is based on computing local fixpoints on one module at a
time; identifying, invalidating, and recomputing only those parts of
the analysis results that are affected by these fine-grain program
changes; and propagating the fine-grained analysis information across
module boundaries.
Our contributions are: extending the incremental (global) fixpoint
algorithm of~\cite{incanal-toplas} with widening
(Sec.~\ref{sec:mon-inc}); providing a formal description of the
modular analysis algorithm of~\cite{mod-an-lopstrbook} with
correctness results (Sec.~\ref{sec:mod}); and providing a new analysis
algorithm that is modular and incremental, also with correctness
results (Secs.~\ref{sec:intermod} and~\ref{sec:th}).
Additionally, we have implemented the proposed approach within the
Ciao/CiaoPP system~\cite{%
  ciaopp-sas03-journal-scp,%
  hermenegildo11:ciao-design-tplp%
} and benchmarked it.
The experimental results observed show good cost-performance
tradeoffs,
in both time and memory consumption,
and suggest that this is an interesting and practically
relevant approach.

\section{Preliminaries and notation}
\label{sec:prelim}

\paragraph{\textbf{CHCs as Intermediate Representation.}}
For generality, we will formulate our algorithm to work on a
block-level intermediate representation of the program, encoded using
(constrained) Horn
clauses.
A Constrained Horn Clause program (CHC), or Constraint Logic Program
(CLP), is a set of \emph{clauses} of the form $H
~\verb+:-+~ A_1, \ldots, A_n$, where $A_1, \ldots, A_n$ are \emph{literals} and $H$ is
an \emph{atom} said to be the \emph{head} of the clause.
For simplicity, and without loss of generality, we assume that each head
atom is normalized, i.e., it is of the form $p(x_1,\ldots,x_m)$ where $p$
is an $m$-ary predicate symbol and $x_1,\ldots,x_m$ are distinct
variables. However, in the examples we will sometimes show programs
unnormalized for brevity.
A set of clauses with the same head is called a \emph{predicate} (procedure). To
refer to predicates we will use normalized atoms and sometimes will call them 
\emph{predicate descriptors}.
A \emph{literal} is an atom or a \emph{primitive constraint} (which we
will also refer to as a \emph{built-in}).
A primitive constraint is defined by the underlying abstract domain(s) and is of
the form $c(e_1,\ldots,e_k)$ where $c$ is a $k$-ary predicate symbol and the
$e_1,\ldots,e_k$ are expressions.
For presentation purposes, the heads of the clauses of each predicate in the
program will be referred to with a unique subscript attached to their predicate
name (the clause number), and the literals of their bodies with dual subscripts
(clause number, body position), e.g., $\pdesc_k \mbox{\tt :-} \pdesc_{k,1}, \ldots
\pdesc_{k,n_{k}}$. The clause may also be referred to as clause $k$ of predicate $\pdesc$.
For example, for the following predicate, \verb+p/3+:

\noindent
\hspace*{10mm} \verb+p+$(X,Y,Z)$ \verb+:-+ $X$\ \verb+=<+\ $0,  Y$\ \verb+=+\ $Z.$ \\
\hspace*{10mm} \verb+p+$(X,Y,Z)$ \verb+:-+ $X$\ \verb+>+\ $0,
        X1$\ \verb+=+\ $X$\verb+-+$1, Y1$\ \verb+=+\ $Y$\verb+*+$X,\ $\verb+p+$(X1,Y1,Z).$ \\ [1mm]
\noindent
\verb+p/3+$_1$ denotes the head of the first clause of \verb+p/3+, and 
\verb+p/3+$_{2,1}$ denotes the first literal of the second clause of
\verb+p/3+, i.e., the constraint $X$\ \verb+>+\ $0$.

We assume that programs are converted to this Horn clause-based
representation, on a modular basis.
The conversion itself is beyond the scope of the paper (and
dependent on the source language).
It is trivially direct in the case of (C)LP programs or (eager) functional
programs, and for imperative programs we refer the reader to, e.g.,~\cite{%
  HGScam06,%
  decomp-oo-prolog-lopstr07,%
  jvm-pe-padl07,%
  big-small-step-vpt2020%
}.
In~\cite{fixpt-javabytecode-FTfJP07} the base algorithms that we extend in this
work were shown to be directly applicable to Java bytecode. In fact, Horn
clauses have since been used successfully as intermediate representations for
many different programming languages and compilation levels (e.g., bytecode,
llvm-IR, ISA, \ldots), in a good number of analysis and verification
tools~\cite{%
  BandaG08,%
  NMHLFM08,%
  resources-bytecode09,%
  DBLP:conf/tacas/GrebenshchikovGLPR12,%
  DBLP:conf/cav/JaffarMNS12,%
  AlbertAGPZ12,%
  DBLP:conf/sas/BjornerMR13,%
  isa-energy-lopstr13-final,%
  DBLP:conf/tacas/AngelisFPP14,%
  DBLP:conf/cav/GurfinkelKKN15,%
  DBLP:conf/birthday/BjornerGMR15,%
  isa-vs-llvm-fopara,%
  DBLP:conf/pldi/MadsenYL16,%
  z3,%
  kafle-cav2016,%
  resource-verification-tplp18,%
  resources-blockchain-sas20%
}
(see Sec.~\ref{sec:related_work} for other related work).
We note that some of these approaches
use the \emph{bottom-up} semantics on the CHC side, and then typically
the \emph{small-step semantics} in the translation, while others,
including ours, exploit the complementary approach of using the
\emph{top-down} semantics on the CHC side, and then
typically the \emph{big-step semantics} in the translation, but some
combine, e.g., big-step with 
bottom-up~\cite{DBLP:conf/cav/GurfinkelKKN15}. 
Big-step and small-step are nicknames often used to refer to, respectively,
Kahn's natural semantics~\cite{Kahn87} and Plotkin's structural
operational semantics~\cite{Plotkin1981,Plotkin04a}.
In the big-step semantics approach, the clause-based encoding is equivalent to a
block-based control flow graph, which is in turn a well-established
intermediate representation for program analysis. Each block is
represented by a clause, constraints or built-ins in a clause
represent the primitives of the language (bytecodes, machine
instructions, commands, etc.), literals represent calls to other
blocks, and predicates with multiple clauses implement alternatives
such as conditionals, case statements, dynamic dispatch, etc.\ (see,
e.g.,~\cite{decomp-oo-prolog-lopstr07,resource-verification-tplp18}).
This approach is particularly well-suited for programs with structured
control flow, although program transformations
allow supporting other program structures.
See~\cite{big-small-step-vpt2020} for a recent overview of the subject. 
In the following we revisit the \emph{top-down} semantics,
and establish our baseline. 

\paragraph{\textbf{Selection of the Concrete Semantics.}}

The semantics of CHC programs that we use as starting point is
goal-dependent (i.e., query-dependent, or ``top-down''), and based on
SLD-resolution~\cite{Robinson65}, and its generalization to Constraint
Logic Programming (CLP)~\cite{jaff87,intro_constraints_stuckey}, where
constraint domains and constraint solving extend the domain of
Herbrand terms with unification.
The traditional description of this resolution
procedure~\cite{Lloyd87,Apt90,jaff87} builds a tree structure in which
the nodes contain \emph{resolvents}. However, when used as a basis for
top-down program analyses, this construction is typically adorned so
that nodes in the resolution tree include representations of the
constraints both before and after completing the branch in which they
appear. These are then called the \emph{call} and \emph{success}
states for that node.
This is because the aim of goal-directed, top-down program analysis is
usually to obtain information on the constraints before and after each
program point.
This idea of storing call and success states is present for example in
the notion of \emph{generalized and trees} of~\cite{bruy91}. However,
such trees only describe the \emph{successful} derivation trees, i.e.,
a query that eventually fails will have an empty tree.
In practice it is useful to generalize this notion to collect also
those parts of the execution trees that lead to false, i.e., the
\emph{calls} made to predicates in the program also during
computations that eventually fail or loop, as
in~\cite{mcctr-fixpt,ai-jlp}.
We will refer to these trees simply as \andtrees.
It is also often interesting to consider trees with also {\sc or}
nodes, i.e., \emph{{\sc and-or}
  trees}, %
rather than considering sets of \andtrees, to capture
analyses such as determinacy~\cite{determinacy-ngc09,KLG06:ICLP},
cardinality~\cite{cardinality-ilps94},
non-failure~\cite{non-failure-iclp97}, etc., but for simplicity we
limit the discussion herein to semantics based on \andtrees.

\vspace*{-4mm}
\paragraph{\textbf{Concrete Semantics.}}

An \andtree represents the execution of a \emph{query}
(corresponding to one of more entry points to the program),
and each node in such a tree represents a call to a predicate,
adorned on the left with the state for that call, and on the right
with the corresponding success state. The concrete semantics of a
program $P$ for a given set of queries $Q$, \csemantics, is the set of
\andtrees\ that represent the execution of the queries in $Q$ for
$P$. Queries are of the form $Q = \langle \pdesc,\theta^c\rangle$ where
$\pdesc$ is a normalized atom corresponding to a predicate in the
program and $\theta^c$ is the \emph{calling} or \emph{initial
  constraint}. Nodes in an \andtree are of the form \andnode[\pdesc],
where $\pdesc$ is the call to a predicate $p$ in $P$, and
$\theta^c, \theta^s$ are, respectively, the call and success constraints over
the variables of $\pdesc$.
Nodes that are part of failing (or looping) branches (i.e., that never
``return'') will have empty success fields: $\langle\pdesc,\theta^c,\emptyset\rangle$.
The \emph{calling context} of a predicate given by the predicate descriptor $\pdesc$
defined in $P$ for a set of queries $\Q$ is the set
$\callingcontext(A,P,\Q)
= \{ \theta^c\ |\ \exists T
\in \csemantics\ s.t.\ \exists \andnode[\pdesc']\ in\ T \wedge
\exists\sigma %
\pdesc' = \sigma(\pdesc)\}$, where $\sigma$ is a \emph{renaming} substitution
over variables in the program, i.e., a substitution that replaces each variable
in the term it is applied to with distinct, fresh variables. In the following we
will use $\sigma$ to denote such renaming substitutions.
We denote by
$\answers(P,\Q)$ the set of answers (success constraints) computed by $P$ for
queries $\Q$, i.e., $\answers(P,\Q)$ is 
$\{ \theta^s |\ s.t.\ \exists T \in \csemantics \wedge \langle\pdesc, \theta^c, \theta^s\rangle = root(T) \}$.

\paragraph{\textbf{Modular Partitions of Programs.}}

A partition of a program is said to be modular when its source code is
distributed in several source units, each defining its interface with
other such units of the program. We will refer to these units as
\emph{modules}.  The interface of a module contains the names of the
predicates it exports and the names of the modules it imports.
Modular partitions of programs may be synthesized, or specified by the
programmer, for example, via a strict module system, i.e., a system in which
modules can only communicate via their interface.
We will use $M$ and $M'$ to denote modules. Given a module $M$:
\begin{itemize}
\item $\exports{M}$ denotes the set of predicate names exported by module $M$,
\item $\imports{M}$ is the set of modules which $M$ imports, and
\item $\mod{A}$ denotes the module in which the predicate
  corresponding to atom $A$ is defined. We sometimes abuse notation
  and denote the module of a query as $\mod{\Q}$, to refer ot the
  module of the predicate called in the query, i.e., if
  $\Q=\apair{\pdesc}{\CP}$ then $\mod{\Q}=\mod{\pdesc}$.
\vspace*{-2mm}
\end{itemize}

\section{Analysis graphs in goal-dependent abstract interpretation}
\label{sec:angraphs}

In this section we present the main abstraction object that is used in
goal-dependent abstract interpretation: the analysis graph. Later
sections will address the procedures for constructing such graphs.

\vspace*{-2mm}
\paragraph{\textbf{Program Analysis by Abstract Interpretation.}}

Abstract Interpretation~\cite{Cousot77} is a technique for static program
analysis in which the execution of the program is simulated on an abstract domain
(\DD) which is simpler than the concrete domain (\CD). Values in the abstract
domain and sets of values in the concrete domain are related via a pair of
monotonic mappings $\langle \alpha, \gamma \rangle$: {\em abstraction} $\alpha:
\CD \rightarrow$ \DD, and {\em concretization} $\gamma:$ \DD $\rightarrow \CD$
which form a Galois connection. An abstract value $d \in \DD$
\emph{approximates} a concrete value $c \in \CD$ if $\alpha(c) \sqsubseteq d$
where $\sqsubseteq$ is the partial ordering on \DD.
We refer to these abstract values interchangeably as \emph{descriptions} or
\emph{patterns}.
The correctness of abstract
interpretation guarantees that the descriptions inferred (by computing a
fixpoint through a Kleene sequence) approximate all the actual values or
traces which occur during any possible execution of the program, and that this
fixpoint computation process will terminate given some conditions on the
description domains (such as being finite, or of finite height, or without
infinite ascending chains) or by the use of a {\em widening}
operator $\nabla$~\cite{Cousot77}.

\paragraph{\textbf{Abstract Domain Operations for the Algorithms.}}

The abstract interpretation-based algorithms that we will present are
all \emph{parametric on the abstract domain}, i.e., they are
independent of the (data-)abstractions used. 
Each such abstract domain is defined by providing: the basic
operations of the domain lattice mentioned above
($\sqsubseteq, \sqcap, \sqcup$ and, optionally, the widening $\nabla$ operator); the abstract
semantics (\emph{transfer functions}, $f^\alpha$) of the constraints (representing
the \emph{built-ins}, or basic operations of the source language); and
the following additional instrumental operations, 
following~\cite{incanal-toplas}:
\begin{itemize}
\item {\tt
    \aproject}$(\lambda,\mathit{Vs})$: restricts the abstract constraint to
  the set of variables $\mathit{Vs}$.
\item {\tt \aextend}$(\pdesc_{k,n},\PP,\AP)$: 
  propagates the information in the success abstract constraint
  $\AP$, which is defined over the
  variables of $\pdesc_{k,n}$, to an abstract constraint $\PP$ that
  includes all the variables of the clause $\pdesc_k$.
\item \acall$(\lambda,\pdesc,\pdesc_k)$: performs the abstract
  unification (conjunction) 
  of predicate descriptor $\pdesc$ with the head of
  clause $\pdesc_k$, including in the new constraint abstract values
  for the variables in the body of clause $\pdesc_k$.
\item {\tt \alub}$(\lambda,\{\lambda_i\})$:
  joins $\lambda$ together with the set of abstract constraints
  $\{\lambda_i\}$. %
  To produce an abstract constraint that is greater or equal than
  $\lambda$. It will either perform the least upper bound ($\sqcup$) or the
  widening operation over the set together with $\lambda$, depending on
  termination or performance needs, typically determined by the
  abstract domain.\footnote{The implementation of the classical
    algorithm includes options for activating or deactivating
    multivariance on calls and also on success. We leave the latter
    out herein for simplicity; however, our results also apply since
    this is equivalent to turning the affected domains into power
    domains.}
\end{itemize}

\paragraph{\textbf{Graphs and paths.}}

We denote by $\graph$ a finite \emph{directed graph} (henceforward called simply
a graph) where $V$ is a set of nodes and $E \subseteq V\times V$ is an edge
relation, denoted with $u \rightarrow v$. A \emph{path} $P$ is a sequence of
edges $(e_1,\ldots,e_n)$ and each $e_i = (x_i,y_i)$ is such that $x_1 = u$, $y_n
= v$, and for all $1 \leq i \leq n - 1$ we have $y_i = x_{i+1}$. We also denote
paths with $u \rightsquigarrow v \in G$. We use $n \in P$ and $e \in P$ to
denote, respectively, that a node $n$ and an edge $e$ appear in a path $P$.

\paragraph{\textbf{Analysis graphs.}}
\label{sec:ags}

We perform \emph{goal-dependent abstract interpretation}, whose result is an
\emph{abstraction} of the \andtree semantics, \csemantics.
The discussion essentially follows the PLAI
algorithm~\cite{mcctr-fixpt,ai-jlp}, using the presentation
of~\cite{incanal-toplas}.
The purpose of this abstraction is to represent as a finite object the (possibly
infinite) set of (possibly infinite) \andtrees in \csemantics. As mentioned
before, the abstract interpretation technique guarantees
that this process terminates and that the \emph{concretization} of the resulting
abstraction will be a safe \mbox{(over-)}approximation of the \andtrees of the
concrete semantics.

The \emph{input} to this abstract interpretation process is a program $P$, an
abstract domain $\DD$, and a set of initial \emph{abstract queries} $\QQ =
\{\CDP{\pdesc_i}{\CP_i}\}$, where each $\pdesc_i$ is a normalized atom, and
$\CP_i \in \DD$.
$\QQ$ defines the (typically infinite) set of concrete queries $\Q$
that the analysis will be valid for. With some abuse of
notation we represent this set as $\gamma(\QQ)$, i.e., 
$\Q = \gamma(\QQ) = \{\CDP{\pdesc}{\theta} \mid\ \theta\in\gamma(\lambda) \wedge \CDP{\pdesc}{\lambda}
\in \QQ\}$.
This also determines the concrete semantics to be safely approximated, which is
then the set of \andtrees $\csemantics = \llbracket P \rrbracket_{\gamma(\QQ)}$.

An \emph{analysis result} is a call graph and a mapping function from predicate
descriptors and call descriptions %
to answer descriptions, both elements of $\DD$.
We also call this structure an {\em analysis graph}.

A \emph{node} in an analysis graph represents that a call to a predicate
($\apair{\pdesc}{\lambda^c}$) 
is possibly made, and it has an associated answer $\AP$, through the
mapping, $\tuple{\pdesc}{\CP}{\AP}$, with $\CP, \AP \in \DD$. 
This represents that \emph{the answer pattern for calls to
  predicate $\pdesc$ with calling pattern $\CP$ is $\AP$},
and it implies that %
for any node in the concrete
trees in \csemantics\ of the form \andnode[\pdesc], there must
exist a node $\tuple{\pdesc}{\CP}{\AP}$ in the analysis graph such
that $\theta^c \in \gamma(\lambda^c)$ and $\theta^s \in \gamma(\lambda^s)$. Therefore, analysis graphs
must capture all the call--success pairs, i.e., all the nodes in the
\andtrees of the concrete semantics (these 
conditions are formulated more precisely in Sec.~\ref{sec:mcorrect}).
For a given predicate $\pdesc$, the analysis graph may contain more
than one node capturing different call situations.
As usual, $\top$ denotes the most general abstract description, which is
equivalent to ``I do not know,'' 
and $\bot$ denotes the abstract description such that $\gamma(\bot)=\emptyset$.
A call mapped to $\bot$ $(\tuple{\pdesc}{\CP}{\bot})$ indicates that all calls
to predicate $\pdesc$ with description $\theta\in\gamma(\CP)$ either fail or
loop, that is, they never succeed.

An \emph{edge} in an analysis graph is of the form
$\arc{\pdesc}{\CP}{k,i}{\qdesc}{\CP'}$.
This represents that \emph{calling predicate $\pdesc$ with calling pattern
  $\CP{}$ may cause predicate $\qdesc$ to be called (via the literal
  $\pdesc_{k,i}$) with calling pattern $\CP'$}.
Correctness with respect to the concrete semantics requires that 
if in any concrete tree in \csemantics\ the clause $\pdesc_k$ is executed
with a calling pattern
$\theta^c$ that causes predicate $\qdesc$
(the literal $\pdesc_{k,i}$) to be called with some calling pattern
$\theta^{c'}$, then there must be an edge in the graph 
$\arc{\pdesc}{\CP}{k,i}{\qdesc}{\CP'}$ and 
$\theta^c \in \gamma(\lambda^c),\ \theta^{c'} \in \gamma(\lambda^{ c'})$.
These edges capture the dependencies between the immediate calls of a
predicates, i.e., given a node in the tree, the immediately following
nodes.
For simplicity, in the rest of the paper we omit $k,i$ when not relevant in the context.

\begin{figure}[t]
    \begin{minipage}{0.23\linewidth}
      \prettylstciao
\begin{lstlisting}
main(Msg, P) :-
    par(Msg, 0, P).

par([], P, P).
par([C|Cs], P~$_0$~, P) :-
    xor(C, P~$_0$~, P~$_1$~),
    par(Cs, P~$_1$~, P).

xor(0,0,0).
xor(0,1,1).
xor(1,0,1).
xor(1,1,0).
\end{lstlisting}
  \end{minipage}
  \hspace{2mm}
  \begin{minipage}{0.3\linewidth}
    \begin{tikzpicture}[>=stealth,on grid,auto]
      \node(top) at (6.5, -1) {$\top$}; \node(par) [below of=top] {$\apar$ (bit)};
      \node(zero) [below left of=par] {$z$ $(0)$}; \node(one) [below right
      of=par] {$o$ $(1)$}; \node(bot) [below right of=zero] {$\bot$}; \draw(top)
      -- (par); \draw(par) -- (one); \draw(par) -- (zero); \draw(zero) -- (bot);
      \draw(one) -- (bot);

      \tikzagconf
      \node (A) at (0,0)[very thick,text width=2cm] {\tuple{{\tt
            main}(\mathit{Msg},P)}{(\mathit{Msg}/\top,P/\top)}{(\mathit{Msg}/\top,P/\apar)}};
      \node (C) at (0,-2)[inner sep=1pt,text width=2.7cm] {\\[-2mm] {\bf {\tiny \fbox{1}}}\\[1mm] \tuple{{\tt
            par}(\mathit{Msg},X,P)}{(\mathit{Msg}/\top,X/z,P/\top)}{(\mathit{Msg}/\top,X/z, P/\apar)}};
      \node (E) at (0,-4.5)[text width=2.7cm] {\tuple{{\tt
            par}(\mathit{Msg},X,P)}{(\mathit{Msg}/\top,X/\apar,P/\top)}{(\mathit{Msg}/\top,X/\apar, P/\apar)}};
      \node (D) at (3.8,-1) [inner sep=4pt] {\tuple{{\tt
            xor}(C,P_0,P_1)}{(C/\top,P_0/z,P_1/\top)}{(C/\apar,P_0/z,P_1/\apar)}};
      \node (F) at (4,-4) [text width=63pt,inner sep=4pt] {\tuple{{\tt
            xor}(C,P_0,P_1)}{(C/\top,P_0/\apar,P_1/\top)}{(C/\apar,P_0/\apar,P_1/\apar)}};

      \path [every node/.style={sloped,anchor=south,auto=false}]
      (A) edge [->] node {1,1} (C)
      (C) edge [->] node {2,1} (D)
      (C) edge [->] node {2,2} (E)
      (E) edge [->] node {2,1} (F)
      (E) edge [->,in=70,out=50,loop,distance=0.6cm] node {2,2} (E);
    \end{tikzpicture}
  \end{minipage}
  \caption{A program that implements a parity function and a possible analysis
    result.\label{fig:mono}}
  \figendsp
\end{figure}

\begin{figure}[t]
  \begin{center}
    \begin{tabular}{cc}
    \includegraphics[width=0.5\textwidth]{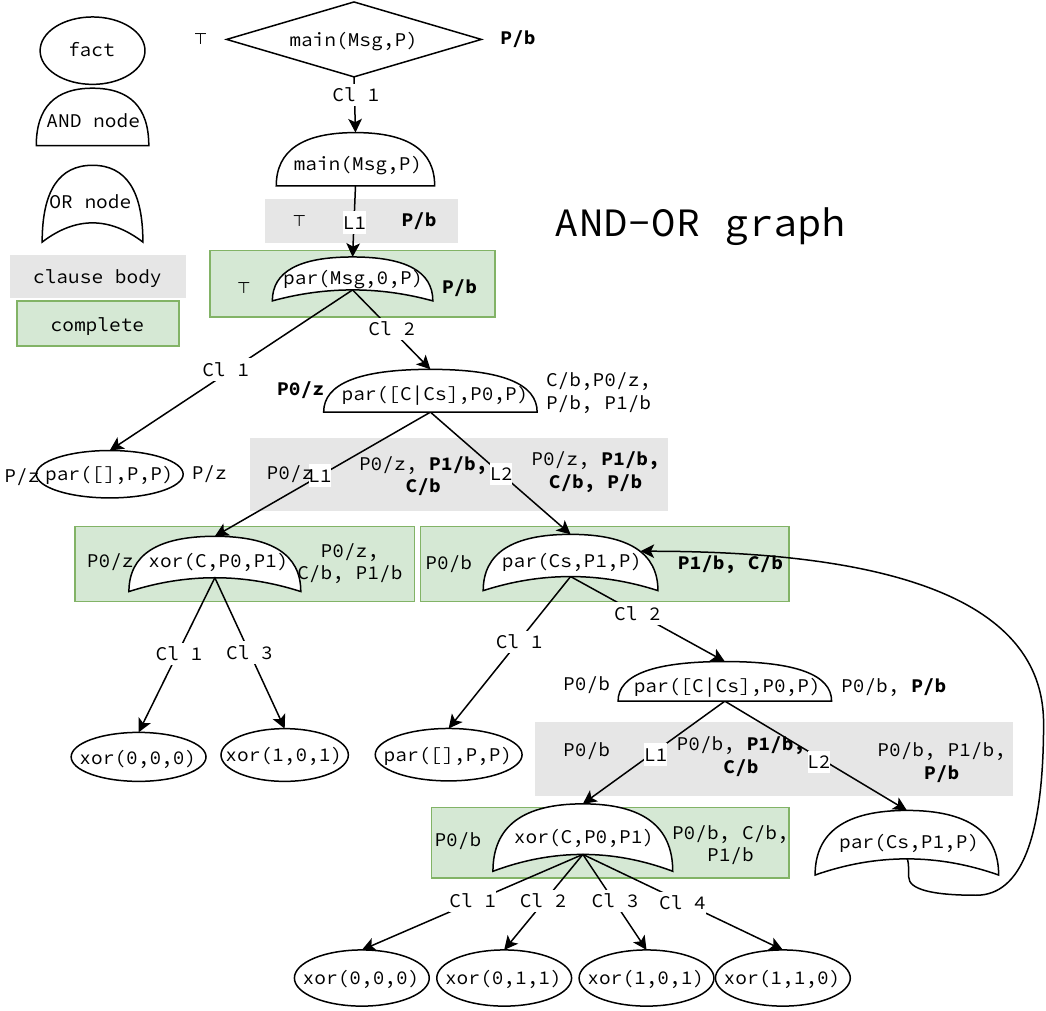}
    & 
    \includegraphics[width=0.5\textwidth]{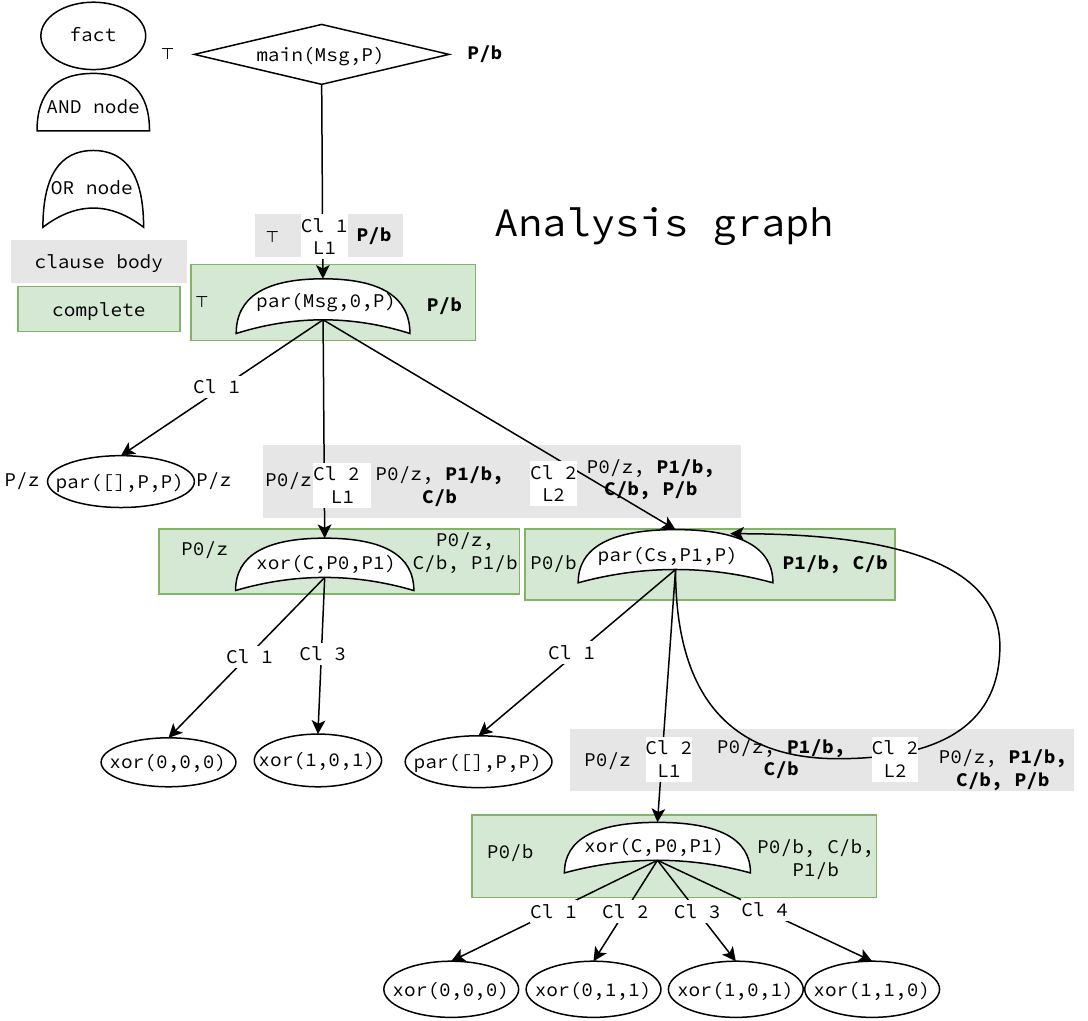}\\
    \end{tabular}
  \figendsp
    \caption{The analysis graph vs.\ the and-or graph -- compacting representation.\label{fig:graphs}}
  \end{center}  
  \figendsp
\end{figure}

\begin{example}
\noindent
  Fig.~\ref{fig:mono} shows an analysis graph (center) for a program
  that computes the parity of a message (left) with an abstract domain that
  infers for each variable whether it takes values of 0 or 1 (right)
  and initial abstract query 
  $\QQ = \{\apair{{\tt main}(\mathit{Msg},P)}{(\mathit{Msg}/\top,P/\top)}\}$. In the examples we
  will mark with a bold outline the initial nodes (i.e., the 
  nodes in $\QQ$). %
  \emph{Node} {\scriptsize \fbox{1}} (\tuple{{\tt par}(\mathit{Msg},X,P)}{(\mathit{Msg}/\top,X/z,P/\top)}{(\mathit{Msg}/\top,X/z,
    P/\apar)}) captures that {\tt par/3} may be
  called with $X$ bound to any in $\gamma(z) = \{0\}$ and, if it succeeds, the third
  argument $P$ will be bound to any of $\gamma(b) = \{1, 0\}$.
  Note that a different node (the one below) captures that there are
  other calls to {\tt par} where $X/z$ holds. 
  The edges in the graph represent the
  $\arc{\pdesc}{\CP}{k,i}{\qdesc}{\CP'}$ relation. For example, two
  such edges exist starting at node {\scriptsize \fbox{1}}, denoting
  (right) that it may call {\tt xor/3} and (below) that it may call
  itself with a different call description.
  Fig.~\ref{fig:graphs} illustrates for the example in Fig.~\ref{fig:mono}, the evolution
  from \emph{{\sc and-or} graphs} (left) to the compact representation
  of the analysis graphs: \emph{{\sc and} nodes} are made implicit
  (right) by keeping the references to the clauses and literals. The information in the \emph{{\sc and-or} graph} can be
  reconstructed by renaming and projecting abstract descriptions of the analysis
  graph, which %
  keeps the information
  only at the predicate and literal level.
  Last, please note that although in this simple example we are
  using a domain with a simple structure of tuples of
  $\emph{Variable}/\emph{AbstractValue}$ pairs, the domain structure
  can be arbitrary and in particular includes \emph{relational}
  domains.
\end{example}

\paragraph{\textbf{Multivariance (a.k.a., context- and path-sensitivity).}}
\label{sec:context-sensitivity}

As seen in the example, these analysis graphs allow representing the
different call patterns encountered during the execution, separating
the cases in which such calls differ, even if some of them subsume
others.
This feature is traditionally referred to as
\emph{multivariance} in the context of logic program analysis,
and, in our context, it serves two purposes:
\begin{enumerate}
\item \emph{Precision:} Different calling patterns to the same
  predicate are stored depending from which exact clause and
  literal this predicate is called from and with which call
  pattern. This idea of storing multiple calling contexts in this way
  is used in recent implementations of context sensitivity in
  imperative program analyses
  (e.g.,~\cite{DBLP:conf/cc/KhedkerK08,Thakur2020}) where it is
  referred to as keeping \emph{multiple value contexts}.
\item \emph{Efficiency:} For the same literal and clause in the
  program, storing different calling patterns allows keeping the
  fixpoint computation localized to only those patterns that change.
\end{enumerate}

While beyond the scope of this paper, note also that
multivariance is a form of \emph{multiple specialization} of
predicates. For example, the graph in Fig.~\ref{fig:mono} contains two
\emph{versions} of predicate \texttt{par/3} and another two of
\texttt{xor/3}, and implies the specialization shown in
Fig.~\ref{fig:versions}.
This is referred to as \emph{materializing} the versions in the
analysis graph~\cite{ai-jlp}. %

\begin{figure}[t]
\begin{center}
\begin{minipage}{0.7\textwidth}
  \prettylstciao

\begin{lstlisting}
%
main(Msg, P) :-
    par_1(Msg, 0, P).

%
par_1([], P, P).                
par_1([C|Cs], P~$_0$~, P) :-    
    true([Msg/T,P/T]),
    xor_1(C, P~$_0$~, P~$_1$~).    
    par_2(Cs, P~$_1$~, P).      

%
par_2([], P, P).                             
par_2([C|Cs], P~$_0$~, P) :-            
    xor_2(C, P~$_0$~, P~$_1$~),  
    par_2(Cs, P~$_1$~, P).            

%
xor_1(0,0,0).                   
xor_1(0,1,1).           %
xor_1(1,0,1).           %
xor_1(1,1,0).           %

%
xor_2(0,0,0).                       
xor_2(0,1,1).
xor_2(1,0,1).
xor_2(1,1,0).
\end{lstlisting}
\end{minipage}
\end{center}
\figendsp
\figendsp
\caption{The program specialization implicit in the analysis, after 
  \emph{version materialization.}\label{fig:versions}}
\figendsp
\end{figure}

\paragraph{\textbf{Reconstructing the paths of concrete executions.}}
\label{sec:path-sensitivity}
The analysis graph, through the \emph{edges}
($\arc{\pdesc}{\CP}{k,i}{\qdesc}{\CP'}$) relation, also provides an abstraction
of the \emph{paths} explored by the concrete executions through the program,
represented by the concrete trees.
In particular, it is possible to reconstruct,
for every node, all possible (and possibly infinite) execution trees
that lead to the call 
pattern described by the node, by following the edges of the analysis
graph. 
The analysis graph thus embodies two different abstractions (two
different abstract domains): the graph itself is a \emph{regular
  approximation} of the paths through the program, using a domain of
regular structures.
Separately, the abstract values (call and success patterns) contained
in the graph nodes are finite representations of the states occurring
at each point in the program paths, by means of the \emph{data
  abstract domain}.
Note that the path abstraction implicit in the graph is more powerful than the
call stack representation in the well known call-strings method introduced
of~\cite{sharir1978two} (see, e.g.,~\cite{DBLP:conf/cc/KhedkerK08,Thakur2020}
for two recent examples of use),
as this method  
only keeps track of the \emph{callers} of the abstracted call, and
typically as a limited-length sequence~\cite{sharir1978two}, whereas
we infer, as a regular tree, all the arbitrarily large sequences of
procedures \emph{executed} before that call, i.e, not only its direct
callers or a limited-depth sequence.
Note also that, as mentioned before, our analysis includes also the call
patterns and paths leading to failure or non-termination in the concrete
semantics (for all of which the answer pattern will be $\bot$ (s.t.\
$\gamma(\bot)=\emptyset$).

\vspace*{-3mm}
\paragraph{\textbf{Notation for and operations on analysis results.}}

The following operations defined over an analysis result $g$ allow us to inspect
and manipulate analysis results.

\begin{longtable}{r p{9.5cm}}
  $\apair{\pdesc}{\CP} \in g:$ & there is a node in the call graph of $g$
  with key $\apair{\pdesc}{\CP}$. \\
  $\tuple{\pdesc}{\CP}{\AP} \in g:$ & there is a node in $g$ with key
  $\apair{\pdesc}{\CP}$ and the answer mapped to that call is $\AP$.\\
  $\arc{\pdesc}{\CP}{}{B}{\CP'} \in g:$ & there are two nodes ($n
  = \apair{\pdesc}{\CP}$ and $n' = \apair{\qdesc}{\CP'}$) in $g$ and there is an edge
                                      from $n$ to $n'$. \\
  {\sf del}$(g,\{n_i\}):$ &  removes from $g$ nodes
  $n_i$ and its incoming and outgoing edges and unsets the element in the mapping function (it becomes
                            undefined for all $n_i$). \\
  \update$(g, \tupleu{\apair{\pdesc}{\CP}}{\AP}):$ & 
  overwrites the value of $\apair{\pdesc}{\CP}$ in the mapping
  function and, if necessary, adds a node to $g$ with key $\apair{\pdesc}{\CP}$. \\
  \update$(g, \{n \rightarrow n'\}):$ & 
  adds an edge from node $n$ to node $n'$  if it did not exist. \\ 
  \update$(g,\{e_i\}):$ & performs {\sf upd}$(g, e_i)$
                               for each element of $\{e_i\}$. \\
\end{longtable}

  \begin{figure}[t]
    \begin{center}
      \begin{tikzpicture}[>=stealth,on grid,auto]
      \tikzagconf
      \node (A) at (0,0)[very thick] {\tuple{{\tt main}(M,P)}{(M/\top,P/\top)}{(\cancel{(M/\top,P/\apar)}) \ \allowdisplaybreaks (M/\top,P/\top)}};
      \node (C) at (0,-2)[] {{\tt (1)} \tuple{{\tt par}(M,X,P)}{(M/\top,X/z,P/\top)}{(M/\top,X/z, P/\apar)}};
      \node (E) at (0,-4)[] {\tuple{{\tt par}(M,X,P)}{(M/\top,X/\apar,P/\top)}{(M/\top,X/\apar, P/\apar)}};
      \node (D) at (3.5,-1) [cross out] {\tuple{{\tt xor}(C,P_0,P_1)}{(C/\top,P_0/z,P_1/\top)}{(C/\apar,P_0/z,P_1/\apar)}};
      \node (F) at (3.5,-3.5) [] {\tuple{{\tt xor}(C,P_0,P_1)}{(C/\top,P_0/\apar,P_1/\top)}{(C/\apar,P_0/\apar,P_1/\apar)}};

      \path [every node/.style={sloped,anchor=south,auto=false}]
      (A) edge [->] node {} (C)
      (C) edge [->,cross out] node {} (D)
      (C) edge [->] node {} (E)
      (E) edge [->] node {} (F)
      (C) edge [->,dashed] node {new} (F)
      (E) edge [->,in=70,out=50,loop,distance=0.6cm] node {} (E);
    \end{tikzpicture}
  \caption{Graph after the modification operations.\label{fig:modgraph}}
  \end{center}  
\end{figure}

\begin{example}
  To illustrate the graph operations we show some examples of operations done to
  the analysis graph of Fig.~\ref{fig:mono}, that we will refer to
  with $\A$.
    \begin{itemize}[leftmargin=5mm,rightmargin=-3mm]
    \item Check if there is a call to \texttt{par/3} with the second argument as 0:\\
      $\apair{{\tt par}(M,X,P)}{(M/\top,X/z,P/\top)} \in \A$. This is true
      (node \fbox{\texttt{1}}).
    \item Check if there is a call to \texttt{main/2}, that, if it
      succeeds the second argument is a \emph{bit}:
      $\tuple{{\tt main}(M,P)}{\lambda^c}{(M/\top,P/\apar)} \in \A$. This is true
      (entry node).
    \item Check if there is a literal with \texttt{xor/3} in any of
      the clauses of \texttt{main/2}:
      $\arc{{\tt main}(M,P)}{\_}{}{{\tt xor}(C,P_0,P_1)}{\_} \in
      \A$. This is false, there is a path from \texttt{main/2} to
      nodes containing \texttt{xor/3} but there is not a direct call.
    \end{itemize}
  These operations do not modify the graph.
\end{example}

\begin{example}
  To illustrate the graph modification operations we show some examples of operations done to
  the analysis graph of Fig.~\ref{fig:mono}, %
  referred to again with $\A$.
  \begin{itemize}[leftmargin=5mm,rightmargin=-3mm]
  \item Remove the node for the abstract call $\apair{{\tt
        xor}(C,P_0,P_1)}{(C/\top,P_0/z,P_1/\top)}$: \\
    {\sf del}$(\A,\{\apair{{\tt xor}(C,P_0,P_1)}{(C/\top,P_0/z,P_1/\top)}\})$.
  \item Update the node for \texttt{main/2} with a more general success pattern: \\
    ${\sf upd}(\A, \tuple{{\tt main}(M,P)}{(M/\top,P/\top)}{(M/\top,P/\top)})$.
    
  \item Add an edge from node \fbox{\texttt{1}} to the remaining node for
    \texttt{xor/3}: \\ ${\sf upd}(\A, \{\apair{{\tt par}(M,X,P)}{(M/\top,X/z,P/\top)}
    \rightarrow \apair{{\tt xor}(C,P_0,P_1)}{(C/\top,P_0/\apar,P_1/\top)}\})$.
  \end{itemize}
  After these operations, the state of the analysis graph is depicted
  in Fig.~\ref{fig:modgraph}.
\end{example}

\section{The baseline analysis algorithms}
\label{sec:base-algs}

The \emph{analysis algorithms} are the fixpoint-calculating procedures
that infer the analysis graphs, described in the previous section, so
that they safely approximate the given program semantics.
\emph{Incremental} algorithms are those that can modify and
recalculate such analysis graphs after program changes, without having
to start the process from scratch.  \emph{Modular} algorithms (in
contrast to \emph{monolithic} algorithms) are those that are capable
of analyzing a modular partition of a program (see
Sec.~\ref{sec:prelim}) without having to load or treat the whole
program at any given step.

In this section we present our baseline algorithms, which already
include some improvements with respect to previous descriptions, while
in Sec.~\ref{sec:intermod} we will present the incremental and modular
algorithm that is our main contribution.

\subsection{The monolithic and incremental fixpoint algorithm}
\label{sec:mon-inc}

We now present our first baseline, the monolithic incremental analysis
algorithm of \cite{incanal-toplas}, extended with widening to ensure
termination in the presence of infinite abstract domains. This 
algorithm (Fig.~\ref{fig:incanal_simp}) takes as input a program $P$,
a set of initial abstract queries $\QQ$, the differences $\Delta$ of
$P$ with respect to a previous version $P'$, and an analysis result
that is \emph{correct} for $P'$. We will refer to this algorithm with
$\A =\ $\textbf{\banalyze}$(P,\QQ,\diff{},\A_0)$. Note that if the
algorithm is called with $\A_0$ an empty analysis, i.e, \emph{from scratch},
then it is the same
as the traditional PLAI algorithm~\cite{ai-jlp}.
As mentioned before, we will refer to these to algorithms as \emph{monolithic}
because they assume that all the predicates executed in the target
program $P$ are provided to the analyzer, i.e., these algorithms treat
only whole programs.

\newcommand{\newfunction}{\vspace{1.2mm}}

\begin{figure}[!tb]
  \algrenewcommand\alglinenumber[1]{\scriptsize #1:}
  \hspace*{-6mm}
  \begin{tabular}{l l}
     \begin{minipage}{0.57\textwidth}
       \textsc{Algorithm} \textbf{\banalyze}$(P,\QQ,\diff{},\A)$
       \begin{algorithmic}[1]
         \newfunction
         \footnotesize

         \ForAll{$\ACP \in \QQ$}
         \State {\sf add-event}$(newcall(\CDP{\pdesc}{\CP}))$
         \EndFor

         \State {\tt delete\_clauses}$(\diff{})$ 
         \State {\tt add\_clauses}$(\diff{})$
         \State {\tt analysis\_loop}$()$
         \State \Return $\A$

         \newfunction
         \Procedure{\tt analysis\_loop}{$ $}
         \While {{\sf events}$() \neq \emptyset$}
         \State $E$ \assign {\sf next-event}$()$
         \State {\tt process}($E$)
         \EndWhile
         \EndProcedure

         \Procedure{\tt add\_clauses}{$\cls$}
         \ForAll{$\phead_k $ {\tt :-} $ \pbody_{k,1}, \ldots, \pbody_{k,n_k} \in \cls$}
         \ForAll{$\CDP{\phead}{\CP} \mapsto \AP \in \A$}
         \State  $\PP$ \assign \acall$(\CP, \pdesc, \phead_k)$
         \State  $\CP_1$ \assign \aproject$(\PP, vars(\pbody_{k,1}))$
         \State {\sf add-event}$(arc(\fulledge{\CDP{\pdesc}{\CP}}{}{\PP}{k}{1}{\CDP{\pdesc_{k,1}}{\CP_1}}))$
         \EndFor
         \EndFor
         \EndProcedure

         \newfunction
         \Procedure{\tt delete\_clauses}{$\cls$}
         \State {$\calls \assign \{\CDP{\pdesc}{\CP} | \CDP{\pdesc}{\CP} \in \A, (\phead_k ~\verb+:-+~ \ldots) \in \cls\}$}
         \State $\mathit{Ns}$ \assign\  $\{N \in \A | N \rightsquigarrow C \in \A, C \in \calls\}$
         \State {\sf del}$(\A, \mathit{Ns})$
         \EndProcedure

         \newfunction
         \Function{\tt lookup\_answer}{$\CDP{\pdesc}{\CP}$}
        \If {$\CDP{\pdesc}{\CP} \mapsto \AP \in \A$}
        \State \Return $\AP$
        \Else
        \State {\sf add-event}($newcall(\CDP{\pdesc}{\CDP{\pdesc}{\CP}})$)
        \State \Return $\bot$
        \EndIf
        \EndFunction
        \Procedure {\tt reanalyze\_updated}{$\CDP{\pdesc}{\CP}$}
        \ForAll{$E \assign \fulledge{\CDP{\qdesc}{\CP{_0}}}{}{\PP}{k}{i}{\CDP{\pdesc}{\CP}} \in \A$}
        \State {\sf add-event}($arc(E)$)
        \EndFor
        \EndProcedure

         \algstore{mainalg}
       \end{algorithmic}
     \end{minipage}
&
  \begin{minipage}{0.6\textwidth}
    \begin{algorithmic}[1]
      \footnotesize
      \algrestore{mainalg}
      \Procedure{\tt process}{$newcall(\CDP{\pdesc}{\CP})$}
      \ForAll{$\phead_k$ :- $\pbody_{k,1}, \ldots, \pbody_{k,n_k} \in \cls$}
      \State $\PP$ \assign \acall$(\CP, \pdesc, \phead_{k})$
      \State $\CP_1$ \assign \aproject$(\PP, vars(\pbody_{k,1}))$
      \State {\sf add-event}($arc(\fulledge{\CDP{\pdesc}{\CP}}{}{\PP}{k}{1}{~\CDP{\pdesc_{k,1}}{\CP_1}}$))
      \EndFor
      \State$\AP$ \assign {\sf initial-guess}($\CDP{\pdesc}{\CP}$)
      \If{$\AP \neq \bot$}
      \State {\tt reanalyze\_updated}($\CDP{\pdesc}{\CP}$)
      \EndIf
      \State \update$(\A, \CDP{\pdesc}{\CP} \mapsto \AP)$
      \EndProcedure

      \newfunction
      \Procedure {\tt process}{$arc(\fulledge{\CDP{\pdesc}{\CP{_0}}}{}{\PP}{k}{i}{\CDP{\qdesc}{\CP_1}})$}
      \State $\calls \assign\{\lambda\ |\ \fulledge{\CDP{\pdesc}{\_}}{}{}{k}{i}{\apair{\qdesc}{\lambda}} \in \A \}$ \label{line:calls}
      \State $\CP \assign$\alub$(\CP_1, \calls)$ \label{line:widencall}

      \If{$B$ is a \emph{built-in}} %
      \State $\AP_0 \assign \ f^\alpha(\CDP{\qdesc}{\CP})$
      \Else $\ \AP_0$ \assign {\tt lookup\_answer}($\CDP{\qdesc}{\CP}$)
      \EndIf
        
      \State \update$(\A, \fulledge{\CDP{\pdesc}{\CP{_0}}}{\RP}{\PP}{k}{i}{\CDP{\qdesc}{\CP}})$
      \State $\RP$ \assign \aextend($\PP,\AP{_0}$)
      \If {$\RP \neq \bot$ and $i \neq n_k$}
      \State $\CP_2$ \assign \aproject$(\RP, vars(\pbody_{k,i+1}))$
      \State {\sf add-event}($arc(\fulledge{\CDP{H}{\CP{_0}}}{}{\RP}{k}{i+1}{\CDP{B}{\CP_2}}$))
      \ElsIf {$\RP \neq \bot$ and $i = n_k$}
      \State $\AP$ \assign \aproject$(\RP, vars(\phead_k))$
      \State {\tt insert\_answer\_info}($\CDP{\pdesc}{\CP{_0}}, \AP$)
        \EndIf
        \EndProcedure

        \newfunction

        \Procedure{\tt insert\_answer\_info}{$\CDP{\pdesc}{\CP}, \AP$}
        \If{$\CDP{\pdesc}{\CP} \mapsto \AP{_0} \in \A$}
        \State $\AP{_1}$ \assign \alub$(\AP{},\{\AP{_0}\})$ \label{line:widensucc}
        \Else \ $\AP{_0} \assign\bot$, $\AP{_1} \assign\AP$
        \EndIf
        \If {$\AP{_0} \neq \AP{_1}$}
        \State \update$(\A, \CDP{\pdesc}{\CP} \mapsto \AP{_1})$
        \State {\sf reanalyze\_updated}($\CDP{\pdesc}{\CP}$)
        \EndIf
        \EndProcedure 
      \end{algorithmic}
    \end{minipage}
  \end{tabular}
  \caption{The monolithic, context-sensitive, incremental fixpoint
    algorithm.\label{fig:incanal_simp}}
\end{figure}

\paragraph{\textbf{\textbf{Operation of the algorithm.}}}

The algorithm is centered around processing two kinds of events: $newcall$
events, which control which predicates and clauses of the program that need reanalysis,
and $arc$ events, which process the body of one clause for a call pattern, starting at
a certain literal.
The algorithm starts by queueing a
$newcall$ event for each of the call patterns that need to be (re)computed. This
triggers \texttt{process}$(newcall(\ACP))$, which processes all the clauses of
predicate $\pdesc$. For each of them the abstract call is performed (\acall, which
includes the renaming) and an $arc$ event is added for the first
literal. The {\sf initial-guess} function returns a guess of the answer, $\AP$,
to $\ACP$. If possible, it reuses the results in $\A$, otherwise returns $\bot$.
Procedure \texttt{reanalyze\_updated} propagates the information of new computed
answers across the analysis graph by creating $arc$ events with the literals
from which the analysis has to be restarted.
\texttt{process}$(arc(\fulledge{\apair{\phead_k}{\CP}}{}{\PP}{k}{i}{\apair{\qdesc}{\CP}}))$
performs a single step of the left-to-right traversal of a clause body.
Since the algorithm is multivariant, an infinite number of different call
patterns may be encountered, even if the domain has finite height. In this case, the
call patterns are generalized, via a widening operator,
denoted by the $\alub$ operation.
Then, if the literal $\pdesc_{k,i}$ is a \textit{built-in}, its transfer
function is applied; otherwise, an edge is added to $\A$ and the $\AP$ is looked
up, which includes creating a $newcall$ event for $\apair{\pdesc}{\CP}$
if the answer is not in the analysis graph. The answer is combined with the
description $\PP$ from the literal immediately before $\pbody_{k,i}$ to obtain
the description (return) for the literal after $\pbody_{k,i}$. This is used
either to generate an $arc$ event to process the next literal, or to update the
answer of the predicate in \texttt{insert\_answer\_info}. This function combines
the new answer with the semantics of the previous answers. To ensure termination
when analyzing with abstract domains with infinite ascending chains, this answer
needs to be generalized, also with a widening operator ($\alub$). Lastly, the new answer is
propagated if needed.

Procedure \texttt{add\_clauses} adds $arc$ events for each of the new
clauses. These trigger the analysis of each clause and the later update of $\A$
by using the edges in the graph.

The \texttt{delete\_clauses} function selects the information to be kept in
order to obtain the most precise semantics of the program, by removing all
information which is potentially inaccurate (all the dependent nodes in the graph).

\paragraph{\textbf{Differences w.r.t. the original incremental algorithm.}}
\label{sec:alg-diff}

The algorithm presented in Fig.~\ref{fig:incanal_simp} differs from
the one described in~\cite{incanal-toplas} only in
lines~\ref{line:calls} and~\ref{line:widencall}, which perform the
widening of the encountered call patterns for the cases in which the
abstract domain has infinite width, and in line~\ref{line:widensucc},
that performs the widening on the success for the cases in which the
abstract domain is of infinite height. The abstract
interpretation technique guarantees that generalization with a widening
operation preserves soundness, and guarantees termination at the expense of losing
of precision.
Since widening may not be necessary for all domains, it may be disabled in the
algorithm by: %
\vspace*{-2mm}
\begin{itemize}\itemsep=0pt
\item removing line~\ref{line:calls}, 
\item replacing line~\ref{line:widencall} by ``$\CP \assign\CP_1$'', 
\item and replacing line~\ref{line:widensucc} by ``$\AP_1 \assign\AP \sqcup \AP_0$''.
\end{itemize}

\subsubsection{Correctness}\label{sec:mcorrect}

We now formulate the correctness results of the algorithm \emph{with
  generalization}, i.e., as presented in Fig.~\ref{fig:incanal_simp}.

\begin{definition}[Correctly approximated calls]\label{def:correct-call}
  Let $P$ be a program, $\Q$ a set of initial concrete queries, and \A an
  analysis graph. We say that $\A$ \emph{correctly approximates the calls in $\csemantics$}  if
  all encountered call patterns during the concrete execution are contained in
  $\A$. That is, for all predicates $\pdesc$ in $P$:
  $$
  \forall \theta^c \in
  \callingcontext(\pdesc,P,\Q). \exists \tuple{\pdesc}{\lambda^c}{\lambda^s}\in \A
  \text{ s.t.}\ \theta^c \in \gamma(\lambda^c).
  $$
\end{definition}

\begin{definition}[Correctly approximated answers]\label{def:correct-succ}
  Let $P$ be a program, $\Q$ a set of initial concrete queries, and \A an
  analysis graph. We say that \emph{the answers in $\A$ correctly approximate the answers
  in $\csemantics$} if they abstract all the answer patterns to the encountered
  call patterns. That is, for all predicates $A$ of $P$:
  $$
  \forall \tuple{\pdesc}{\lambda^c}{\lambda^s} \in \A, \forall
    \theta^c\in\gamma(\lambda^c) \text{ if } \theta^s \in \answers(P,
    \{\CDP{\pdesc}{\theta^c}\}) \text{ then }\theta^s\in\gamma(\lambda^s).
    $$
\end{definition}

\begin{definition}[Correct global analysis]\label{def:correct-ana}
  Let $P$ be a program, $\Q$ a set of initial concrete queries, and $\A$ an
  analysis graph. $\A$
  \emph{is correct for $P, \Q$} if
  \begin{itemize}[leftmargin=5mm,rightmargin=-3mm]
  \item[a)] \A correctly approximates the calls for $P$, $\Q$
    (Def.~\ref{def:correct-call}) and
  \item[b)] \A correclty approximates the answers for $P$, $\Q$ (Def.~\ref{def:correct-succ}).
  \end{itemize}
\end{definition}

Given these definitions, the following Theorems \ref{th:mon-scratch},
\ref{th:mon-add}, and~\ref{th:mon-del} from~\cite{incanal-toplas} hold,
because, as stated earlier, generalization via a widening guarantees
correctness:

\begin{theorem}[Correctness of \banalyze from scratch]\label{th:mon-scratch}
  Let $P$ be a program, and $\QQ$ a set of abstract queries.
  \noindent
  The analysis result $\A=\banalyze(P,\QQ,\emptyset,\emptyset)$ for $P$ with
  $\QQ$ is \emph{correct} for $P$ and $\gamma(\QQ)$.
\end{theorem}

\begin{theorem}[Correctness of \banalyze adding clauses]\label{th:mon-add}
  Let $P$ and $P'$ be two programs such that s.t.\ $\diff{} = (C_{add}, \emptyset)$, $P
  = (P' \cup C_{add})$, and $\QQ$ a set of abstract queries.
  If $\A_0 = \banalyze(P',\QQ,\emptyset,\emptyset)$, then
  the analysis result $\A = \banalyze(P,\QQ,\diff{},\A_0)$ for $P$
  with $\QQ$ \emph{correct} for $P$ and $\gamma(\QQ)$.
\end{theorem}

\begin{theorem}[Correctness of \banalyze deleting clauses]\label{th:mon-del}
  Let $P$ and $P'$ be two programs such that s.t.\ $\diff{} = (\emptyset, C_{del})$, $P
  = P' \setminus C_{del}$, and $\QQ$ a set of abstract queries.
  If $\A_0 = \banalyze(P',\QQ,\emptyset,\emptyset)$, then
  the analysis result $\A = \banalyze(P,\QQ,\diff{},\A_0)$ for $P$
  with $\QQ$ \emph{correct} for $P$ and $\gamma(\QQ)$.
\end{theorem}

We introduce a new theorem that generalizes Theorems~\ref{th:mon-scratch},
\ref{th:mon-add}, and~\ref{th:mon-del}.

\begin{theorem}[Correctness of \banalyze starting from a partial
  analysis]\label{th:inc-part}
  Let $P$ be a program, $\QQ$ a set of abstract queries, and $\A_0$ \emph{any
    analysis graph}.
  Let $\A = \banalyze(P,\QQ,\emptyset,\A_0)$. $\A$ is \emph{correct} for $P$ and
  $\gamma(\QQ)$ if for all concrete queries $Q \in \gamma(\QQ)$ all nodes $N$
  from which there is a path in the concrete execution $Q \rightsquigarrow N$ in
  \csemantics, that are abstracted in the analysis $\A_0$
  are included in $\QQ$, i.e.:
  $$
  \forall \Q,N. \Q \in \gamma(\QQ) \wedge Q \rightsquigarrow N \in
  \llbracket P \rrbracket_{Q}, \forall N_\alpha \in \A_0. N \in \gamma(N_\alpha)
  \implies N_\alpha \in \QQ.
  $$
\end{theorem}

Intuitively, the algorithm is correct for any query $Q$ not already abstracted in
$\A_0$. If $\A_0$ contains already information about $Q$, it needs to be
rechecked by recomputing the analysis of all the nodes in which $\Q$ depends by
including them in $\QQ$. Theorem~\ref{th:inc-part} is a generalization because, implicitly,
procedures {\tt add\_clauses} and {\tt delete\_clauses} are doing exactly, this:
either removing the analysis so that it is computed from scratch again or adding
the necessary queries (directly by creating the corresponding $newcall$ events)
to guarantee that the analysis is correct.

\begin{proof}
  This follows from the creation of a $newcall$ event for each of the queries
  $\QQ$. The processing of the events trigger the recomputation and later update
  of all the nodes of the analysis graph that are potentially under the
  fixpoint.
\end{proof}

Note that %
$\A_0$ is not assumed to be the (correct) output of a
previous analysis, it can be any analysis (below, above, or incomparable with the
fixpoint). Also
note that if all nodes in the analysis graph are included, together with the
original queries, in $\QQ$ the result
is guaranteed to be correct.

\subsubsection{Precision}

If generalization is removed from the algorithm, as indicated in
Sec.~\ref{sec:alg-diff}, and assume that {\sf initial-guess} returns a value
below the least fixed point, the following precision result
from~\cite{incanal-toplas} is preserved when analyzing with finite abstract
domains:

\begin{theorem}[Precision of \banalyze]\label{th:mana-prec}
  Let $P, P'$ be programs, such that $P$ differs from $P'$ by $\diff{}$, let
  $\QQ$ a set of abstract queries, and $\A_0 =
  \banalyze(P',\QQ,\emptyset,\emptyset)$ an analysis graph. The following hold:
  \begin{itemize}[leftmargin=5mm,rightmargin=-3mm]
  \item If $\A = \banalyze(P,\QQ,\emptyset,\emptyset)$, then $\A$ is the \emph{least
    program analysis graph} for $P$ and $\gamma(\QQ)$, and 
  \item $\banalyze(P,\QQ,\diff{},\A_0) = \banalyze(P,\QQ,\emptyset,\emptyset)$.
  \end{itemize}
\end{theorem}

\noindent
That is, when analyzing from scratch, always the most precise result is produced,
and when reusing a least program analysis graph in the incremental analysis, the
new result is the least program analysis graph as well. This means that there is
no analysis graph with smaller call or answer patterns that correctly
over-approximates the behavior of the program.

Theorem~\ref{th:mana-prec} shows that, if the $\A_0$ is a correct and precise
analysis, then the incremental analysis result is correct and precise. However,
the conditions on $\A_0$ can be relaxed if we strengthen the conditions on the
queries and still guarantee the same precision/correctness results. The
following new theorem states the general condition for guaranteeing precision
when (re)starting from a partial analysis result.

\begin{theorem}[Precision of \banalyze starting from a partial
  analysis]\label{th:prec-inc}
  Let $P$ be a program, $\QQ$ a set of abstract queries, $\A_0$ an analysis graph
  \emph{below the least fixed point (lfp)}, i.e., $\forall \tuple{A}{\CP}{\AP_0}
  \in \A_0.\tuple{A}{\CP}{\AP} \in \A \wedge \AP_0 \sqsubseteq \AP$, and the
  conditions on $\QQ$ of Theorem~\ref{th:inc-part} hold then:
  $$
  \banalyze(P,\QQ,\emptyset,\emptyset) = \banalyze(P,\QQ,\emptyset,\A_0).
  $$
\end{theorem}

\vspace*{-3mm}
\begin{proof}
  The abstract interpretation technique \cite{Cousot77}
  guarantees that the fixed point of a set of monotonic equations can be
  computed by repeatedly applying each of the equations in a chaotic iteration
  manner. %
  If the iteration is started at $\bot$, it is guaranteed that the least fixed
  point of the equations is found. In our case, the equations are the Horn
  clauses that encode the (concrete) semantics of the program $P$.
  Let $f_P(X)$ be one step of the chaotic iteration, i.e., \emph{applying
    semantics of one clause of $P$ to the current value of the sequence}. When
  starting from an empty analysis, \banalyze will compute the $\lfp$ by applying
  $f_P(X)$ a number of times:
  \[ \bot \sqsubseteq f_P(\bot) \sqsubseteq f_P(f_P(\bot)) \sqsubseteq f_P^3(\bot) \sqsubseteq
    \ldots \sqsubseteq f_P^k(\bot) = \ldots = f_P^{k+n}(\bot) = \lfp(P)
  \]
  
  \noindent
  In the sequence above, the fixpoint value is reached in the $k$-th step of the
  iteration. However, this value is not confirmed yet to be the fixpoint. The
  chaotic iteration process needs to continue until all the equations have been
  exhaustively applied and the value of the fixpoint is kept, this is represented by
  the $n$ steps after $f_P^k$.
  Note that the number of steps $k$ and $n$ will depend highly on the strategy
  for the chaotic iteration. In our case, we safely reduce them by keeping the
  dependencies between clauses.
  
  Starting from a partial analysis is equivalent to computing the Kleene
  fixpoint of the original program including a new equation, which is a
  constant, representing the initial results. Let us call this equation
  \newcommand{\initeq}{\ensuremath{\A_0}\xspace}%
  \initeq. %
  Our goal is to prove that chaotic iteration of $f_P$ with $\A_0$ also results
  in the $\lfp(P)$ if $\A_0 \sqsubseteq \lfp(P)$.

  By definition, for any $k$-th step of the iteration $f_P^k(\bot) \sqsubseteq
  \lfp(P)$, also, by hypothesis, $\A_0 \sqsubseteq \lfp(P)$. Therefore, for any
  $k$ and applying any random clause, $\A_0 \sqcup f_P^k(\bot) \sqsubseteq
  \lfp(P)$. So, if we ``plug in'' the initial analysis $\A_0$ at any point of
  the chaotic iteration over $f_P$, because the equations of $P$ are monotonic,
  for any $k$, $f_P(\A_0 \sqcup f_P^k(\bot)) \sqsubseteq f_P(\lfp(P))$, and
  precision is preserved. Concretely, this also implies that precision is
  preserved if we start from $f_P(\A_0)$.

  The condition imposed on the set of queries guarantees that the
  chaotic iteration includes all the equations that the iteration
  needs to be rerun with (see Theorem~\ref{th:inc-part}).
  This justifies not reprocessing the equations that are not affected
  by the changes in the algorithm, since the corresponding steps can
  be skipped safely.
\end{proof}

Note that these precision results imply also correctness since the
\emph{lfp} is obtained, which was already proved in Sec.~\ref{sec:mcorrect}. Nevertheless,
precision has been included separately because it does not hold in the presence of
generalization: using widening, as required for dealing with infinite domains, 
implies not being able to guarantee that the least fixed 
point is obtained, and given that this operator is not assumed to be
associative, also does not guarantee the analysis result will
be the same (i.e., that the same imprecision is obtained), as this
depends on how the processing of the events is scheduled.

\subsection{The modular fixpoint algorithm}\label{sec:mod}

We now present %
the reference algorithm for analyzing modular programs, 
described in~\cite{mod-an-lopstrbook}. As expected, the
approach consists in analyzing partitions of programs making assumptions
about the code that is external to each partition. Several
possibilities were proposed in that work for making such assumptions,
including, e.g., assuming that nothing is known about the answer
($\top$), computing the ``topmost'' abstraction of the call (as before
but taking into account any local information available), or
strategies with better precision but, in general, more costly, such as
assuming $\bot$ temporarily for the unknown answers and later
reanalyzing whenever a better abstraction of the answer is
available. In this work we fix the strategy to the latter one in order
to obtain the best precision. Also, module analysis order may affect
the speed at which the fixpoint computation converges.  Some
scheduling policies were studied in~\cite{modbenchmarks-lopstr05}.
We provide a new pseudocode for the algorithm of \cite{mod-an-lopstrbook},
specialized for the case in which the maximum precision is
aimed for. Then, we provide new formal results about correctness and
precision of this algorithm.
Also, both for generality and reusability, although not required for
our results, we propose a formulation of the algorithm
that is parametric on the analysis used within each modular partition,
which in our case is instantiated to \banalyze. 

\paragraph{\textbf{Modular analysis results.}}

To store the overall analysis result of the program and keep track of fine-grain
dependencies between modules, we propose to use also an analysis graph
structure at the inter-modular level. 
One can see this as a sort of ``projection'' of the \textit{monolithic} analysis graph,
described in Sec.~\ref{sec:ags}, in which only the information about the
predicates in the boundaries of the modules is kept. Nodes represent calls to
predicates and edges capture the relations between the predicates in the
boundaries of the partitions (exported/imported predicates) with arcs
\arc{A}{\CP}{}{B}{\CP'} meaning \emph{a call to $A$ in \mod{A} with description
  \CP\ may cause a call to $B$ with description $\CP'$ and $\mod{B} \in
  \imports{\mod{A}}$}. From this point on, we will use $\GAG$ to denote the
modular (global) analysis graph, and $\LAG$ to denote the analysis of a single
module (local analysis graph).

Fig.~\ref{fig:mod} shows a modular version of the program and analysis results
of Fig.~\ref{fig:mono}. The nodes of this (global) analysis graph 
encode that calling the exported predicate {\tt main/1} of module
{\tt main} may cause a call to {\tt xor/3} exported by module {\tt bitops} with
two different call descriptions (two edges).

\newcommand{\rcolumn}{3}
\newcommand{\lcolumn}{0}
\newcommand{\vlinep}{1.4}

\begin{figure}[t]
  \begin{minipage}{0.37\linewidth}
    \prettylstciao  
\begin{lstlisting}
:- module(main, [main/1]).

:- use_module(bitops).
main(Msg, P) :-
    par(Msg, 0, P).

par([], P, P).
par([C|Cs], P~$_0$~, P) :-
    xor(C, P~$_0$~, P~$_1$~),
    par(Cs, P~$_1$~, P).
\end{lstlisting}

\begin{lstlisting}
:- module(bitops, [xor/3]).

xor(0,0,0).
xor(0,1,1).
xor(1,0,1).
xor(1,1,0).
\end{lstlisting}
  \end{minipage}
  \hspace{+3mm}
  \begin{minipage}{0.3\linewidth}
    \begin{tikzpicture}[>=stealth,on grid,auto]
      \tikzagconf

      \node (A) at (\lcolumn+2,0) [rectangle, very thick] {\tuple{{\tt main}(M,P)}{(M/\top,P/\top)}{(M/\top, P/\apar)}};
      \node (D) at (\lcolumn,-2) [rectangle,text width=80pt] {\tuple{{\tt xor}(C,P_0,P_1)}{(C/\top,P_0/z,P_1/\top)}{(C/\apar,P_0/z,P_1/\apar)}};
      \node (F) at (\lcolumn+4,-2) [rectangle,text width=80pt] {\tuple{{\tt xor}(C,P_0,P_1)}{(C/\top,P_0/\apar,P_1/\top)}{(C/\apar,P_0/\apar,P_1/\apar)}};
      \path [->] (A) edge (D);
      \path [->] (A) edge (F);
    \end{tikzpicture}
  \end{minipage}
  \caption{Modular version of Fig.~\ref{fig:mono} with a possible modular
    analysis result.\label{fig:mod}}
\end{figure}

\begin{figure}[t]
  \textsc{Algorithm} \textbf{\manalyze}$(P = \{M_i\},\QQ)$
  \begin{algorithmic}[1]
    \State {\sf add-entries}$(\{ k \in \QQ\ |\ k \not\in \GAG\})$, {\sf
      upd}$(\GAG, \{ \tuplek{k}{\bot} \ | \ k \in \QQ \})$ \label{line:mod-init}
    \While {{\sf entries}$(\GAG,\QQ) \neq \emptyset$} \label{line:mod-while}
    \State $(M, E) \assign${\sf next-entries}$(\GAG,\QQ)$
    \State $\LAG \assign\emptyset$
    \State {\sf upd}$(\LAG, \{\tuple{A}{\CP}{\AP} \in \GAG \ | \
    \mod{A} \in \imports{M}\})$ \Comment{\alglinetitle{PreloadImported}} \label{line:imported}
    \State $\LAG := \banalyze(M,E,\emptyset,\LAG)$
    \For{$\tuplek{\apair{P}{\CP}}{\AP_l}
      \in \LAG . \tuplek{\apair{P}{\CP}}{\AP_g} \in \GAG\ \implies \AP_l \neq
      \AP_g$}
    \State $\AP \assign\alub(\AP_l,\{\AP_g\})$ \label{line:mod-widensucc}
    \State {\sf upd}$(\GAG,\tuplek{\apair{P}{\CP}}{\AP})$
    \Comment{\alglinetitle{StoreAnswers}} \label{line:mod-storea}
    \State {\sf add-entries}($\{k \ | \ \arck{k}{\apair{P}{\CP}} \in \GAG\}$)
    \EndFor
    \State {\sf del}$(\GAG, \{ \arck{\apair{P}{\CP}}{k'} \in \GAG\})$
    \Comment{\alglinetitle{UpdateDependencies}}
    \State $R = \{\arck{\apair{P}{\CP}}{k'} \ | \ \exists \ \apair{P}{\CP}
        \rightsquigarrow k \in \LAG, k' = \apair{A}{\CP}, \mod{A} \neq
        M\})$
        \For{$\arck{k}{\apair{Q}{\CP_t}} \in R$} \label{line:mod-reana}
        \State $\calls  \assign\{\lambda\ |\ \apair{\pdesc}{\lambda} \in \GAG
        \}$ \label{line:mod-cwidencalls}
        \State $\CP \assign\alub(\AP_t,\calls)$ \label{line:mod-widencalls}
        \If{$\apair{Q}{\CP} \not\in \GAG$} \Comment{\alglinetitle{ScheduleNewCalls}}
        \State {\sf add-entries}$(\apair{Q}{\CP}\})$ \label{line:mod-end}
        \EndIf
        \State {\sf upd}$(\GAG, \{\arck{k}{\apair{Q}{\CP}}\})$ \label{line:mod-stored}
    \EndFor

    \EndWhile
    \State \Return $\GAG$
  \end{algorithmic}
\caption{Modular fixpoint algorithm.\label{fig:mod-base}}
\figendsp
\end{figure}

\paragraph{\textbf{Operation of the algorithm.}}

The algorithm takes as input a (partitioned) program $P = \{M_i\}$, some
initial queries $\QQ$ to any exported predicate of the program, i.e., any
$\apair{\pdesc}{\CP} \in \QQ, A \in \exports{\mod{\pdesc}}$.
If there are recursive dependencies between modules, the modules in each clique
will be grouped and analyzed as a whole module (after doing the necessary
renamings).
This decision is based on the observation that,
if we choose to not group modules that are in the same recursive
clique, then, after program changes within the clique, we will have to
delete all the internal analysis information, as we will see later,
and this is essentially equivalent to considering the clique a single
module.  Alternatively, it would be possible in principle to pass more
detailed information across modules, but then again this is
essentially equivalent to doing monolithic incremental for the clique.

The pseudocode of the algorithm is detailed in Fig.~\ref{fig:mod-base}. Each of
the modules in the program will be analyzed independently, and possibly several
times. The algorithm keeps a queue of all the call patterns that need to be
(re)analyzed for each module.
To distinguish between the queries defined by the user and the intermediate
queries done internally by the modular analysis algorithm, we will call the
latter \emph{entries} and they will be referred to with $E$.
The queue is initialized with an entry for each of the abstract queries.
Modular analysis is controlled by this queue that contains the call patterns
with possibly incomplete answers (added with procedure {\sf
  add-entries}). At each iteration of the loop a module is reanalyzed
independently for its set of annotated entries ($E$) extracted from the queue.
This is done by procedure {\sf next-entries} which extracts from
the queue entries that are reachable from the initial $\QQ$ in $\GAG$.
In every iteration modules are analyzed \emph{from scratch}. This means that, in
principle, the analysis of module $M$ with entries $E$ should be performed by
$\LAG = \banalyze(M,E,\emptyset,\emptyset)$. However, \banalyze assumes that all
code is available for analysis. Since this is not so in this modular
case, \banalyze needs to be
provided with an abstraction of the predicates imported by $M$. To this end,
in line~\ref{line:imported} (\alglinetitle{PreloadImported}), the nodes and
answers of the global graph $\GAG$ of predicates imported by $M$ are added to
$\LAG$.
After this, \GAG is updated, by propagating the newly computed answers
(\alglinetitle{StoreAnswers}), provided that a generalization is made before to
ensure termination and updating the dependencies of the predicates in the
boundary of the modules (\alglinetitle{UpdateDependencies}), adding entries for
the newly encountered call patterns (\alglinetitle{ScheduleNewCalls}), also
generalizing them if necessary.

\subsubsection{Correctness}

We now formalize the notion of \emph{correct modular analysis}. 
Let \linebreak $\mathtt{first\_ext\_calls}(E,\csemantics)$ be a function that, given a set
of execution trees $\csemantics$ returns the set of calls reachable from any $e \in E$ that are the first
reachable predicate that is imported by $\mod{e}$, together with $E$. That is:
$$ \{C \ | \exists e \in E \text{ and } (e \rightsquigarrow C) \in \csemantics. %
\forall l \in
(e \rightsquigarrow C). \mod{l} = \mod{e} \wedge \mod{C} \neq
\mod{e} \} \cup E $$

\begin{definition}[Correctly approximated intermodular
  calls]\label{def:mod-correct-call}
  Let $P$ be a program and $\Q$ a set of concrete queries, $\GAG$ an analysis
  graph, and $E$ a set of entries, and let $I$ be the transitive closure of
  $\mathtt{first\_ext\_calls}(E,\csemantics)$. We say that $\GAG$ \emph{correctly
    approximates the intermodular calls of $\csemantics$} if it abstracts all the call
  patterns in the transitive closure of $I$. That is:
  $$
  \forall \apair{\pdesc}{\theta^c} \in I.\exists
  \apair{\pdesc}{\CP} \in \GAG \wedge \theta^c \in \gamma(\CP).
  $$
\end{definition}

\noindent
That is, $\GAG$ contains all the calls of the exported predicates that were
originated from a different module in which they are defined, and that are
reachable from $\Q$. Note that this set in the concrete execution may be
infinite, e.g., in the case in which an imported predicate is called inside a
loop.

\begin{definition}[Correct modular analysis]\label{def:cmod}
  Given a program $P$, split in modules $M_i$, and initial concrete queries $\Q$, we
  say a modular analysis graph \emph{$\GAG$ is correct for $P, \Q$} if:
  \begin{enumerate}[leftmargin=5mm,rightmargin=-3mm]
  \item[a)] it approximates the intermodular calls correctly
    (see Def.~\ref{def:mod-correct-call}) and
  \item[b)] it approximates the answers correctly (see
    Def.~\ref{def:correct-succ}).
  \end{enumerate}
\end{definition}

As mentioned earlier, $\banalyze$ assumes that either the procedures executed by
a program are defined in the clauses provided to the analyzer, or they are
basic, built-in operations of the language, i.e., they are interpreted applying
their corresponding transfer function. This is not the case when analyzing
programs module by module, and assumptions need to be made about the imported
code. The following lemma states that the analysis graph inferred by \banalyze is
correct assuming the answers of $\LAG_0$ if it only contains abstractions of the
imported predicates. In other words, if $\LAG_0$ correctly over-approximates the
behavior of the imported predicates, then the analysis of the module is correct.

\begin{lemma}[Correctness of \banalyze modulo imported predicates]\label{th:cmod-one}
  Let $M$ be a module of program $P$, $E$ a set of abstract queries.
  Let $\LAG_0$ be an analysis graph such that $\forall \apair{A}{\CP} \in
  \LAG_0. \mod{A} \in \imports{M}$. The analysis result
  $$
  \LAG = \banalyze(M,E,\emptyset,\LAG_0)
  $$
  is \emph{correct} (see Def.~\ref{def:correct-ana}) for $M$ and $\gamma(E)$ assuming $\LAG_0$.
\end{lemma}

\begin{proof}
  By Theorem~\ref{th:mon-scratch}, \banalyze produces a correct analysis whenever
  the initial analysis graph is empty. Since, $\LAG_0$ contains only information
  about the imported predicates, the analysis graph inferred is correct for
  all the predicates in $M$, assuming that the original information in $\LAG_0$
  is correct.
\end{proof}

\begin{theorem}[Correctness of \manalyze]\label{th:mod_basic}%
  Let $P$ be a modular program, and $\QQ$ a set of abstract queries.
  The modular analysis graph:
  $$
  \GAG = \manalyze(P,\QQ)
  $$
  is \emph{correct} (Def.~\ref{def:cmod}) for $P$ and $\gamma(\QQ)$.
\end{theorem}
\begin{proof}
  By induction on the number of modular partitions. If there is only one
  partition, the conditions in Def.~\ref{def:mod-correct-call} hold trivially
  because the only intermodular call patterns are the $\QQ$ (added in
  line~\ref{line:mod-init}).
  Since $\LAG$ is correct by Theorem~\ref{th:mon-scratch} and the results are
  updated in line~\ref{line:mod-storea} the conditions in
  Def.~\ref{def:correct-succ} hold. And no further iteration is required.
  
  If the program $P$ is partitioned into $n$ modules, we need to prove that if
  analyzing $n-1$ modules finishes, then analyzing all $n$ modules also finishes.
  Assuming that the analysis of the first $n-1$ modules finishes and is
  correct, the result of these $n-1$ modules could be seen as one module,
  reducing this general case to the case of $2$ modules.
  To prove this the following invariant of the algorithm is used:

  \emph{Before extracting from the queue via {\sf next-entries}
    (line~\ref{line:mod-while}), either the results in $\GAG$  are
    correct, or the queue is not empty.}
  
  This invariant trivially holds immediately after initializing the queue with
  the queries in line~\ref{line:mod-init}.
  Then, at each iteration of the while loop,
  since there are only 2 modules, when one is extracted from the
  queue, the queue is empty. After analyzing (line 4), we know $\LAG$
  is correct if $\GAG$ was correct. If no answers changed w.r.t.\ $\GAG$, no
  modules are added and the fixed point was reached.
  If the results change, every answer that changed is generalized and updated in
  $\GAG$, which results in adding an entry to it (line~\ref{line:mod-storea}).
  Then, since there are only two modules, there can be at most one module in the
  queue, since the one being processed is extracted. If after processing one
  module the nodes and answers (excluding the answers to $\QQ$) stay the same,
  no new events will be added to the queue. In this case, then the analysis is
  already correct, by Lemma~\ref{th:cmod-one}, because \banalyze was
  performed assuming already correct information.
  Else, if new answers were encountered it means that the previous information
  was incomplete, these answers are stored (line~\ref{line:mod-storea}), and the entries that
  depend on these answers are added to the queue, so the invariant holds.
  If new call patterns were encountered, then it means that the analysis was not
  completed yet. The algorithm, after generalization, schedules them to be
  reanalyzed (line~\ref{line:mod-end}), and therefore the invariant holds as well.
\end{proof}

As mentioned earlier, the goal of this algorithm was not to perform incremental
analysis but rather to reduce the working set of the basic (monolithic) analyzer.
In fact, in \cite{mod-an-lopstrbook}, the authors neither provide a clear
strategy of how to tackle the problem of reusing the analysis result after making
modifications to the program nor perform experiments.

\subsubsection{Precision}\label{sec:mod-prec}
We now show the precision guarantees when analyzing with finite abstract domains
if the generalization step is removed, i.e., by:
\begin{itemize}
\item replacing line~\ref{line:mod-widensucc} by $\AP \assign\AP_l \sqcup
  \AP_g$,
\item removing line~\ref{line:mod-cwidencalls}, and
\item replacing line~\ref{line:mod-widencalls} by $\CP \assign\AP_t$.
\end{itemize}

\begin{lemma}[Precision of \banalyze modulo imported predicates]
  Let $M$ be a module of program $P$, $E$ a set of abstract queries.
  Let $\LAG_0$ be an analysis graph such that $\forall \apair{A}{\CP} \in
  \LAG_0. \mod{A} \in \imports{M}$ if $\LAG_0$ contains \emph{the least
    fixed point} as defined in Theorem~\ref{th:prec-inc}. The analysis result
  $$
  \LAG = \banalyze(M,E,\emptyset,\LAG_0)
  $$
  is the \emph{least program analysis graph} for $M$ and $\gamma(E)$ assuming
  $\LAG_0$.
\end{lemma}

\begin{proof}
  Since all values reused are the least fixed point, no imprecision is
  introduced by $\LAG_0$. Correctness follows from Lemma~\ref{th:cmod-one}.
\end{proof}

\begin{theorem}[Precision of \manalyze]
  Let $P$ be a modular program and $\QQ$ a set of abstract queries. The modular
  analysis result
  $$
  \GAG = \manalyze(P,\QQ)
  $$
  is the \emph{least modular analysis graph} for $P$ and $\gamma(\QQ)$.
\end{theorem}

\begin{proof}
  Since no imprecision is introduced during the modular processing, and all
  answers are started assuming $\bot$ (line~\ref{line:mod-init}), %
  each of the calls to \banalyze
  will produce results that are below or exactly the least fixed point.
  Correctness follows from Theorem~\ref{th:mod_basic}.
\end{proof}

\secbeg
\section{The Algorithm for Incremental and Modular Context-sensitive Analysis}
\secend
\label{sec:intermod}

We now propose an algorithm that performs a goal-directed, top-down, incremental abstract
interpretation of modular Horn clause programs.
The analyzer takes a program (target), a set of initial call states, and,
optionally, analysis results of a previous version of the program, and
information about the changes w.r.t.\ the target program.
The analyzer will annotate the program with information about the current state
of the variables at each clause and literal whenever they are reached when executing
the calls described by the initial call states, reusing as much of the provided
analysis results as possible.

\paragraph{\textbf{Analysis graphs for modular and incremental analysis.}} 
To have an algorithm that processes partitions of programs modularly but, at the
same time, is able to update localized information we propose to keep, in
addition to $\GAG$, a local analysis graph per modular partition $M$, referred
to with $\LAG_M$. The analysis result then consists on a set of graphs $\{\GAG,
\{\LAG_i\}\}$. An example of an analysis result of this shape is depicted in
Fig.~\ref{fig:mod-inc}. The information of the local analysis graphs is drawn in
black and with nodes as ellipses. The left box corresponds to the \texttt{main}
module, $\LAG_{\mathtt{main}}$, and the box on the left to the \texttt{bitops}
module, $\LAG_{\mathtt{bitops}}$. The nodes in blue, dashed, and with rectangles show the
information in the global analysis graph $\GAG$, which coincides with
Fig.~\ref{fig:mod}.

\newcommand{\rmcolumn}{4}
\begin{figure}[t]
  \begin{minipage}{0.33\linewidth}
    \prettylstciao  
\begin{lstlisting}
:- module(main, [main/1]).

:- use_module(bitops).
main(Msg, P) :-
    par(Msg, 0, P).

par([], P, P).
par([C|Cs], P~$_0$~, P) :-
    xor(C, P~$_0$~, P~$_1$~),
    par(Cs, P~$_1$~, P).
\end{lstlisting}

\begin{lstlisting}
:- module(bitops, [xor/3]).

xor(0,0,0).
xor(0,1,1).
xor(1,0,1).
xor(1,1,0).
\end{lstlisting}
  \end{minipage}
  \hspace{+3mm}
  \begin{minipage}{0.45\linewidth}
    \centering
    \textcolor{blue}{$\GAG$: global analysis graph}

    \begin{tikzpicture}[framed,>=stealth,on grid,auto]
      \tikzagconf
      \node (l1) at (\lcolumn, 1) [draw=none, minimum height=0pt] {$\LAG_{\tt main}$};
      \node (l2) at (\rmcolumn,0.3) [draw=none, minimum height=0pt] {$\LAG_{\tt bitops}$};
  \node (A) at (\lcolumn,0) [rectangle,very thick,dashed,draw=blue] {\tuple{{\tt main}(M,P)}{M/\top,P/\top}{(M/\top, P/\apar)}};
  \node (x) at (\lcolumn,0) [text width=40pt] {};
  \node (C) at (\lcolumn,-2)[] {\tuple{{\tt par}(M,X,P)}{(M/\top, X/z, P/\top)}{(M/\top, X/z, P/\apar)}};
  \node (E) at (\lcolumn,-4)[] {\tuple{{\tt par}(M,X,P)}{(M/\top, X/\apar, P/\apar)}{(M/\top, X/\apar, P/\apar)}};
  \node (D) at (\rmcolumn,-1) [rectangle, dashed,draw=blue,text width=3cm] {\tuple{{\tt xor}(C,P_0,P_1)}{(C/\top, P_0/z, P_1/\top)}{(C/\apar,P_0/z,P_1/\apar)}};
  \node (x1) at (\rmcolumn,-1) [text width=55pt] {};
    \node (F) at (\rmcolumn,-3) [rectangle,dashed,draw=blue,text width=3cm] {\tuple{{\tt xor}(C,P_0,P_1)}{(C/\top, P_0/\apar, P_1/\top)}{(C/\apar,P_0/\apar,P_1/\apar)}};
    \node (x2) at (\rmcolumn,-3) [text width=55pt] {};
    \path [->] (A) edge [dashed,draw=blue] (D);
    \path [->] (A) edge [dashed,draw=blue] (F);
    \path [->] (A) edge [bend right=60] (C);
    \path [->] (C) edge (x1);
    \path [->] (C) edge (E);
    \path [->] (E) edge (x2);
    \path [->] (E) edge [in=60,out=30,loop,distance=0.6cm] (E);
  
  \draw [] (\vlinep+0.5,1.2) -- (\vlinep+0.5,-4.8);
\end{tikzpicture}
\end{minipage}
\caption{A modular version of Fig.~\ref{fig:mono} keeping a local analysis graph
  per module.\label{fig:mod-inc}}
\end{figure}

\subsection{Operation of the algorithm}

\newcommand{\colornew}[1]{\textcolor{blue}{#1}}
The algorithm takes as input a (partitioned) program $P = \{M_i\}$, some initial
queries $\QQ$, a previous correct analysis result $\{\GAG, \{\LAG_i\}\}$, and a set of
program edits in the form of additions and deletions
$(\diff{M_i})$, which collect the differences w.r.t.\ the previous state for
each module.
The pseudocode of the algorithm is detailed in Fig.~\ref{fig:incanal_mod}. The
steps required to perform local analysis incrementally are presented in
\colornew{\textbf{blue}}, i.e., those steps that were added or modified in the
modular, non-incremental algorithm depicted in Fig.~\ref{fig:mod-base}.
Before starting the analysis process, the entries of edited modules and new
queries are marked to be (re)analyzed.
Each of the scheduled modules will be analyzed independently, and possibly
several times. %
Modular analysis is, again, controlled by a queue to which entries with possibly
incomplete answer descriptions are added (with the procedure {\sf add-entries}).
At each iteration of the loop a module is reanalyzed independently for its set of
annotated entries ($E$) extracted from the queue.
This is done by procedure {\sf next-entries} which extracts from
the queue entries that are reachable from the initial $\QQ$ in $\GAG$.
Incrementally analyzing a module consists of updating the information about the
calls to imported predicates in $\LAG_M$, by removing possibly inaccurate
results and adding the newly computed ones, and calling \banalyze.
Finally, \GAG is updated, which includes updating the newly computed answers,
updating the dependencies of the predicates in the boundary of the modules, and
adding to the queue to reanalyze the dependent predicates and call patterns.
The operations performing local incremental analysis are:
\newcommand{\newdiff}{\Delta}

\begin{figure}[t]
  
  \textsc{Algorithm} \textbf{\mianalyze}$(P = \{M_i\},\QQ,\GAG,\{\LAG_i\}, \newdiff)$
  \begin{algorithmic}[1]
    \newfunction
    \State \colornew{{\sf add-entries}$(\{ k =
      \apair{A}{\CP} \ | \ k \in \GAG, \newdiff_{\mod{A}} \neq \emptyset\})$}
    \Comment{\alglinetitle{\colornew{AnalyzeOutdated}}}
    \State \colornew{{\sf add-entries}$(\{ k \in \QQ \ | \ k \not\in \GAG\})$,  {\sf
      upd}$(\GAG, \{ \tuplek{k}{\bot} \ | \ k \in \QQ \})$}
    \Comment{\algline{\colornew{AnalyzeNew}}} 

    \While {{\sf entries}$(\GAG,\QQ) \neq \emptyset$}
    \State $(M, E)$ := {\sf next-entries}$(\GAG,\QQ)$
    \State \colornew{$I \assign\{\tuple{A}{\CP}{\AP} \in \LAG_M \ | \
      \mod{A} \in \imports{M} \}$} \Comment{\alglinetitle{\colornew{Imported}}} \label{line:beg_loc_an}
  \State \colornew{$I_{p} \assign \{ k \ |\ \tuplek{k}{\AP} \in I,
    \tuplek{n}{\AP'} \in \GAG, \AP \not\sqsubseteq \AP' \}$}
    \Comment{\alglinetitle{\colornew{ImpreciseImported}}}
    \State \colornew{$I_{c} \assign\{k' \ | \ k' \rightsquigarrow k \in \LAG_M, \ \tuplek{n}{\AP} \in I,
    \tuplek{n}{\AP'} \in \GAG, \AP' \sqsubset \AP\})$}
    \\ \hfill \Comment{\alglinetitle{\colornew{IncorrectImported}}}
    \State \colornew{{\sf del}$(\LAG_M, \{k\ |\ k_c \in I_{p}, k \rightsquigarrow k_c \in
    \LAG_M$ or $(n_a \rightsquigarrow k_c\in \LAG_M \wedge k_a \rightsquigarrow k
    \in \LAG_M)\})$}
    \\ \hfill \Comment{\alglinetitle{\colornew{DelImprecise}}}
    \State {\sf upd}$(\LAG_M, \{\tuple{A}{\CP}{\AP} \in \GAG \ | \ \mod{A} \in
    \imports{M}\})$ \Comment{\alglinetitle{PreloadImported}}
    \State \colornew{$\LAG_M \assign\banalyze(M,E\cup I_c,\diff{M},\LAG)$,
    $\newdiff_{M} \leftarrow \emptyset$} \label{line:end_loc_an} %
  \State \colornew{{\sf del}$(\LAG_M, \{ \apair{A}{\CP} \ |
    \not\exists k', \arck{k'}{\apair{A}{\CP}} \in \LAG_M, \mod{A} \neq M\})$} \Comment{\algline{\colornew{RemoveUnused}}}
    \For{$\tuplek{\apair{A}{\CP}}{\AP_l}
      \in \LAG_M . \tuplek{\apair{A}{\CP}}{\AP_g} \in \GAG\ \implies \AP_l \neq
      \AP_g$, $\mod{A} \neq M$}
    \State $\AP \assign\alub(\AP_l,\{\AP_g\})$
    \State {\sf upd}$(\GAG,\tuplek{\apair{A}{\CP}}{\AP})$ \Comment{\alglinetitle{StoreAnswers}}
    \State {\sf add-entries}($\{k \ | \ \arck{k}{\apair{A}{\CP}} \in \GAG\}$)
    \EndFor
    \State {\sf del}$(\GAG, \{ \arck{\apair{A}{\CP}}{k'} \in \GAG\})$ \Comment{\alglinetitle{UpdateDependencies}}
    \State $R = \{\arck{\apair{A}{\CP}}{k'} \ | \ \exists \ \apair{P}{\CP}
        \rightsquigarrow k \in \LAG_M, k' = \apair{A}{\CP}, \mod{A} \neq
        M\})$
    \For{$\arck{k}{\apair{Q}{\CP_t}} \in R$} %
    \State $\calls  \assign\{\lambda\ |\ \apair{\pdesc}{\lambda} \in \GAG \}$
    \State $\CP \assign\alub(\AP_t,\calls)$
    \If{$\apair{Q}{\CP} \not\in \GAG$} \Comment{\alglinetitle{ScheduleNewCalls}}
    \State {\sf add-entries}$(\apair{Q}{\CP}\})$
    \EndIf
    \State {\sf upd}$(\GAG, \{\arck{k}{\apair{Q}{\CP}}\})$ \label{line:mana-end-while}
    \EndFor
    
    \EndWhile
    \State \Return $\GAG, \{\LAG_i\}$
    \end{algorithmic}
\caption{Incremental and modular fixpoint algorithm.\label{fig:incanal_mod}}
\figendsp
\end{figure}
\begin{description}
\item [\alglinetitle{AnalyzeOutdated}] Adds to the analysis queue the entries
  of modules that changed, i.e, those whose diff ($\diff{}$) is not empty.
\item [\alglinetitle{AnalyzeNew}] Adds to the analysis queue the entries
  that have not been analyzed yet.
\item [\alglinetitle{Imported}] Collects the current approximations made about
  the predicates imported by the module to be analyzed.
\item [\alglinetitle{IncorrectImported}] Collects in $I_c$ the nodes of the
  $\LAG_M$ that are incorrect (below the fixpoint), i.e., the ones whose approximation in
  the $\LAG_M$ was smaller than in the $\GAG$ to reanalyze them later.
\item [\alglinetitle{ImpreciseImported}] Collects in $I_p$ all the imported
  nodes that are potentially imprecise, i.e., those in which in the abstraction
  $\LAG_M$ that are \emph{bigger} than the current stored in $\GAG$.
\item [\alglinetitle{DelImprecise}]
  Deletes from $\LAG_M$ the nodes that relied on assumptions that depend on
  $I_p$, because they are potentially imprecise.
\item [\alglinetitle{Analyze}] The \banalyze function is called with entries
  for: the calls scheduled by the modular analyzer ($E$), the nodes that
  depended on imported information that may be below the fixpoint ($I_c$). Note
  that no entries will be added for the nodes that were imprecise as its
  information will be removed up to the entries that they were triggered by,
  which guarantees that the analysis will be correct and precise for those
  entries.
\item [\alglinetitle{RemoveUnused}] Removes the call patterns of the imported
  predicates that were not reached, i.e., if there are no edges in the call
  graph to them.
\end{description}

\noindent
For the remaining operations we refer the reader to the description of
Fig.~\ref{fig:mod-base}.

\paragraph{\textbf{Enhancing the deletion strategy.}}\label{sec:SCC-del}

The proposed deletion strategy is quite pessimistic. Updating imprecise
information about imported predicates most of the times means reusing only a few
answers that did not depend on the changes per module.
However, it may occur that the analysis does not change after these changes occur,
or that some nodes/edges are still correct and precise.
A solution is to partially reanalyze the program without removing these
potentially useful results. Our proposed algorithm allows performing such
a partial reanalysis, by
partitioning the desired module into smaller partitions, for example,
using information on {\em strongly connected components}.
This can be achieved within the algorithm by replacing line
\alglinetitle{DelImprecise} with Fig.~\ref{fig:del_bu}. This runs the
algorithm with a partition of the current
module as input program, %
which is split using the (static) SCCs of the clauses
({\sf split-sources-scc}). This includes also partitioning the results ({\sf
  split-in-scc}) to initialize $\GAG$ using $\LAG_M$, and setting as \QQ\ the
initial $E$ of this modular analysis. The reanalysis of this partitioned module
will be given in a modular form, so it has to be \emph{flattened} back for it to
be compatible with the rest of the analysis results. This process consists in 
merging all the graphs in which the analysis was performed into one graph that
contains all the nodes and edges except the edges in $\GAG''$.

\begin{figure}
  \begin{algorithmic}
    \State \algline{DelImprecise}{} \par
    \hskip\algorithmicindent $ Calls = \{k \ | \ \tuplek{k}{\AP} \in \LAG_{M}, k \in I_p\}$ \par 
    \hskip\algorithmicindent $\{M'_i\},I_{M'_i}= $ {\sf split-sources-scc}$(M,I_p)$ \par
    \hskip\algorithmicindent $\{\GAG', \{\LAG'_i\}\} =$
    {\sf split-in-scc}$(\LAG_M)$ \par
    \hskip\algorithmicindent $\{\GAG'' \{\LAG'_i\}\}:= \mianalyze(\{M'_i\},I_{M'_i},\{\GAG',
    \{\LAG'_i\}\},\emptyset)$ \par
    \hskip\algorithmicindent $\LAG_M := $ {\sf flatten}$(\{\GAG'' \{\LAG'_i\}\})$ \par
    \hskip\algorithmicindent $\newdiff_{M} \leftarrow \emptyset$
    \end{algorithmic}
\figendsp
\caption{Enhanced modular deletion strategy.\label{fig:del_bu}}
\figendsp
\end{figure}

\subsection{Running examples of the algorithm}

 \begin{figure}[t]
   \centering
   \prettylstciao
   \begin{minipage}{0.4\linewidth}
     $M$: (unchanged) main module
\begin{lstlisting}
:- module(main, [main/2]).

:- use_module(bitops).
main(Msg, P) :-
    par(Msg, 0, P).

par([], P, P).
par([C|Cs], P~$_0$~, P) :-
    xor(C, P~$_0$~, P~$_1$~),
    par(Cs, P~$_1$~, P).
\end{lstlisting}
     $B_0$: initial state of {\tt bitops}
\begin{lstlisting}
:- module(bitops, [xor/3]).
xor(0,0,0).
\end{lstlisting}     
   \end{minipage}
   \hspace{5mm}
   \begin{minipage}{0.4\linewidth}
     $B_1$: clauses added to {\tt bitops}
     \begin{lstlisting}
:- module(bitops, [xor/3]).
xor(0,0,0). %
xor(0,1,1).
xor(1,0,1).
xor(1,1,0).
\end{lstlisting}
     $B_2$: a clause is deleted from {\tt bitops}
     \begin{lstlisting}
:- module(bitops, [xor/3]).
xor(0,0,0).
xor(0,1,1).
xor(1,0,1).
%
\end{lstlisting}
   \end{minipage}
   \caption{Different program states.}
   \label{fig:program}
 \end{figure}

To show the algorithm in action we now analyze incrementally
 different versions of the program that computes the parity (some of which are
 incomplete). The different states of the sources are shown in
 Fig.~\ref{fig:program}. Initially we have the analysis result of $P_0= \{M,
 B_0\}$, $\A_0$ in Fig.~\ref{fig:running_ex}. This was the result of running the
 algorithm from scratch $\A_0 =$ \mianalyze$(P_0, \QQ, \emptyset,
 (\emptyset,\emptyset))$, with initial query $\QQ = \{\apair{{\tt
     main}(M,P)}{(M/\top, P/\top)}\}$. In this version it was inferred that if
 {\tt main(M, P)} succeeded then {\tt P} is 0 $(\gamma(z))$.

 \renewcommand{\rmcolumn}{3.5} 
\begin{figure}[t]
  \hspace{-5mm}
  \begin{minipage}{0.4\textwidth}
  $\underline{\A_0}$ ($P = \{M, B_0\}$)
  
  \begin{tikzpicture}[framed,>=stealth,on grid,auto]
    \tikzagconf
    \node (l1) at (0.9,-1) [draw=none, minimum height=0pt]{$\LAG_{\tt main}$};
    \node (l2) at (\rmcolumn,0) [draw=none, minimum height=0pt] {$\LAG_{\tt bitops}$};
    \node (A) at (\lcolumn,0) [very thick,text width=70pt,dashed,rectangle,draw=blue] {\tuple{{\tt main}(M,P)}{(M/\top, P/\top)}{(M/\top, P/z)}};
    \node (x) at (\lcolumn,0) [text width=50pt] {};
    \node (C) at (\lcolumn,-2)[text height=3pt,text width=60pt] {\tuple{{\tt par}(M,X,P)}{(M/\top, X/z, P/\top)}{(M/\top, X/z, P/z)}};
    \node (D) at (\rmcolumn,-1) [rectangle, dashed,draw=blue,text width=90pt] {\tuple{{\tt xor}(C,P_0,P_1)}{(C/\top, P_0/z, P_1/\top}{(C/z,P_0/z,P_1/z)}};
    \node (x1) at (\rmcolumn,-1) [text height=15pt,text width=60pt] {};
    \path [->] (A) edge [dashed,draw=blue] (D);
    \path [->] (A) edge [bend right=60] (C);
    \path [->] (C) edge (x1);
    \path [->] (C) edge [in=120,out=60,loop,distance=0.8cm] (C);
    
    \draw [] (\vlinep+0.27,0.5) -- (\vlinep+0.27,-2.8);
  \end{tikzpicture}
\end{minipage}
\hspace{15mm}
\begin{minipage}{0.5\textwidth}
  $\underline{\A_0'} = \A_0$ + inc. analyzed {\tt bitops} ($B_1$)
  
  \begin{tikzpicture}[framed,>=stealth,on grid,auto]
    \tikzagconf
    \node (l1) at (0.9,-1) [draw=none, minimum height=0pt] {$\LAG_{\tt main}$};
    \node (l2) at (\rmcolumn,0.3) [draw=none, minimum height=0pt] {$\LAG_{\tt bitops}$};
    \node (A) at (\lcolumn,0) [rectangle,text width=70pt,very thick,dashed, draw=blue] {\tuple{{\tt main}(M,P)}{(M/\top, P/\top)}{(M/\top, P/z)}};
    \node (x) at (\lcolumn,0) [minimum height=30pt,text width=50pt] {};
    \node (C) at (\lcolumn,-2)[text width=60pt] {\tuple{{\tt par}(M,X,P)}{(M/\top, X/z,P/\top)}{(M/\top, X/z, P/z)}};
    \node (D) at (\rmcolumn,-1) [rectangle,dashed,draw=blue,text width=90pt] {\tuple{{\tt xor}(C,P_0,P_1)}{(C/\top,P_0/z, P_1/\top)}{\xcancel{(C/z,P_0/z,P_1/z)}} $(C/\apar,P_0/z,P_1/\apar)$};
    \node (x1) at (\rmcolumn,-1) [text height=22pt,text width=60pt] {};
    \path [->] (A) edge [dashed,draw=blue] (D);
    \path [->] (A) edge [bend right=60] (C);
    \path [->] (C) edge (x1);
    \path [->] (C) edge [in=120,out=60,loop,distance=0.8cm] (C);
  
    \draw [] (\vlinep+0.27,0.5) -- (\vlinep+0.27,-2.8);
  \end{tikzpicture}
\end{minipage}

\vspace{5mm}

\hspace{-5mm}
\begin{minipage}{0.5\textwidth}
  $\underline{\A_0''} = A_0'$ + inc. analyzed {\tt main} ($M$)
  
  \begin{tikzpicture}[framed,>=stealth,on grid,auto]
    \tikzagconf
    \node (l1) at (\lcolumn,-1) [draw=none, minimum height=0pt] {$\LAG_{\tt main}$};
    \node (l2) at (\rmcolumn,0.3) [draw=none, minimum height=0pt] {$\LAG_{\tt bitops}$};
    \node (A) at (\lcolumn,0) [rectangle,text width=70pt,very thick,dashed,draw=blue] {\tuple{{\tt main}(M,P)}{(M/\top, P/\top)}{(M/\top, \xcancel{P/z}\ P/\apar)}};
  \node (x) at (\lcolumn,0) [text width=50pt] {};
  \node (C) at (\lcolumn,-2)[text width=65pt] {\tuple{{\tt par}(M,X,P)}{(M/\top, X/z, P/\top)}{(M/\top, X/z, \xcancel{P/z}\ P/\apar)}};
  \node (E) at (\lcolumn,-4)[text width=60pt] {\tuple{{\tt par}(M,X,P)}{(M/\top, X/\apar, P/\top)}{(M/\top, X/\apar, P/\apar)}};
  \node (D) at (\rmcolumn,-1) [rectangle,dashed,draw=blue,text width=90pt] {\tuple{{\tt xor}(C,P_0,P_1)}{(C/\top,P_0/z, P_1/\top)}{(C/\apar,P_0/z,P_1/\apar)}};
    \node (x1) at (\rmcolumn,-1) [text width=60pt] {};
    \node (F) at (\rmcolumn,-3) [rectangle,dashed,draw=blue,text width=90pt,text height=9pt] {\tuple{{\tt xor}(C,P_0,P_1)}{(C/\top,P_0/\apar,P_1/\top)}{\bot}};
    \node (x2) at (\rmcolumn,-3) [minimum height=26pt,text width=60pt] {};

    \path [->] (A) edge [dashed,draw=blue] (D);
    \path [->] (A) edge [dashed,draw=blue] (F);
  \path [->] (A) edge [bend right=60] (C);
  \path [->] (C) edge (x1);
  \path [->] (C) edge (E);
  \path [->] (E) edge (x2);
  \path [->] (E) edge [in=60,out=30,loop,distance=0.6cm] (E);
  
  \draw [] (\vlinep+0.4,0.5) -- (\vlinep+0.4,-4.7);
\end{tikzpicture}
\end{minipage}
\hspace{5mm}
\begin{minipage}{0.5\textwidth}
  $\underline{\A_1}$ ($P =\{M, B_1\}$)
  
  \begin{tikzpicture}[framed,>=stealth,on grid,auto]
    \tikzagconf
    \node (l1) at (\lcolumn, -1) [draw=none, minimum height=0pt] {$\LAG_{\tt main}$};
    \node (l2) at (\rmcolumn,0.3)[draw=none, minimum height=0pt] {$\LAG_{\tt bitops}$};
    \node (A) at (\lcolumn,0) [rectangle,very thick,dashed,draw=blue,text width=70pt] {\tuple{{\tt main}(M,P)}{(M/\top, P/\top)}{(M/\top, P/\apar)}};
    \node (x) at (\lcolumn,0) [text width=50pt] {};
    \node (C) at (\lcolumn,-2)[text width=60pt] {\tuple{{\tt par}(M,X,P)}{(M/\top,X/z, P/\top)}{(M/\top,X/z, P/\apar)}};
    \node (E) at (\lcolumn,-4)[text width=60pt] {\tuple{{\tt par}(M,X,P)}{(M/\top,X/\apar, P/\top)}{(M/\top,X/\apar, P/\apar)}};
    \node (D) at (\rmcolumn,-1) [rectangle,dashed,draw=blue,text width=90pt] {\tuple{{\tt xor}(C,P_0,P_1)}{(C/\top,P_0/z, P_1/\top)}{(C/\apar,P_0/z,P_1/\apar)}};
    \node (x1) at (\rmcolumn,-1) [minimum height=26pt,text width=60pt] {};
    \node (F) at (\rmcolumn,-3) [rectangle,dashed,draw=blue,text width=90pt] {\tuple{{\tt xor}(C,P_0,P_1)}{(C/\top, P_0/\apar, P_1/\top)}{\xcancel{\bot}} $(C/\apar,P_0/\apar,P_1/\apar)$};
    \node (x2) at (\rmcolumn,-3) [text width=60pt] {};
    \path [->] (A) edge [dashed,draw=blue] (D);
    \path [->] (A) edge [dashed,draw=blue] (F);
    \path [->] (A) edge [bend right=60] (C);
    \path [->] (C) edge (x1);
    \path [->] (C) edge (E);
    \path [->] (E) edge (x2);
    \path [->] (E) edge [in=60,out=30,loop,distance=0.6cm] (E);
    \draw [] (\vlinep+0.38,0.5) -- (\vlinep+0.38,-4.7);
  \end{tikzpicture}
\end{minipage}
\caption{Analysis results in several reanalysis steps.\label{fig:running_ex}}
\end{figure}

\begin{example}[Adding clauses]
  If some clauses are added to {\tt bitops} resulting in $B_1$, the program to
  be (re)analyzed becomes $P_1 = \{M, B_1\}$.
Incremental analysis by running \mianalyze$(P, \QQ, \A_0, (\{{\tt
  xor_2, xor_3, xor_4}\}, \emptyset))$ proceeds as follows. The entries
of {\tt bitops} are added to the queue and it is analyzed
with $E = \{\apair{{\tt xor}(C,P_0,P_1)}{P_0/z}\}$ and the analysis result changes
to $(C/\apar,P_0/z,P_1/\apar)$ (shown in $\A'_0$). This change needs to be
propagated to module {\tt main}, which is analyzed next in the queue. Following
the steps of the algorithm:

\begin{description}
\item [\alglinetitle{AnalyzeOutdated}] The entries to the module \texttt{main}
  are added.
\item [\alglinetitle{AnalyzeNew}] No entries are added because there are no new
  queries.
\item [\alglinetitle{Imported}]
  $I = \{\apair{{\tt xor}(C,P_0,P_1)}{P_0/z}\}$
\item [\alglinetitle{IncorrectImported}]
  $I_c = \{\apair{{\tt xor}(C,P_0,P_1)}{P_0/z}\}$
\item [\alglinetitle{ImpreciseImported}]
  $I_p = \emptyset$ since the only imported node was below the fixed point.
\item [\alglinetitle{Analyze}] The analyzer is called with $E = \{\apair{{\tt
      main}(M,P)}{(M/\top, P/\top)}\}$ and $I_c$ as described.
\item [\alglinetitle{RemoveUnused}] All imported call patterns in
  $\LAG_{\tt main}$ are reached (there is an edge to them) and nothing is removed.
\item [\alglinetitle{StoreAnswers}] $\tuple{{\tt main}(M,
    P)}{(M/\top, P/\top)}{(M/\top, P/z)}\}$ is updated in $\GAG$, no (parent) entries need
  to be added to the queue because it is the initial query.
\item [\alglinetitle{UpdateDependencies}] All the edges of \GAG from nodes of
  {\tt main} to {\tt bitops} are removed. \\
  $R = \{ \arc{{\tt main}(M,P)}{(M/\top, P/\top)}{}{{\tt
      xor}(C,P_0,P_1)}{(C/\top, P_0/z, P_1/\top)},$ \\ $\arc{{\tt main}(M,P)}{(M/\top,
    P/\top)}{}{{\tt xor}(C,P_0,P_1)}{(C/\top, P_0/b, P_1/\top)}\}$
\item [\alglinetitle{ScheduleNewCalls}] 
A newly encountered call description is added in {\sf add-entries}, \\
$\apair{{\tt xor}(C,P_0,P_1)}{(C/\top, P_0/b, P_1/\top}$, and all the edges in
$R$ are added to \GAG.
\end{description}
\noindent
Next, module {\tt bitops} needs to be analyzed again, only for the pending call
description $\apair{{\tt xor}(C,P_0,P_1)}{(C/\top,P_0/b,P_1/\top)}$, the new answer
($C/\apar,P_0/\apar,P_1/\apar$) will be updated in \GAG, adding again an entry
for predicate {\tt main}. The next iteration of the analysis loop, the answer
will be updated but it will not imply any changes in the analysis result of the
module, therefore the algorithm reached a fixed point ($\A_1$ in
Fig.~\ref{fig:running_ex}).
\end{example}

\begin{example}[Deleting clauses]%
  The {\tt bitops} module is edited from $B_1$ to $B_2$, and the program to be
  analyzed is $P_2 = \{M, B_2\}$. Incremental analysis by \mianalyze$(P_2, \QQ,
  \A_1, (\emptyset, \{{\tt xor_4}\}))$ proceeds as follows.
Module {\tt bitops} was changed, so it is analyzed with $E = \{\apair{{\tt
    xor}(C,P_0,P_1)}{(C/\top,P_0/z,P_1/\top)}, \apair{{\tt xor}(C,P_0,P_1)}{(C/\top,P_0/b,P_1/\top)}\}$. The answers are
recomputed from scratch, however, the overall result of the module does not
change, so nothing needs to be done in \GAG, and it is not necessary to
recompute the analysis graph of module {\tt main}, and $\A_2 = \A_1$.
\end{example}

\section{Fundamental results of the algorithm}\label{sec:th}

In this section we provide the correctness and precision guarantees of the
proposed algorithm.
We first provide some notation that will be instrumental in the proofs of the
theorems.
We build the domain of analysis results (parametric on $\DD$) as sets of
$(\mathtt{pred\_name}, \DD, \DD)$. The set of predicate names may be infinite in
general but in each program it is finite. We do not represent the dependencies
(edges in the graph), as they are needed only for efficiency. We define the
partial order in this domain $\AGDD$ that compares the answer patterns to the
call patterns abstracted by the analysis graph. That is:
$$
g_1,g_2 \in \AGDD, g_1 \sqsubseteq_\AGDD g_2 \text{ if } \forall
\tuplek{k}{\AP_1} \in g_1 . \exists \tuplek{k}{\AP_2} \in g_2 \ \wedge \AP_1
\sqsubseteq \AP_2.
$$
For simplicity, \DD\ in the following represents this domain of
analysis results, and all domain operators refer to this domain.

The incremental analysis of a module within the algorithm (in the body
of the while loop in the pseudocode lines~\ref{line:beg_loc_an}
to~\ref{line:end_loc_an}) is denoted with the function:
$$
\LAG_{M'} = \lanalyze(M',E,\GAG,\LAG_M,\Delta_M),
$$
\noindent
where $M'$ is a module, $\LAG_M$ is the analysis result of $M$ for $E$,
$\Delta_M$ are the differences between $M'$ and $M$, to which $\LAG_M$ corresponds,
to get $\LAG_{M'}$, the analysis result of $M'$, and \GAG contains the (possibly
temporary) information for the predicates imported by $M'$.

Lastly, we represent performing an iteration of the while loop
(lines~\ref{line:beg_loc_an} to~\ref{line:mana-end-while}) as the high-level
operation of updating the newly computed information in $\GAG$:
$$
\mathit{MA}(M',E,\GAG,\LAG_M,\diff{M}) = {\sf upd}(\GAG, \lanalyze(M',E,\GAG,\LAG_M,\Delta_M))
$$
Note that, after a number of (chaotic) iterations, $\mathit{MA}$ is monotonic, and
ultimately stationary due to the use of the widening operator.

\subsection{Correctness}

The following lemma shows that if a module $M$ is analyzed for entries $E$
assuming some $\GAG$ obtaining $\LAG_{M}$, if the assumptions change to
$\GAG'$, incrementally updating these assumptions produces an analysis graph
$\LAG'_M$ that is correct assuming $\GAG'$.

\begin{lemma}[Correctness updating $\LAG$ modulo $\GAG$]\label{th:updlag}
  Let $M$ be a module of program $P$ and $E$ a set of entries. Let
  $\GAG$ be a previous state of the global analysis graph, if $\LAG_{M}$
  is correct for $M$ and $\gamma(E)$ assuming $\GAG$.
  If $\GAG$ changes to $\GAG'$ the analysis result
  $$
  \LAG'_{M} = \lanalyze(M,E,\GAG',\LAG_{M},\emptyset)
  $$
  is \emph{correct} (see Def.~\ref{def:correct-ana}) for $M$ and $\gamma(E)$ assuming $\GAG$.
\end{lemma}

\begin{proof}
  To prove this we need to show that all the answers that differ from $\GAG$ to
  $\GAG'$ for the calls to predicates imported by $M$ are included in $E$. Since
  these are the requisites in Theorem~\ref{th:inc-part} to guarantee that the
  result is correct.
  The \alglinetitle{ImpreciseImported} are collected and removed, therefore it
  is guaranteed that all the entries in $E$ that depended on these will be
  correct (the analysis is empty).
  When collecting the \alglinetitle{IncorrectImported} only those nodes are
  added to the entries. However, because $\LAG_{M}$ was assumed to be correct,
  it is guaranteed that adding these entries is enough, because $\LAG_{M}$
  correctly over-approximates the parts of \csemantics\ that were already in
  $\LAG_{M}$ and the ones that are missing are guaranteed to be correct because
  they are missing.
\end{proof}

The following Theorem~\ref{th:mod_scratch} captures the correctness of the algorithm
when starting from an empty analysis result. I.e., starting with an empty $\GAG$
and $\LAG_{M_i}$. Note that this is not the same as running the traditional
modular analysis, as information is reused when iterating
between modules, whereas in \manalyze{} every iteration the $\LAG$'s are clean.

\begin{theorem}[Correctness of \mianalyze{} from scratch]\label{th:mod_scratch}
  Let $P$ be a modular program, and $\QQ$ a set of abstract queries.
  Then, if:
  $$
  \{\GAG, \{\LAG_{M_i}\}\} = \mianalyze(P,\QQ,\emptyset,\emptyset)
  $$
  $\GAG$ is \emph{correct} (see Def.~\ref{def:cmod}) for $P$ and $\gamma(\QQ)$.
\end{theorem}

\begin{proof}
  Correctness follows using the same argument as in Theorem~\ref{th:mod_basic}, with
  the difference that instead of applying Lemma~\ref{th:cmod-one} to \banalyze, we
  apply Lemma~\ref{th:updlag} to \lanalyze.
\end{proof}

\begin{theorem}[Correctness of \mianalyze{}]\label{th:mod_inc}
  Let $P, P'$ be modular programs that differ by $\diff{}$, $\QQ$ a set of queries, and
  $\A = \mianalyze(P,\QQ,\emptyset,( \emptyset, \emptyset))$, then if:
  $$
  \{\GAG', \{\LAG'_{M_i}\}\} = \mianalyze(P',\QQ,\A,\Delta)
  $$
  $\GAG'$ is \emph{correct} (see Def.~\ref{def:correct-ana}) for $P$ and $\gamma(\QQ)$.
\end{theorem}

\begin{proof}
  
  By induction on the number of modular partitions. It is true for any partition
  of program $P$ in $n$ modules with no recursive dependencies on predicates
  between modules.
This condition ensures that if removing for some clause in $\LAG_M$ is needed
{\em all} the dependent information for recomputing is indeed removed
(nothing imprecise is reused from some other $\LAG_{M'}$).
\begin{itemize}[leftmargin=5mm,rightmargin=-3mm]
\item If program $P$ has one module, it is the same case as having one module in
  the modular analysis. So it follows from Theorem~\ref{th:mod_basic}.
\item As in the proof of Theorem~\ref{th:mod_basic}, if program $P$ is
  partitioned into $n$ modules, we need to prove that if we finish analyzing
  $n-1$ modules, then we finish analyzing all $n$ modules. Assuming that the
  analysis of the first $n-1$ modules finishes and it is correct, this $n-1$
  result could be seen as one module, reducing this general case to the case of
  $2$ modules.
\item If program $P=\{M_a, M_b\}$ is partitioned into $2$ modules let us assume
  that $M_a$ imports $M_b$. Let us assume that we reanalyze $M_b$ first. We
  study the reanalysis cases of
  $\GAG' = \mathit{MA}(M_b, E,\GAG,\LAG_{M{_b}}, \diff{M_b})$:
  \begin{enumerate}
  \item If $\GAG' = \GAG$ the procedure is equivalent if program $P$ has one
    module.
  \item If $\GAG \sqsubset \GAG'$, then analysis results need to be propagated
    to $A$. Once the results of $A$ are updated, the analysis iterations of $A$
    and $B$ will be equivalent as when analyzing from scratch, only new
    call patterns may appear.
  \item If $\GAG' \sqsubset \GAG$ these analysis results need to be propagated
    to the analysis of $A$, which will be reanalyzed. Once $A$ and $B$ have
    updated their incompatible information the further (re)analyses can only
    become smaller, but since $MA$ is monotonic and \DD~is finite, a fixpoint is
    reached, which is correct for $P$, since the computation of each of the
    modules is correct.
  \item Else, the information is incompatible. This can only happen if there
    were additions and deletions. This information needs to be propagated to
    $M_a$ and the reanalysis of $M_a$ will only lead to cases 1, 2, or 3.
  \end{enumerate}
\end{itemize}
\end{proof}

\noindent
Note that the correctness of the proposed enhanced deletion strategy follows
from Theorem~\ref{th:mod_inc}.

\subsection{Precision}

As in the previous sections, we now show the precision properties of the
algorithm when analyzing with finite abstract domains, if we remove the
generalization via widening as stated in Secs.~\ref{sec:alg-diff}
and~\ref{sec:mod-prec}. First note that for the $\mathit{MA}$ function, since
the \emph{lfp} is monotonic w.r.t. the initial assumptions and {\sf upd} is
monotonic, if generalization is disabled then $\GAG$ will be the \emph{least
  program analysis graph}, as the \textit{lfp} of each of the individual modules
was computed.

The following lemma shows that if a module $M$ is analyzed for $E$ assuming
some $\GAG$ obtaining $\LAG_{M}$, then if the assumptions change to $\GAG'$,
incrementally updating these assumptions will produce an analysis graph $\LAG'_M$
that is the same as analyzing $M$ with assumptions $\GAG$ from scratch. That is,
the least analysis graph for module $M$.

\begin{lemma}[Precision updating $\LAG$ modulo $\GAG$]\label{th:prec-lag}
  Let $M$ be a module contained in program $P$, $E$ a set of entries. Let $\GAG$
  be a previous state of the global analysis graph, if $\LAG_{M} =
  \lanalyze(M,E,\GAG,\emptyset,\emptyset)$. If $\GAG$ changes to $\GAG'$ the
  analysis result:
  $$
  \lanalyze(M,E,\GAG',\LAG_{M},\emptyset) = \lanalyze(M,E,\GAG',\emptyset,\emptyset)
  $$
  is the same as analyzing from scratch, i.e., the \emph{lfp} of $M$, $E$.
\end{lemma}

\begin{proof}
  The proof of this lemma follows from the proof of Lemma~\ref{th:updlag} and
  the guarantee that $\LAG_{M}$ is the least analysis graph if the
  generalization is removed from \banalyze (Theorem~\ref{th:prec-inc}).
\end{proof}

\begin{theorem}[Precision of \mianalyze{} from scratch]\label{th:prec-mianalyze-scracth}
  Let $P$ be a modular program and $\QQ$ a set of abstract queries.
  The analysis result
  $$
  \A = \mianalyze(P,\QQ,\emptyset,\emptyset) = \manalyze(P,\QQ)
  $$
  such that $\A = \{\GAG, \{\LAG_{M_i}\}\}$, then $\GAG = \GAG'$.
\end{theorem}

\begin{proof}
  Since the \lfp is monotonic w.r.t. the initial assumptions and ${\sf upd}$ is
  monotonic, $\mathit{MA}$ is monotonic. Therefore, chaotic iteration of $MA$
  with the different modules of a program will reach a fixpoint which is the
  {\em least fixed point}, because the separated \lfp of each of the modules is
  computed. Chaotic iteration is guaranteed in the same way as correctness in
  Theorem~\ref{th:mod_scratch}. Termination is guaranteed because $MA$ is
  monotonic and \DD~is finite.
\end{proof}

If $P$ is changed to $P'$ by editions $\Delta$ and it is reanalyzed
incrementally, the algorithm will return a \GAG that encodes the same {\em
  global analysis result} as if $P'$ is analyzed from scratch. I.e., the least
program analysis graph.

\begin{theorem}[Precision of \mianalyze]\label{th:prec-mianalyze}%
  Let $P, P'$ be modular programs that differ by $\diff{}$, $\QQ$ a set of queries, and
  $\A = \mianalyze(P,\QQ,\emptyset,( \emptyset, \emptyset))$, then
  \[
    \mianalyze(P',\QQ,\emptyset,(\emptyset, \emptyset)) =
    \mianalyze(P',\QQ,\A,\diff{}).
  \]
\end{theorem}

\begin{proof}
  This is proved by following the same strategy as in Theorem~\ref{th:mod_inc},
  replacing the termination condition that relied on the widening operator with
  the guarantees that the abstract domain is finite and that $\mathit{MA}$ is
  monotonic, and the guarantee of Lemma~\ref{th:prec-lag} that no imprecision is
  introduced analyzing each individual module.
\end{proof}

\secbeg
\section{Experiments}
\secend
\label{sec:experiments}

We have implemented the proposed algorithm within the \texttt{Ciao}/\texttt{CiaoPP}
system~\cite{%
  hermenegildo11:ciao-design-tplp,%
  ciaopp-sas03-journal-scp%
}, which can be found in \url{https://github.com/ciao-lang/ciaopp}.
We use a selection of well-known benchmarks from previous studies of incremental
analysis, e.g., {\tt ann} (a parallelizer) and {\tt boyer} (a theorem prover
kernel), are programs with a relatively large number of clauses located in a
small number of modules. In contrast, e.g., {\tt bid} is a more modularized
program (see Table~\ref{tab:bench_data} for more details, and
\url{https://github.com/ciao-lang/ciaopp\_tests/tree/master/tests/incanal} for
the source code of the benchmarks).
We used the original modular structure as modular partition,
and evaluated five strategies:
\begin{itemize}
\item \texttt{mon}: the baseline non-modular, non-incremental
  algorithm~\cite{ai-jlp}, i.e., \banalyze described in
  Sec.~\ref{sec:mon-inc} with initial results always empty.
\item \texttt{mon-inc}: the monolithic incremental
  algorithm~\cite{incanal-toplas} as described in
  Sec.~\ref{sec:mon-inc}.
\item \texttt{mon-scc}: the monolithic incremental
  algorithm~\cite{incanal-toplas} as described in Sec.~\ref{sec:mon-inc} with
  the bottom-up deletion strategy of~\cite{incanal-toplas}.
\item \texttt{mod}: as a coarse-grain modular algorithm, which consists on our
  proposed algorithm, without keeping each of the local analysis graphs. Note
  that this is not the same as the algorithm of Sec.~\ref{sec:mod}, as it did
  not consider modifying the modules.
\item \texttt{mod-inc}: our proposed algorithm that is modular and
  incremental (Sec.~\ref{sec:intermod}).
\item \texttt{mod-scc}: for the experiments that include deleting clauses,
  the alternative strategy for updating the analysis following the strongly
  connected components of the program (Sec.~\ref{sec:SCC-del}).

\end{itemize}

\begin{table}[t]
  \centering
  \begin{tabular}{|r|c|r|r|r|}
    \cline{1-5}
    Bench & \# Modules & \#  Predicates 
                       & \# Clauses & LOC \\
    \cline{1-5}
    \texttt{hanoi} & 2 & $4$ & $6$ & 46\\
    \texttt{aiakl} & 4 & $8$ & $15$ & 71\\
    \texttt{qsort} & 3 & $8$  & $17$ & 49\\
    \texttt{progeom} & 2 & $10$ & $18$& 73\\
    \texttt{bid} & 7 & $21$ & $48$ & 207\\
    \texttt{rdtok} & 5 & $15$ & $57$ & 293\\
    \texttt{cleandirs} & 3 & $36$ & $81$ & 528 \\
    \texttt{read} & 3 & 25 & 94 & 352\\
    \texttt{warplan} & 3 & $37$ & $114$ & 281\\
    \texttt{boyer} & 4 & $29$ & $145$ & 279 \\
    \texttt{peephole} & 3 & $33$ & $169$& 377\\
    \texttt{witt} & 4 & $69$ & $176$& 618\\
    \texttt{ann} & 3 & $69$ & $229$ & 641\\
    \texttt{manag\_proj} & 8 & $105$ & $143$ & 805\\
    \texttt{check\_links} & 4 & $220$ & $504$ & 2042\\
    \cline{1-5}
 \end{tabular}
 \caption{Benchmark characteristics sorted by lines of code.\label{tab:bench_data}}
\end{table}

We performed experiments with four different abstract domains: a simple
reachability domain (\texttt{pdb}), a groundness domain (\texttt{gr}),
a dependency tracking via propositional clauses domain~\cite{free-def-comb}
(\texttt{def}), and the %
sharing and freeness abstract domain~\cite{%
  freeness-iclp91%
} (pointer sharing and uninitialized
pointers, \texttt{shfr}). 
We use the exported predicates from the main module (with $\top$ call
pattern) as the set of initial queries (i.e., no additional information is
provided in the program).

We ran all experiments on a Linux machine (kernel 4.9.0-8-amd64) with Debian
9.0, a Xeon Gold 6154 CPU, and 16 GB of RAM. However, running the test in a
standard laptop shows similar performance.

\paragraph{\textbf{Analyzing from scratch.}}

We first study the analysis from scratch of all the benchmarks for all
approaches, to observe the overhead introduced by the bookkeeping of the
algorithms. The analysis times in milliseconds are shown in
Table~\ref{tab:from_scratch}. For each benchmark, four rows are shown,
corresponding to the four analysis algorithms mentioned earlier: monolithic
(\texttt{mon}), monolithic incremental ({\tt mon-inc}), modular ({\tt mod}),
and, lastly, modular incremental ({\tt mod-inc}), i.e., the proposed approach.
In the monolithic setting, the overhead introduced is negligible.
Interestingly, the incremental modular analysis performs better overall than
simply modular even in analysis from scratch. This is due to the reuse of local
information specially in complex benchmarks such as {\tt ann}, {\tt peephole},
{\tt warplan}, or {\tt witt}.
In the best cases (e.g., {\tt witt}, \texttt{cleandirs}, or {\tt check\_links}
analyzed with \texttt{shfr}) the performance of incremental modular competes
with monolithic thanks to the incremental updates, dropping from $20s$ to
$3s$, from $1.2s$ to $0.8s$, and from $2.5s$ to $1.2s$ respectively.

Note that a smaller program does not necessarily imply that the analyzer
will run faster, it depends on the structure of the code and
the kind of data that the program operates with.
Also, the cost of performing a modular analysis highly depends on the module
scheduling policy, and whether the modular partitions were correctly produced. In
this case, if the programmer divided the program in a reasonable manner.
For example, analyzing \texttt{boyer} (with any domain) modularly comes at no
cost, while in the case of \texttt{cleandirs} it is 3 times slower
than doing it monolithically. 

\begin{table}[t]
  \footnotesize
  \begin{tabular}{ll}
  \begin{minipage}{0.2\textwidth}
  \begin{tabular}{|l|l|r|r|r|r|}
    \cline{1-6}
  \multicolumn{2}{|l|}{Benchmark} & \texttt{pdb} & \texttt{gr} & \texttt{def} & \texttt{shfr} \\
  \cline{1-6}
    \multirow{4}{*}{\texttt{hanoi}} &
    \texttt{mon} & 5.2 & 2.9 & 2.2 & 10.7 \\
                                  & \texttt{mon-inc} & 5.3 & 3.0 & 2.2 & 10.2 \\
                                  & \texttt{mod} & 12.3 & 7.2 & 5.8 & 22.1 \\
                                  & \texttt{mod-inc} & 10.0 & 6.2 & 4.6 & 18.3 \\
\cline{1-6}
 \multicolumn{2}{|l|}{\multirow{4}{*}{\texttt{aiakl}}}
& 5.9 & 6.4 & 4.6 & 7.9 \\
\multicolumn{2}{|l|}{} & 7.6 & 7.3 & 5.8 & 8.4 \\
\multicolumn{2}{|l|}{} & 15.0 & 18.9 & 14.0 & 18.0 \\
\multicolumn{2}{|l|}{} & 16.0 & 15.5 & 13.0 & 16.4 \\
\cline{1-6}
 \multicolumn{2}{|l|}{\multirow{4}{*}{\texttt{qsort}}}
& 7.8 & 8.3 & 4.0 & 9.5 \\
\multicolumn{2}{|l|}{} & 8.0 & 8.5 & 4.3 & 10.5 \\
\multicolumn{2}{|l|}{} & 21.7 & 21.6 & 13.5 & 24.9 \\
\multicolumn{2}{|l|}{} & 19.5 & 20.3 & 10.0 & 20.1 \\
\cline{1-6}
 \multicolumn{2}{|l|}{\multirow{4}{*}{\texttt{progeom}}}
& 5.4 & 5.4 & 5.1 & 6.4 \\
\multicolumn{2}{|l|}{} & 6.1 & 5.4 & 5.4 & 7.1 \\
\multicolumn{2}{|l|}{} & 24.1 & 23.6 & 21.6 & 28.7 \\
\multicolumn{2}{|l|}{} & 18.4 & 18.7 & 14.8 & 20.3 \\
\cline{1-6}
 \multicolumn{2}{|l|}{\multirow{4}{*}{\texttt{bid}}}
& 18.8 & 14.8 & 9.9 & 22.9 \\
\multicolumn{2}{|l|}{} & 17.7 & 15.4 & 10.2 & 26.8 \\
\multicolumn{2}{|l|}{} & 61.0 & 55.1 & 39.1 & 68.3 \\
\multicolumn{2}{|l|}{} & 42.4 & 42.1 & 32.4 & 55.7 \\
\cline{1-6}
 \multicolumn{2}{|l|}{\multirow{4}{*}{\texttt{rdtok}}}
& 33.5 & 44.0 & 15.7 & 63.3 \\
\multicolumn{2}{|l|}{} & 51.3 & 29.2 & 17.8 & 66.0 \\
\multicolumn{2}{|l|}{} & 85.1 & 61.3 & 40.2 & 122.0 \\
\multicolumn{2}{|l|}{} & 52.3 & 53.5 & 36.6 & 90.9 \\
\cline{1-6}
 \multicolumn{2}{|l|}{\multirow{4}{*}{\texttt{cleandirs}}}
& 33.2 & 27.6 & 26.2 & 384.1 \\
\multicolumn{2}{|l|}{} & 31.7 & 29.1 & 27.7 & 389.0 \\
\multicolumn{2}{|l|}{} & 145.5 & 123.8 & 140.3 & 1189.2 \\
\multicolumn{2}{|l|}{} & 93.8 & 77.2 & 80.5 & 778.3 \\
\cline{1-6}
  \end{tabular}
\end{minipage}
             &
               \hspace{4cm}
               
\begin{minipage}{0.3\textwidth}
  \begin{tabular}{|l|l|r|r|r|r|}
\cline{1-6}
    \multicolumn{2}{|l|}{Benchmark} &   \texttt{pdb}   & \texttt{gr} & \texttt{def} & \texttt{shfr}  \\
  \cline{1-6}
       \multicolumn{2}{|l|}{\multirow{4}{*}{\texttt{read}}}
& 217.5 & 116.5 & 47.8 & 399.0 \\
\multicolumn{2}{|l|}{} & 172.3 & 105.0 & 35.0 & 400.7 \\
\multicolumn{2}{|l|}{} & 192.0 & 118.4 & 45.1 & 422.4 \\
\multicolumn{2}{|l|}{} & 189.9 & 126.9 & 45.5 & 472.4 \\
\cline{1-6}
 \multicolumn{2}{|l|}{\multirow{4}{*}{\texttt{warplan}}}
& 46.0 & 24.5 & 20.1 & 63.3 \\
\multicolumn{2}{|l|}{} & 41.0 & 26.6 & 16.6 & 64.3 \\
\multicolumn{2}{|l|}{} & 71.7 & 52.7 & 35.8 & 180.7 \\
\multicolumn{2}{|l|}{} & 57.0 & 37.1 & 24.1 & 102.9 \\
\cline{1-6}
 \multicolumn{2}{|l|}{\multirow{4}{*}{\texttt{boyer}}}
& 38.3 & 24.1 & 14.9 & 50.0 \\
\multicolumn{2}{|l|}{} & 37.0 & 31.5 & 17.4 & 51.5 \\
\multicolumn{2}{|l|}{} & 48.3 & 39.3 & 21.5 & 68.2 \\
\multicolumn{2}{|l|}{} & 44.9 & 37.3 & 19.1 & 65.4 \\
\cline{1-6}
 \multicolumn{2}{|l|}{\multirow{4}{*}{\texttt{peephole}}}
& 67.0 & 43.2 & 19.2 & 157.6 \\
\multicolumn{2}{|l|}{} & 64.1 & 45.6 & 21.2 & 156.4 \\
\multicolumn{2}{|l|}{} & 155.8 & 75.6 & 43.6 & 392.8 \\
\multicolumn{2}{|l|}{} & 115.1 & 62.4 & 40.2 & 267.0 \\
\cline{1-6}
 \multicolumn{2}{|l|}{\multirow{4}{*}{\texttt{witt}}}
& 183.4 & 11.6 & 33.5 & 2490.4 \\
\multicolumn{2}{|l|}{} & 186.0 & 16.4 & 38.8 & 2491.2 \\
\multicolumn{2}{|l|}{} & 1134.6 & 6.7 & 120.8 & 20550.3 \\
\multicolumn{2}{|l|}{} & 414.8 & 9.8 & 71.1 & 3222.9 \\
\cline{1-6}
 \multicolumn{2}{|l|}{\multirow{4}{*}{\texttt{ann}}}
& 84.5 & 58.4 & 35.0 & 120.5 \\
\multicolumn{2}{|l|}{} & 85.4 & 64.1 & 38.5 & 123.7 \\
\multicolumn{2}{|l|}{} & 264.1 & 174.5 & 89.7 & 296.6 \\
\multicolumn{2}{|l|}{} & 145.3 & 127.0 & 60.6 & 241.5 \\
\cline{1-6}
 \multicolumn{2}{|l|}{\multirow{4}{*}{\texttt{manag\_proj}}}
& 111.0 & 24.1 & 51.3 & 18049.2 \\
\multicolumn{2}{|l|}{} & 98.3 & 28.3 & 48.8 & 17967.3 \\
\multicolumn{2}{|l|}{} & 291.3 & 54.7 & 150.3 & 37184.9 \\
\multicolumn{2}{|l|}{} & 221.8 & 44.4 & 104.7 & 34595.0 \\
\cline{1-6}
 \multicolumn{2}{|l|}{\multirow{4}{*}{\texttt{check\_links}}}
& 701.7 & 301.6 & 167.5 & 803.3 \\
\multicolumn{2}{|l|}{} & 678.9 & 251.5 & 178.5 & 819.2 \\
\multicolumn{2}{|l|}{} & 1292.6 & 680.8 & 600.5 & 2530.3 \\
\multicolumn{2}{|l|}{} & 776.1 & 360.8 & 267.2 & 1162.5 \\
\cline{1-6}

  \end{tabular}
\end{minipage}
  \end{tabular}
  \caption{Analysis times from scratch (ms).}
  \label{tab:from_scratch}
\end{table}

\paragraph{\textbf{Clause addition/deletion experiment.}}
\label{sec:exp_add}
As a stress test for the proposed algorithm, we measured the cost of
re-analyzing the program incrementally adding (or removing) one clause at a
time, until the program is completed (or empty). That is, for the addition
experiment, the analysis was first run for the first clause only. Then the next
clause was added and the resulting program (re)analyzed. This process was
repeated until all the clauses in all the modules were added.
For the deletion experiment, starting from an already analyzed
program, the last clause was deleted and the resulting program (re)analyzed.
This process was repeated until no clauses were left. The experiment was
performed for all the approaches using the initial (top-down) deletion strategy
(\texttt{mod-inc}) and the SCC-partition deletion strategy of Sec.
\ref{sec:SCC-del} (\texttt{mod-scc}).

\newcommand{\includedetgraphlarge}[3]
{
\begin{minipage}{0.5\textwidth}
  \includegraphics[width=1\textwidth]{#1_#2_#3_details}
\end{minipage}
}

\newcommand{\detrowlarge}[2]
{
  \includedetgraphlarge{#1}{pdb}{#2}
  &
  \includedetgraphlarge{#1}{gr}{#2}
  \\ 
  \includedetgraphlarge{#1}{def}{#2}
  &
  \includedetgraphlarge{#1}{shfr}{#2}
  \\ 
}

\subsection{Discussion}

In our experiments we observed that the analyses performed with the \texttt{gr}
and \texttt{def} domains were the most relevant to evaluate the usefulness of
the algorithm. This is due to the domain operations being fairly simple (when
compared with the cost of executing the fixpoint algorithm), so that, the complexity
of the algorithm is not completely hidden by the complexity of the domain
operations (e.g. \texttt{shfr}). At the same time, they are complex enough that there is some
fixpoint iteration (which does not occur in, e.g., \texttt{pdb}, since $\top$ is
assumed for every call pattern). Therefore, in this section we focus mainly in
the analysis with the \texttt{def} domain, but include as well some discussion
about \texttt{gr}.
Nevertheless, for the results for the remaining domains we refer the reader to
the~\ref{app:benchmarks}.

\begin{figure}[t]
  \begin{tabular}{l l}
    \includedetgraphlarge{warplan}{def}{add} &
                                               \includedetgraphlarge{warplan}{def}{del}
  \end{tabular}
  \caption{Analysis time (ms) for \texttt{warplan} with \texttt{def} for both experiments.}
  \label{fig:warplan-details}
\end{figure}

Fig.~\ref{fig:warplan-details} shows the addition and deletion experiments for
the \texttt{warplan} benchmark analyzed with \texttt{def}. Each point represents
the time taken to reanalyze the program after incrementally adding/deleting one
clause. The horizontal axis denotes the number of clauses added/deleted at that
point of the experiment.
We observe that the proposed incremental algorithm outperforms overall
the non-incremental settings when the time needed to reanalyze is large.
We find that for smaller benchmarks our algorithm performs up to 8 times faster
than the traditional monolithic, non-incremental algorithm, and, in the worst
cases performs as fast as the traditional modular algorithm. The detailed
analysis times per iteration for the remaining benchmarks are available in
Figs.~\ref{fig:app-small-details},~\ref{fig:app-details-a},
and~\ref{fig:app-details-b}. 

We observe that, even when analyzing takes less time, i.e., when the program has
fewer clauses, the analysis time of the algorithm proposed is faster overall.
Moreover, as the analysis grows in complexity, the cost our approach grows
significantly slower than that of the traditional algorithm.
In the case of the deletion experiments, we observe also clear
advantages, specially when using the strategy of partitions in SCC
presented in Sec.~\ref{sec:SCC-del}.

\newcommand{\includeaccgraphbig}[2]
{
  \includegraphics[width=1\textwidth]{#1_#2_acc}
}

\newcommand{\includeaccgraphsmall}[2]
{
\begin{minipage}{0.7\textwidth}
  \includegraphics[width=1\textwidth]{#1_#2_acc}
\end{minipage}
}

\newcommand{\benchsplarge}{\hspace*{2pt}}
\newcommand{\benchspsmall}{\hspace*{4pt}}

\newcommand{\benchnameslarge}{\scriptsize\texttt{
    hanoi \benchsplarge
    aiakl \benchsplarge
    qsort \benchsplarge
    progeom \benchsplarge
    bid \benchsplarge
    rdtok \benchsplarge
    cleandirs \benchsplarge
    read \benchsplarge
    warplan \benchsplarge
    boyer \benchsplarge
    peephole \benchsplarge
    witt \benchsplarge
    ann \benchspsmall
    manag\_proj \benchspsmall
    check\_links 
  }}

\newcommand{\benchnamessmall}{\tiny\texttt{
    \hspace*{1.3cm} hanoi \benchspsmall \hspace*{1pt}
    aiakl \benchspsmall 
    qsort \benchspsmall
    progeom \benchspsmall \hspace*{1pt}
    bid \benchspsmall  \hspace*{1pt}
    rdtok \hspace*{1pt} %
    cleandirs %
    read \benchspsmall
    warplan  \hspace*{2pt}%
    boyer \benchspsmall
    peephole \benchspsmall
    witt \benchspsmall  \hspace*{1pt}
    ann \benchspsmall  %
    manag\_proj %
    check\_links 
  }}

\newcommand{\acctablebig}[3]
{
  \hspace{-25mm}
  \begin{tabular}{c}
    {\large \texttt{#2}} \\
    \hspace{8mm} \includeaccgraphbig{#2}{#1} \\
    \hspace{15mm} \benchnameslarge\\
    \ \\
    {\large \texttt{#3}} \\
    \hspace{8mm} \includeaccgraphbig{#3}{#1} \\
    \hspace{15mm} \benchnameslarge\\
 \end{tabular}%
}

\newcommand{\acctablesmall}[1]
{
  \hspace{-25mm}
  \begin{tabular}{c c}
    \texttt{pdb} & \texttt{gr} \\
    \includeaccgraphsmall{pdb}{#1} & \hspace{-10mm} \includeaccgraphsmall{gr}{#1} \\
    \hspace{4mm} \benchnamessmall &  \hspace{-5mm} \benchnamessmall \\
    \texttt{def} & \texttt{shfr} \\
    \includeaccgraphsmall{def}{#1}  & \hspace{-10mm} \includeaccgraphsmall{shfr}{#1} \\
    \hspace{4mm} \benchnamessmall & \hspace{-5mm} \benchnamessmall \\
  \end{tabular}%
}

\paragraph{\textbf{Analysis time per action.}}

\newcommand{\colm}[1]{\texttt{m#1}}

\begin{table}[t]
  \footnotesize
  \centering

\begin{tabular}{|>{\tt}l|rrr|rrr|rrr|rrr|}
\toprule
  {} & \multicolumn{3}{l|}{\texttt{mon}} & \multicolumn{3}{l|}{\texttt{mon-inc}} & \multicolumn{3}{l|}{\texttt{mod}} & \multicolumn{3}{l|}{\texttt{mod-inc}} \\
  \midrule
  bench & \colm{ean} &   \colm{ax} & \colm{in} &    \colm{ean} &   \colm{ax} & \colm{in} & \colm{ean} &   \colm{ax} & \colm{in} &  \colm{ean} &   \colm{ax} & \colm{in} \\
\midrule
aiakl       &  2.2 &   5.6 & 1.0 &     1.8 &   5.9 & 1.3 &  1.8 &  14.9 & 0.5 &     1.7 &  14.1 & 0.5 \\
ann         &  7.2 &  76.8 & 1.1 &     3.3 &  51.8 & 1.6 &  2.5 & 156.8 & 0.3 &     2.3 & 133.9 & 0.6 \\
bid         &  3.4 &  18.1 & 1.5 &     2.6 &  11.9 & 1.9 &  2.5 &  32.7 & 0.3 &     2.3 &  26.6 & 0.5 \\
boyer       & 21.5 &  46.4 & 1.5 &     3.2 &  13.0 & 1.4 &  6.0 &  15.1 & 0.4 &     1.9 &  14.8 & 0.5 \\
check\_links & 29.3 & 608.6 & 2.6 &    14.5 & 571.5 & 5.5 & 21.6 & 889.4 & 0.4 &     7.1 & 664.1 & 0.8 \\
cleandirs   &  9.1 &  28.9 & 1.5 &     4.5 &  18.6 & 1.9 &  6.2 &  96.5 & 0.9 &     3.7 &  63.4 & 0.9 \\
hanoi       &  2.4 &   4.7 & 0.7 &     1.8 &   5.3 & 1.1 &  2.9 &  11.5 & 0.4 &     2.1 &  10.0 & 0.4 \\
manag\_proj  & 19.7 & 100.8 & 4.4 &     8.3 &  35.2 & 4.8 &  6.9 &  97.7 & 0.3 &     4.6 &  69.5 & 0.4 \\
peephole    &  9.6 &  64.8 & 1.5 &     2.7 &  30.2 & 1.6 &  2.1 &  95.8 & 0.4 &     1.9 &  76.5 & 0.6 \\
progeom     &  2.3 &   5.4 & 1.1 &     1.6 &   3.4 & 0.7 &  2.3 &  10.2 & 0.8 &     2.2 &   9.7 & 0.8 \\
read        & 38.9 & 177.4 & 1.1 &    13.4 & 151.3 & 1.3 & 18.5 & 214.2 & 0.8 &     9.9 & 165.2 & 0.6 \\
qsort       &  2.8 &  11.0 & 1.1 &     1.8 &   4.4 & 1.1 &  2.8 &  10.5 & 0.4 &     2.4 &   9.5 & 0.5 \\
rdtok       &  8.2 &  31.1 & 1.2 &     6.2 &  57.6 & 1.3 &  6.0 &  27.2 & 0.3 &     6.0 &  59.3 & 0.4 \\
warplan     & 10.2 &  35.0 & 0.8 &     4.2 &  18.5 & 1.5 &  5.4 &  30.7 & 0.7 &     2.5 &  21.3 & 0.9 \\
witt        & 15.2 & 177.7 & 1.5 &     5.5 & 142.0 & 2.2 & 11.1 & 653.2 & 0.3 &     4.8 & 323.4 & 0.4 \\
\bottomrule \end{tabular}

   \caption{Analysis times (ms) per action of the clause \emph{addition} experiment with \texttt{def}.}\label{tab:add-abs-time}
\end{table}

\begin{table}[t]
  \scriptsize
  \hspace{-14mm}
\begin{tabular}{|>{\tt}l|rrr|rrr|rrr|rrr|rrr|rrr|}
\toprule
  {} & \multicolumn{3}{l|}{\texttt{mon}} & \multicolumn{3}{l|}{\texttt{mon-inc}} & \multicolumn{3}{l|}{\texttt{mon-scc}} & \multicolumn{3}{l|}{\texttt{mod}} & \multicolumn{3}{l|}{\texttt{mod-inc}} & \multicolumn{3}{l|}{\texttt{mod-scc}} \\
  \midrule
bench & \colm{ean} &   \colm{ax} & \colm{in} &    \colm{ean} &   \colm{ax} & \colm{in} &    \colm{ean} &   \colm{ax} & \colm{in} & \colm{ean} &     \colm{ax} & \colm{in} &    \colm{ean} &   \colm{ax} & \colm{in} &    \colm{ean} &   \colm{ax} & \colm{in} \\
\midrule
aiakl       &  2.6 &   7.2 & 1.3 &     1.9 &   6.9 & 1.3 &     1.4 &   6.1 & 0.9 &  2.0 &    15.2 & 0.5 &     1.8 &  13.4 & 0.4 &     1.5 &  13.7 & 0.3 \\
ann         &  7.9 &  85.8 & 1.2 &     4.0 &  80.4 & 1.8 &     3.3 &  80.3 & 1.5 &  8.8 &   197.1 & 0.3 &     3.4 & 153.6 & 0.6 &     2.8 & 153.1 & 0.6 \\
bid         &  3.0 &  17.1 & 1.6 &     2.5 &  19.4 & 1.6 &     1.9 &  18.3 & 1.4 &  3.4 &    59.8 & 0.5 &     2.9 &  53.1 & 0.4 &     2.6 &  50.3 & 0.4 \\
boyer       & 21.1 &  51.5 & 1.3 &    11.4 &  35.3 & 1.3 &     2.2 &  36.2 & 1.1 &  6.3 &    44.9 & 0.4 &     5.3 &  45.3 & 0.6 &     1.6 &  45.1 & 0.3 \\
check\_links & 29.0 & 635.6 & 1.5 &    28.3 & 674.8 & 6.4 &    20.9 & 598.2 & 5.5 & 22.6 &   946.0 & 0.3 &    17.0 & 720.1 & 0.7 &    13.3 & 716.0 & 0.5 \\
cleandirs   &  9.2 &  27.5 & 1.6 &     5.6 &  32.1 & 1.7 &     3.3 &  30.1 & 1.4 &  6.2 &   128.9 & 0.9 &     4.4 &  90.8 & 0.7 &     3.5 &  86.5 & 0.6 \\
hanoi       &  2.6 &   5.3 & 0.8 &     1.8 &   5.6 & 0.9 &     1.5 &   5.9 & 0.7 &  3.5 &    13.8 & 0.5 &     2.7 &  12.4 & 0.7 &     2.5 &  12.6 & 0.5 \\
manag\_proj  & 20.9 & 103.6 & 4.5 &     9.3 &  98.3 & 4.5 &     7.7 &  90.1 & 4.1 &  7.9 &   259.3 & 0.3 &     5.7 & 212.1 & 0.3 &     5.8 & 217.7 & 0.3 \\
peephole    & 10.3 & 100.6 & 1.7 &     3.4 &  67.4 & 1.5 &     2.3 &  65.4 & 1.3 &  5.6 &   138.6 & 1.1 &     3.7 & 113.4 & 1.0 &     2.7 & 116.2 & 0.9 \\
progeom     &  2.3 &   5.3 & 1.1 &     1.8 &   5.5 & 1.0 &     1.2 &   6.3 & 0.9 &  2.7 &    22.0 & 0.7 &     2.5 &  19.3 & 0.7 &     2.1 &  18.6 & 0.6 \\
read        & 39.0 & 184.7 & 1.1 &    33.8 & 189.1 & 1.1 &     3.9 & 171.9 & 1.0 & 17.9 &   185.7 & 0.7 &    19.4 & 191.3 & 0.7 &     5.8 & 187.2 & 0.6 \\
qsort       &  2.5 &   7.5 & 1.1 &     2.2 &   8.5 & 1.2 &     1.4 &   9.3 & 0.8 &  3.6 &    22.3 & 0.4 &     3.4 &  20.4 & 0.6 &     3.1 &  20.8 & 0.5 \\
rdtok       &  8.2 &  29.7 & 1.5 &     7.3 &  31.8 & 1.3 &     2.2 &  31.8 & 1.1 &  6.6 &    61.8 & 0.3 &     8.2 &  54.1 & 0.6 &     4.8 &  54.8 & 0.3 \\
warplan     & 11.1 &  52.7 & 1.0 &     6.2 &  40.2 & 1.3 &     2.0 &  39.3 & 1.0 &  5.5 &    63.2 & 0.8 &     4.1 &  56.0 & 1.0 &     2.7 &  53.7 & 0.7 \\
witt        & 14.6 & 174.8 & 1.2 &     7.1 & 179.3 & 2.1 &     4.5 & 185.2 & 1.9 & 15.0 & 1,021.7 & 0.3 &     7.9 & 461.3 & 0.4 &     7.1 & 423.9 & 0.4 \\
\bottomrule \end{tabular}

   \caption{Analysis times (ms) per action of the clause \emph{deletion} experiment with \texttt{def}.}\label{tab:del-abs-time}
\end{table}

In order to get an overall idea of the cost in terms of the time taken
by analysis we
have included Tables~\ref{tab:add-abs-time} and~\ref{tab:del-abs-time} for the
addition and deletion experiment respectively. They show, split by benchmark and
analysis configuration, the \texttt{mean},
\texttt{max}imum, and \texttt{min}imum analysis times after each
modification made in the experiment, for each program.
The objective is to provide intuition for 
the ``response times'' of the analyses after each such modification.
We center our attention on the costlier \emph{instances} of
the benchmarks, i.e., the (re)analysis runs which take the longest
after a modification is 
performed in the program
for the traditional, \emph{monolithic} analysis.
In absolute terms these are \texttt{check\_links}, with the largest
analysis time ($608.6$ms), followed by \texttt{witt} ($177.7$ms),
\texttt{read} ($177.4$ms), and \texttt{manag\_proj} ($100.8$ms). In
terms of overall cost (\emph{mean}) of the reanalysis %
we have \texttt{read} ($38.5$ms), \texttt{check\_links} ($29.3$ms),
and \texttt{boyer} ($21.5$ms).
These high differences between the mean and maximum analysis times 
are due to the very small %
values for the first additions,
in which the program is very small and there are no iterations.
The analysis times for the remaining experiments are available in
detailed analysis times for each step are provided in
Fig.~\ref{app:abs-times}.
These should be in principle the benchmarks that we should focus on
incrementalizing, as more time is saved. This applies not only to the
monolithic analysis, but also to the modular analysis.
To observe the increase in performance obtained, Table~\ref{tab:speedup} shows the
speedup of our algorithm with respect to: \texttt{mon-inc}, in the case of the addition
experiments, and \texttt{mod-scc}, in the case of the deletion experiment.
The analysis times for the remaining speedups are provided
in~\ref{app:speedups}.

\begin{figure}
  \scriptsize
  \centering
  
  \hspace*{-3mm}
\begin{tabular}{|>{\tt}l|r|r|r|}
  \toprule
  \multicolumn{1}{|l|}{\multirow{2}{*}{\texttt{bench}}} & \multicolumn{3}{c|}{\texttt{mod-inc}}  \\
                                                        & vs.  & vs.  & vs.  \\
                                                        & \texttt{mon} & \texttt{mon-inc} &  \texttt{mod} \\
\midrule
             aiakl &              1.3 &                  1.1 &              1.0 \\
               ann &              3.1 &                  1.4 &              1.1 \\
               bid &              1.5 &                  1.1 &              1.1 \\
             boyer &             11.6 &                  1.7 &              3.2 \\
       check\_links &              4.1 &                  2.1 &              3.1 \\
         cleandirs &              2.5 &                  1.2 &              1.7 \\
             hanoi &              1.2 &                  0.9 &              1.4 \\
 manag\_proj &              4.3 &                  1.8 &              1.5 \\
          peephole &              5.0 &                  1.4 &              1.1 \\
           progeom &              1.1 &                  0.7 &              1.1 \\
       read &              3.9 &                  1.4 &              1.9 \\
             qsort &              1.2 &                  0.7 &              1.2 \\
             rdtok &              1.4 &                  1.0 &              1.0 \\
           warplan &              4.1 &                  1.7 &              2.2 \\
              witt &              3.1 &                  1.1 &              2.3 \\
\bottomrule \end{tabular}
\begin{tabular}{|>{\tt}l|r|r|r|r|r|}
  \toprule
  \multicolumn{1}{|l|}{\multirow{2}{*}{\texttt{bench}}} & \multicolumn{5}{c|}{\texttt{mod-scc}}  \\
                                                        & vs. & vs. & vs. & vs. & vs. \\
                                                        &  \texttt{mon} &  \texttt{mon-inc} &  \texttt{mon-scc} &  \texttt{mod} &  \texttt{mod-inc} \\
\midrule
            aiakl &              1.7 &                  1.2 &                  0.9 &              1.3 &                  1.2 \\
              ann &              2.9 &                  1.4 &                  1.2 &              3.2 &                  1.2 \\
              bid &              1.2 &                  1.0 &                  0.7 &              1.3 &                  1.1 \\
            boyer &             13.0 &                  7.1 &                  1.4 &              3.9 &                  3.3 \\
      check\_links &              2.2 &                  2.1 &                  1.6 &              1.7 &                  1.3 \\
        cleandirs &              2.6 &                  1.6 &                  1.0 &              1.8 &                  1.3 \\
            hanoi &              1.1 &                  0.7 &                  0.6 &              1.4 &                  1.1 \\
manag\_proj &              3.6 &                  1.6 &                  1.3 &              1.4 &                  1.0 \\
         peephole &              3.8 &                  1.3 &                  0.9 &              2.1 &                  1.4 \\
          progeom &              1.1 &                  0.9 &                  0.6 &              1.3 &                  1.2 \\
      read &              6.8 &                  5.9 &                  0.7 &              3.1 &                  3.4 \\
            qsort &              0.8 &                  0.7 &                  0.5 &              1.2 &                  1.1 \\
            rdtok &              1.7 &                  1.5 &                  0.5 &              1.4 &                  1.7 \\
          warplan &              4.2 &                  2.3 &                  0.8 &              2.1 &                  1.6 \\
             witt &              2.1 &                  1.0 &                  0.6 &              2.1 &                  1.1 \\
\bottomrule \end{tabular}
   
  \caption{Speedups of the clause \emph{addition} (left) and \emph{deletion} (right) experiments with \texttt{def}.}\label{tab:speedup}
\end{figure}

\paragraph{\textbf{Accumulated analysis time.}}

\begin{figure*}[t]
  \hfill \hspace{6cm}
  \includegraphics[width=0.15\textwidth]{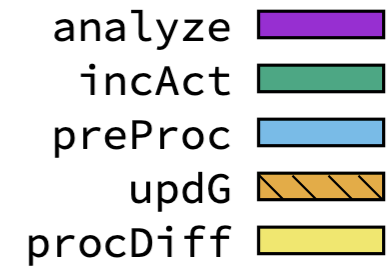}

  \vspace{-1cm}
  \hspace{-7mm}

  \begin{tabular}{l}
    \multicolumn{1}{c}{\texttt{gr}} \\
    \includeaccgraphbig{gr}{add}  \vspace{-5mm}\\
    \benchnamessmall \\
    \multicolumn{1}{c}{\texttt{def}} \\
    \includeaccgraphbig{def}{add} \vspace{-5mm}\\
    \benchnamessmall \\
  \end{tabular}
   \caption{Accumulated analysis time (normalized w.r.t \texttt{mon}) adding
    clauses. The order inside each set of bars is:
  |{\tt mon}|{\tt mon-inc}|{\tt mod}|{\tt mod-inc}|.}
  \label{fig:add-acc}
\end{figure*}

\begin{figure*}[t]
  \hfill \hspace{6cm}
  \includegraphics[width=0.15\textwidth]{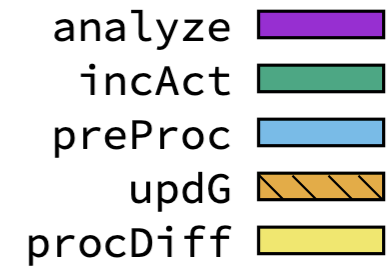}

  \vspace{-1cm}
  \hspace{-7mm}

  \begin{tabular}{l}
    \multicolumn{1}{c}{\texttt{gr}} \\
    \includeaccgraphbig{gr}{del} \vspace{-5mm} \\
    \benchnamessmall \\
    \multicolumn{1}{c}{\texttt{def}} \\
    \includeaccgraphbig{def}{del} \vspace{-5mm} \\
    \benchnamessmall \\
  \end{tabular}
  \caption{Accumulated analysis time (normalized w.r.t \texttt{mon}) deleting
    clauses. The order inside each set of bars is:
    |{\tt mon}|{\tt mon-inc}|{\tt mon-scc}|{\tt mod}|{\tt mod-inc}|{\tt
      mod-scc}|.}
  \label{fig:del-acc}
\end{figure*}

To observe in a more detailed manner 
how the analyzers behave we present in Figs.~\ref{fig:add-acc} and
\ref{fig:del-acc} the accumulated analysis times, i.e., the analysis time 
of all the experiments aggregated by benchmark, \emph{divided by how
  much time was spent in the different parts of the algorithm}.
The results are for 
the \texttt{gr} and \texttt{def} abstract domains, and %
each of the bars shows of the full set of addition and deletion experiments.

Fig.~\ref{fig:add-acc} shows the accumulated analysis time for the addition
experiments. As mentioned before the bars are split to show the time
taken in each operation: {\tt 
  analyze} is the time spent in the module analyzer, {\tt incAct} is the time
spent updating the local analysis results, {\tt preProc} is the time spent
processing clause relations (e.g., calculating the SCCs), {\tt updG} is the time
spent updating \GAG, and {\tt procDiff}, the time spent applying the changes to
the analysis. This last parameter only appears in the incremental settings.
The bars are normalized with respect to the monolithic non-incremental
(\texttt{mon}) algorithm, which always takes ``1'' to execute. E.g., if
analyzing {\tt rdtok} with the \texttt{gr} domain for the monolithic
non-incremental setting is taken as 1, the modular incremental
(\texttt{mod-inc}) setting takes approx.\ $0.6$, so it is approx.\ $1.67$ times
faster.

As before, the benchmarks are sorted by number of LOC. Because of this, it
can be observed that the incremental analysis does tend to be more useful as 
program size grows. Overall, the incremental settings (\texttt{mon-inc},
\texttt{mod-inc}) are always faster than the corresponding non-incremental
settings (\texttt{mon}, \texttt{mod}). Furthermore, while the traditional
modular analysis is sometimes slower than the monolithic one (for the small
benchmarks: {\tt hanoi} and {\tt qsort}), our modular incremental algorithm
always outperforms both, obtaining $10\times$ overall speedup over monolithic
in the best cases (\texttt{boyer} analyzed with \texttt{def} or {\tt peephole}
analyzed with \texttt{shfr}). Furthermore, in the larger benchmarks modular
incremental outperforms even the monolithic incremental approach.

Fig.~\ref{fig:del-acc} shows the results of the deletion experiment. The
analysis performance of the incremental approaches is in general better than the
non-incremental approaches, except some cases for small programs. Again, our
proposed algorithm shows very good performance, in the best case $10\times$
speedup (\texttt{read} analyzed with \texttt{shfr}), and overall $5\times$
speedup ({\tt ann}, {\tt peephole}, and \texttt{witt}), competing with
monolithic incremental {\tt scc} and outperforming in general monolithic
incremental {\tt td}.
The SCC-guided deletion strategy seems to be more efficient than the
top-down deletion strategy. This confirms that the top-down deletion
strategy tends to be quite pessimistic when deleting information, and
modular partitions limit the scope of deletion.
For the accumulated analysis time of the remaining domains, please
see Figs.~\ref{fig:app-add-acc} and~\ref{fig:app-del-acc}.

\begin{figure}[t]
  \centering
  \includegraphics[width=0.49\textwidth]{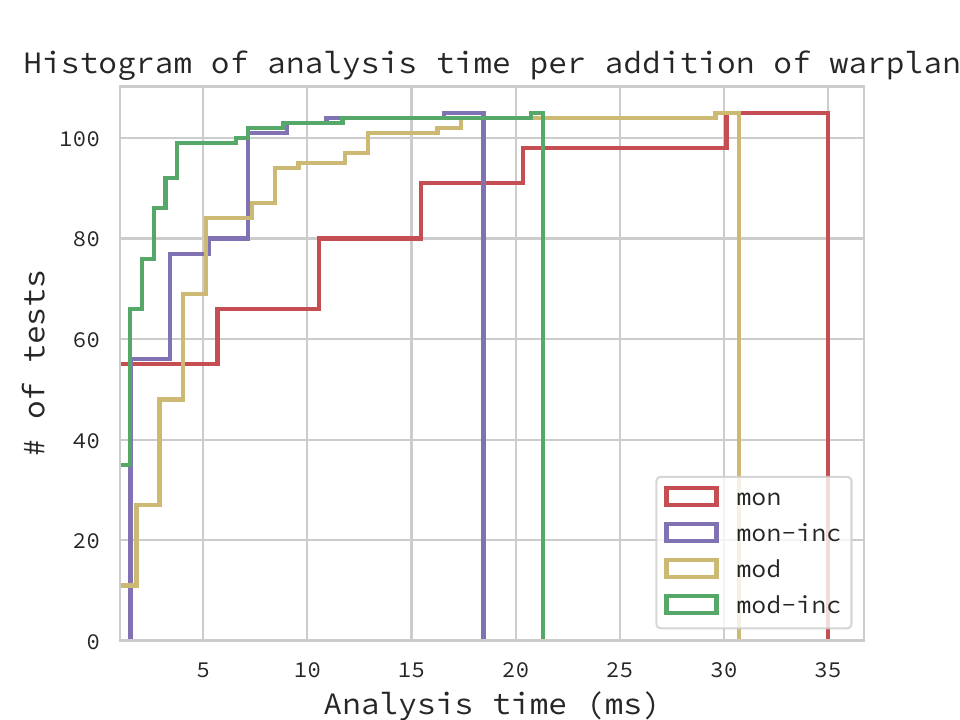}
  \includegraphics[width=0.49\textwidth]{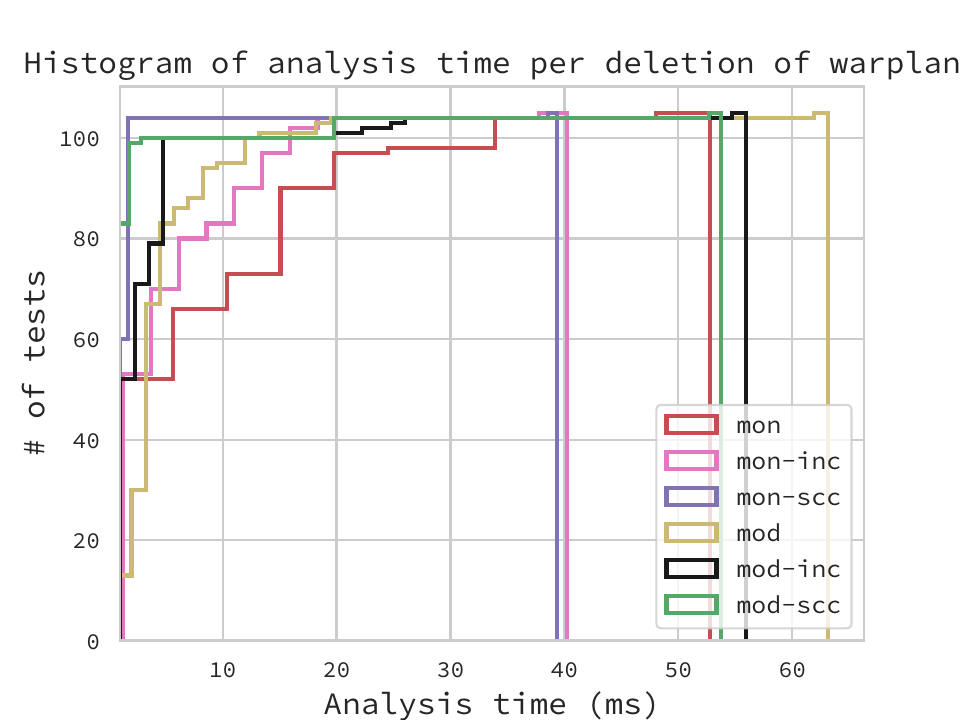}
  \caption{Distribution over time of instances of the addition (left) and
    deletion (right) experiments for \texttt{warplan} with \texttt{def}.}
  \label{fig:hist-def-warplan}
\end{figure}

\begin{figure}[t]
  \centering
  \includegraphics[width=0.49\textwidth]{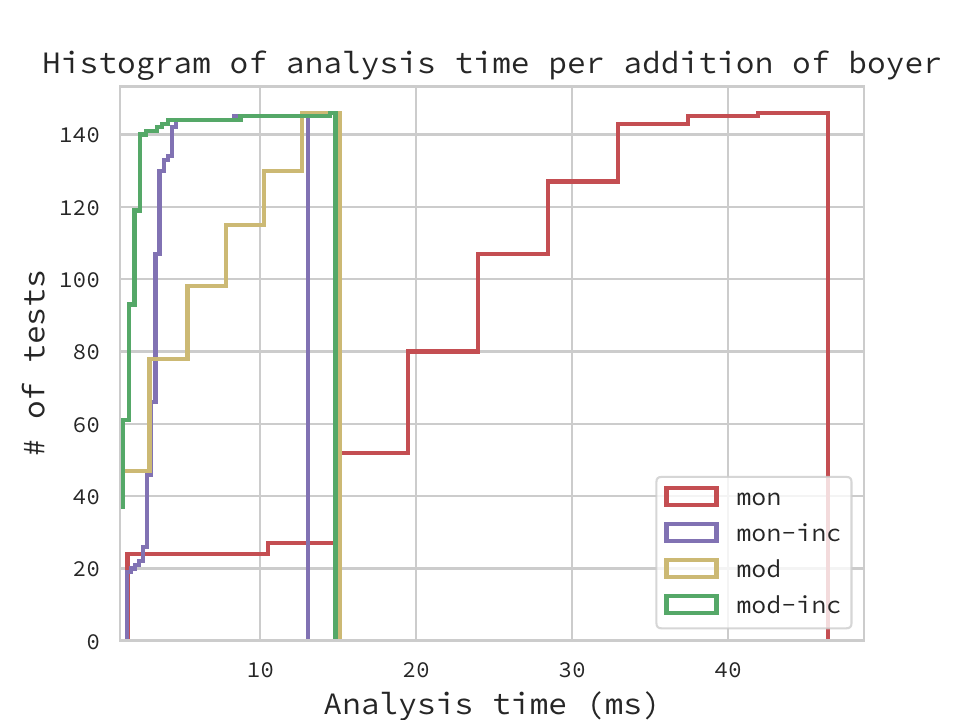}
  \includegraphics[width=0.49\textwidth]{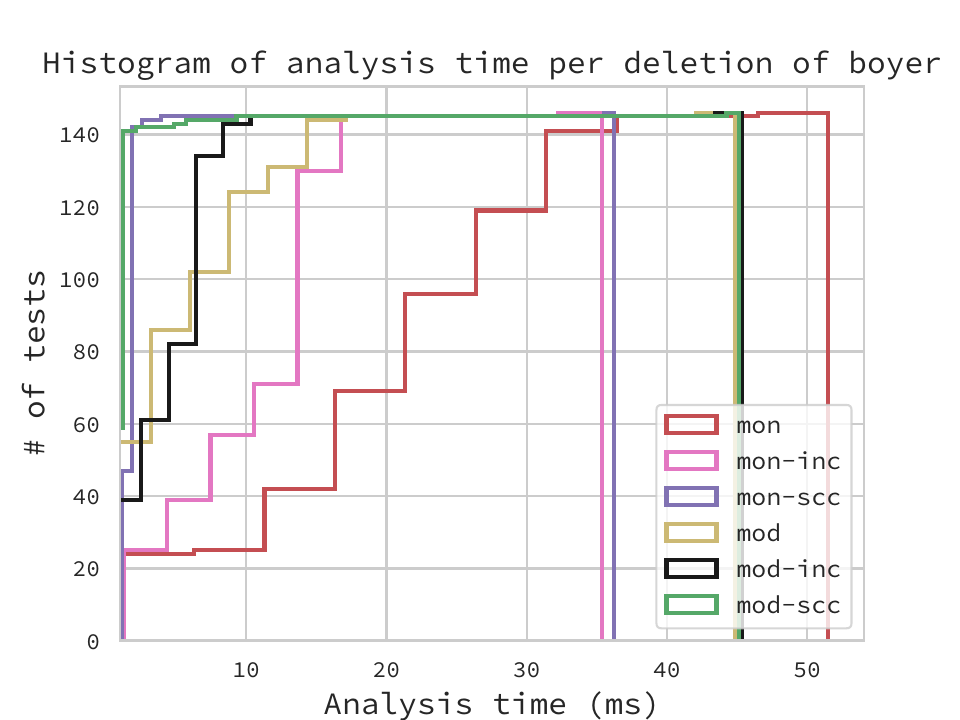}

  \caption{Distribution over time of instances of the addition (left) and
    deletion (right) experiments for \texttt{boyer} with \texttt{def}.}
  \label{fig:hist-def-boyer}
\end{figure}

\paragraph{\textbf{Distribution of analysis times.}} Next, we study how the
analysis time of the experiments is distributed.
Figs.~\ref{fig:hist-def-warplan} and~\ref{fig:hist-def-boyer} show histograms
that illustrate the number of analyzed instances of the experiments with respect
to the analysis time, regardless of the order in which the experiments were
performed, and for each configuration. In the vertical axis we plot the number
of tests, i.e., \emph{how many different instances of the addition or deletion
  experiment that were performed could be analyzed in that time or less}. In the
horizontal axis we represent the analysis time. For example, on the left-hand
side of Fig.~\ref{fig:hist-def-warplan}, for $5 ms$ in the vertical axis,
starting from the bottom of the graph we first find the red line corresponding
to the monolithic analysis (\texttt{mon}). This means that approx.\ 55 of the
analyses performed in \texttt{warplan} finished in $5 ms$ or less. Then we find
the yellow line (\texttt{mod} analysis): for this setting, 70 of the instances
of the addition experiment were analyzed in $5 ms$ or less. The next line that
we find is the purple line, corresponding to the \texttt{mon-inc} configuration.
In this case 78 instances were analyzed in $5 ms$ or less. Finally, we have our
configuration, \texttt{mod-inc}, that was able to analyze 99 instances of the
addition experiment in $5 ms$.
Figs.~\ref{fig:hist-def-warplan} and~\ref{fig:hist-def-boyer} show
that, overall, the analysis time of the proposed algorithm is faster
than that of the previous configurations.

\begin{figure}[t]
  \centering
  \includegraphics[]{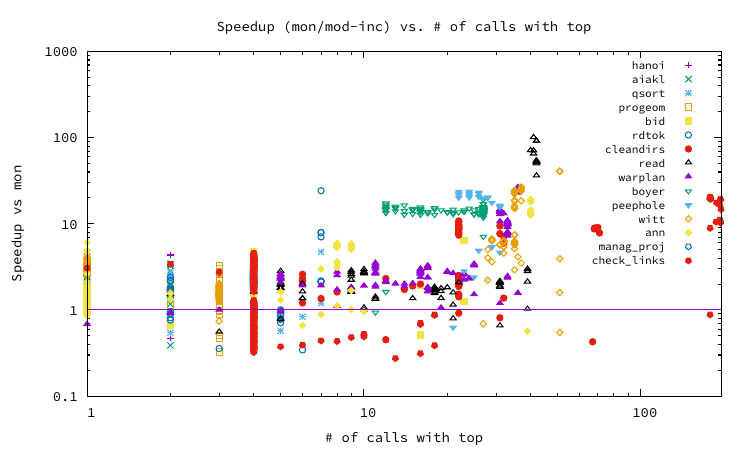}
  \vspace*{-3mm}
  \caption{Speedup vs.\ monolithic depending on the number of nodes in the analysis graph.}
  \label{fig:speed-def-mon-top-nodes}
\end{figure}

\begin{figure}[t]
  \centering
  \vspace*{-5mm}
  \includegraphics[]{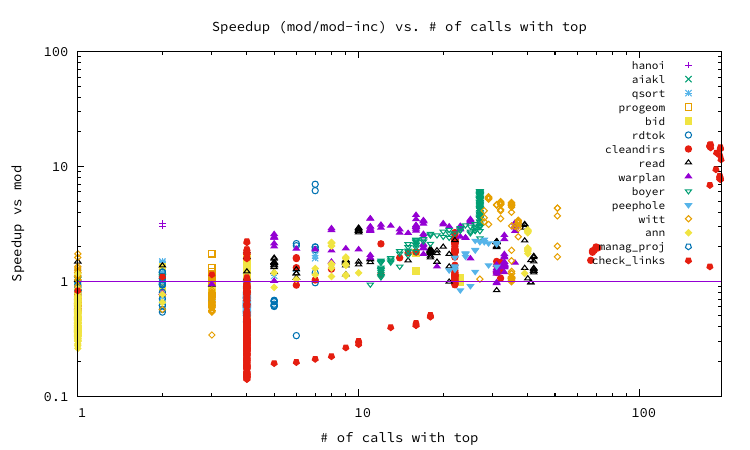}
  \vspace*{-3mm}
  \caption{Speedup vs.\ modular depending on the number of calls to $\top$.}
  \label{fig:speed-def-mod-top-calls}
\end{figure}

\paragraph{\textbf{Correlations to benchmark and analysis graph characteristics.}}

We also looked for correlations between benchmark characteristics 
and the speedups observed. While this topic would require a study of
its own, we have observed some correlations with benchmark-related
analysis characteristics.
Figs.~\ref{fig:speed-def-mon-top-nodes}
and~\ref{fig:speed-def-mod-top-calls} show scatter plots of the
speedup obtained with respect to two such characteristics: the number
of nodes in the analysis graph an the number of calls to $\top$.  The
plots show that there is some correlation between the size of the
analysis graph and the speedup obtained: 
we observed that the incremental and modular analysis proposed obtains
is beneficial for larger analysis graphs. The sizes of the analysis
graphs depend themselves on the complexity of the abstract domain (due
to the algorithm being multivariant), the size of the program, and the
size of the strongly connected components of the program.
Also, we observed that the slowdowns encountered correspond to very small
runtimes of the algorithms, e.g., for smaller programs, which are likely to be
due to the overhead of the additional bookeeping required by the algorithm.
Which is not very concerning, as they are small.

\paragraph{\textbf{Memory Usage.}}

\begin{table}
  \small
  \centering
  \begin{tabular}{|>{\tt}l|>{\raggedleft}p{15mm}|>{\raggedleft}p{15mm}|>{\raggedleft}p{15mm}|
                           >{\raggedleft}p{15mm}|>{\raggedleft}p{6mm}cl|}
    \toprule
    bench &  
            \texttt{mon} & \texttt{mon-inc} & \texttt{mod} & \texttt{mod-inc} & \multicolumn{3}{c|}{reduction} \\
          & & & & & \multicolumn{3}{c|}{(\texttt{mon-inc} vs \texttt{mod-inc})}\\
\midrule
    hanoi &  
            16K & 16K & 12K & 12K & & 0.75 & (25\%)\\
    aiakl &  
            28K & 28K & 8K  & 16K & & 0.57 & (43\%)\\
    qsort &
            28K & 32K & 12K & 16K & & 0.50 & (50\%)\\
    progeom &  
            24K & 32K & 20K & 20K & & 0.63 & (37\%)\\ %
    bid &  
            80K & 80K & 36K & 40K & & 0.50 & (50\%)\\
    rdtok &  
           100K &112K & 68K & 80K & & 0.71 & (29\%)\\ %
    cleandirs &  
           200K &204K &144K &152K & & 0.75 & (25\%)\\
    read &
           304K &308K &260K &268K & & 0.87 & (13\%)\\
    warplan &  
           144K &156K &116K &128K & & 0.82 & (18\%)\\
    boyer &  
           140K &144K & 76K & 80K & & 0.55 & (44\%)\\
    peephole &
           200K &208K &108K &116K & & 0.56 & (44\%)\\
    witt &  
           504K &524K &352K &364K & & 0.69 & (30\%)\\
    ann &  
           316K &324K &120K &132K & & 0.41 & (59\%)\\ %
    manag\_proj &  
           464K &460K &248K &268K & & 0.58 & (42\%)\\ %
    check\_links &
           2.3M &2.3M &1.8M &1.8M & & 0.78 & (22\%)\\ %
    \bottomrule
\end{tabular}
\caption{Maximum memory usage for the edition experiments with \texttt{def} in
  bytes.}
\label{tab:mem}
\end{table}

We also studied the memory usage for the structures needed for
analysis, that is, the analysis graphs, and the other structures
needed for memoizing.
Table~\ref{tab:mem} contains the maximum memory needed for these
structures for any of the modifications analyzed for each benchmark,
i.e., the memory high water mark. For the monolithic case, this is the
maximum memory necessary to keep the analysis results, and for the
modular case, the maximum size of the analysis results of a module and
the intermodular information.
We do not show any distinction between the different deletion strategies of the
incremental algorithm as the necessary bookkeeping of both is the same.

First, note that, since the incremental algorithms
(\texttt{mon-inc} and \texttt{mod-inc}) need to perform additional
bookkeeping, they always need more memory than the corresponding non
incremental ones (\texttt{mon} and \texttt{mod}). However, this
difference is small and arguably a very reasonable price to pay for the
significant reductions in analysis times.
Also note that the modular analyses (\texttt{mod} and
\texttt{mod-inc}) always bring a reduction in the memory required to
be able to complete every analysis instance.
This is of course important because, while it is always possible to
wait a bit longer for an analysis result, if the analysis does not fit
in the available memory, either the performance will be much worse, due
to swapping, or the analysis simply cannot be completed, if virtual
memory is depleted.

More importantly, we observe that we obtain a reduction in the memory
use of the proposed modular incremental algorithm, \texttt{mod-inc},
with respect to the original monolithic incremental algorithm,
\texttt{mon-inc}. This is shown in the last column of
Table~\ref{tab:mem}. The memory usage reduction obtained ranges
between 59\% for the \texttt{ann} benchmark and 13\% for the
\texttt{read} benchmark.
Ideally, we would like to achieve a reduction of memory proportional
to the number of modules in which the program is distributed, but
on one hand there is overhead
due to the fact that each module needs to keep information for the calls to
predicates imported from other modules, 
and in addition a very large reduction in maximum memory usage
requires the partitions to be of similar size, quite independent, and
with similarly-sized analysis graphs.
Our benchmarks instead typically contain a module with the main
functionality and some libraries with simpler code, so that the
distribution of code among the modules is not even, and so the
correlation between memory usage reduction and number of modules in
the program is not direct.
However, we expect that in actual applications, which tend to have a
much larger number of modules and use a good number of libraries, the
memory usage reduction will be much larger.
\secbeg
\vspace*{-2mm}
\section{Related work}
\label{sec:related_work}
\secend

\vspace{1mm}  
\noindent\textbf{Classical data-flow analysis}:
Since the first algorithm for incremental analysis was proposed
by~\cite{%
  Ros-81%
}, there has been considerable research and proposals
in this topic (see the bibliography of~\cite{%
  Ra-Re%
}).
Depending on how data flow equations are solved, these algorithms can be
separated into those based on variable elimination, which include~\cite{%
  Bur-90%
},~\cite{%
  Car-Ryd-88%
}, and~\cite{Ryd-88}; and
those based on iteration methods which include~\cite{%
  Coo-Ken-84%
}
and~\cite{Pol-Sof}. A hybrid approach is described in~\cite{%
  Mar-Ryd-90%
}.
Our algorithms are most closely related to those using iteration.
Early incremental approaches such as~\cite{%
  Coo-Ken-84%
} were based on
{\em restarting iteration}. That is, the fixpoint of the new program's
data flow equations
is found by starting iteration from the fixpoint of the old program's
data flow equations. This is always safe, but may lead to unnecessary
imprecision if the old fixpoint is not below the \lfp of
the new equations~\cite{Ryd-Mar-Pau-88}.
{\em Reinitialization approaches} such as~\cite{Pol-Sof} improve
the accuracy of this technique by reinitializing
nodes in the data flow graph to bottom if they are potentially
affected by the program change. Thus, they are as precise as if the
new equations had been analyzed from scratch.
These algorithms are generally not based on abstract interpretation.
\textsc{Reviser}~\cite{%
  DBLP:conf/icse/ArztB14%
} extends the more
generic IFDS~\cite{DBLP:conf/popl/RepsHS95} framework to support
incremental program changes. However IFDS is limited to distributive flow
functions (related  %
to \emph{condensing} domains) %
while our approach does not impose any restriction on the domains.

\vspace{1mm}  
\noindent\textbf{Other work on CHCs and constraint logic programs}: Apart from
the work that we
extend~\cite{%
  incanal-iclp95,%
  inc-fixp-sas,%
  incanal-toplas%
}, incremental
analysis was 
proposed (just for incremental addition) in the Vienna abstract
machine model~\cite{Vienna-abs-mach,VAM}. It was studied also in
compositional analysis of modules in (constraint) logic
programs~\cite{%
  Co-De-Gi,%
  BGLM94}, but it did not consider incremental
analysis at the level of clauses.
More recently, FLIX~\cite{%
  DBLP:conf/pldi/MadsenYL16%
} uses a bottom-up semi-naïve strategy to solve \emph{Datalog
  programs} extended with lattices and monotone transfer
functions. This approach is similar to
CLP analysis via bottom-up abstract interpretation. However it has not
been extended to support incremental updates.
\emph{Incremental tabling}~\cite{%
  swift2014incremental%
} offers a
straightforward method to design incremental
analyses~\cite{%
  Eichberg2007%
}, when they can be expressed as tabled logic programs. However, while
these methods are much closer to our incremental algorithm, they may
suffer similar problems than generic incremental computation due to
the lack of fine-grained control.

\vspace{1mm}  
\noindent\textbf{Other work on modular analysis}~\cite{%
  CousotModular02%
} is based on splitting large 
programs into smaller parts (e.g., based on the source code
structure). Exploiting modularity has proved essential in
industrial-scale analyzers~\cite{%
  ccfmmr09,%
  clousot-2010,%
  infer-nfm-2011%
}. %
Despite the fact that separate analysis provides only coarse-grained
incrementality, there have been surprisingly few results studying its
combination with fine-grained incremental analysis.

\vspace{1mm}  
\noindent
\textbf{Verification.} Incremental algorithms have also been
proposed to perform formal verification of programs. They reuse results of prior
verification, given some properties to be verified
\cite{DBLP:conf/cav/ConwayNDE05,fedyukovich2016property,DBLP:conf/sas/RothenbergDH18},
which do not take advantage of modular structures of programs.
\cite{DBLP:conf/fmcad/SeryFS12} use a compositional approach to obtain
properties for each procedure, checking whether each property holds for the
successive versions of the program.
These approaches are not directly comparable with the work that we have
presented since we do not rely on any specifications of the program behavior to
run the analysis.

\vspace{1mm}  
\noindent\textbf{Generic incremental computation frameworks}:
Obviously, the possibility exists of using a general
incrementalized execution algorithm.
Incremental algorithms compute an updated output from a previous
output and a difference on the input data, which the hope that the
process is (computationally) cheaper than computing from scratch a new
output for the new input.
The approach of~\cite{%
  DBLP:conf/kbse/SzaboEV16%
} takes advantage of an
underlying incremental evaluator, IncQuery, and implements modules via
the monolithic approach. 
There exist other frameworks such as self-adjusting
computation~\cite{%
  DBLP:conf/pepm/Acar09%
} that provides libraries for incrementalizing non-incremental
algorithms. Although in some cases (like \emph{tree contraction}) it
can reproduce the performance of specialized incremental algorithms,
experiments show that in general there is a significant overhead
(between 4 and 10) over non-incremental algorithms.  We believe that
it is a promising approach but not yet ready for replace incremental
algorithms designed, proved, and implemented from scratch.

\secbeg
\vspace*{-6mm}
\section{Conclusions}
\label{sec:conclusions}
\secend

We have described, implemented, and evaluated a context sensitive,
fixpoint analysis algorithm that performs efficient context-sensitive
analysis incrementally on modular partitions of programs.
We provided a unified view of the algorithms that we built upon,
providing a formal description of their correctness and precision
guarantees that also covers widening.
Our algorithm takes care of propagating the fine-grain change
information across module boundaries and implements all the actions
required to recompute the analysis fixpoint incrementally after
additions and deletions in the program.
We have shown that the algorithm is correct and computes the most
precise analysis for finite abstract domains, while supporting 
widening for dealing with infinite domains. 
We have also provided some new results for the
baseline algorithms. 
We have also implemented and benchmarked the proposed approach within the
Ciao/CiaoPP system.
Our preliminary results %
show promising speedups for programs of medium and larger size when
compared
to existing non-modular, fine-grain incremental analysis techniques,
as well as improvements in memory consumption.
In addition, the finer granularity of the proposed modular incremental
fixpoint algorithm also %
brings improvements with respect to
modular analysis alone (which only preserved analysis results at the
module boundaries), producing better results even in the limit case of
analyzing the whole program from scratch.
Finally, we have also observed some correlations between obtainable
speedups and certain benchmark-related analysis characteristics, such
as the number of nodes in the analysis graph an the number of calls to
$\top$ abstractions.
Going deeper in this direction is a clear avenue for future work.

\ \\
\noindent
Competing interest declaration: The authors declare none.

\vspace*{-4mm}
\begin{small}
  \bibliographystyle{acmtrans}
  \bibliography{extracted}

\begin{thebibliography}{}

\bibitem[\protect\citeauthoryear{Acar}{Acar}{2009}]{DBLP:conf/pepm/Acar09}
{\sc Acar, U.~A.} 2009.
\newblock Self-adjusting computation: (an overview).
\newblock In {\em Proceedings of the 2009 {ACM} {SIGPLAN} Symposium on Partial
  Evaluation and Semantics-based Program Manipulation, {PEPM} 2009, Savannah,
  GA, USA, January 19-20, 2009}, {G.~Puebla} {and} {G.~Vidal}, Eds. {ACM},
  1--6.

\bibitem[\protect\citeauthoryear{Albert, Arenas, Genaim, Puebla, and
  Zanardini}{Albert et~al\mbox{.}}{2012}]{AlbertAGPZ12}
{\sc Albert, E.}, {\sc Arenas, P.}, {\sc Genaim, S.}, {\sc Puebla, G.}, {\sc
  and} {\sc Zanardini, D.} 2012.
\newblock Cost {A}nalysis of {O}bject-{O}riented {B}ytecode {P}rograms.
\newblock {\em Theoretical Computer Science (Special Issue on Quantitative
  Aspects of Programming Languages)\/}~{\em 413,\/}~1, 142--159.

\bibitem[\protect\citeauthoryear{Albert, Correas, Puebla, and
  Rom\'{a}n-D\'{i}ez}{Albert et~al\mbox{.}}{2012}]{albertcpr12}
{\sc Albert, E.}, {\sc Correas, J.}, {\sc Puebla, G.}, {\sc and} {\sc
  Rom\'{a}n-D\'{i}ez, G.} 2012.
\newblock {I}ncremental {R}esource {U}sage {A}nalysis.
\newblock In {\em Proceedings of the 2012 ACM SIGPLAN Workshop on Partial
  Evaluation and Program Manipulation, PEPM 2012, Philadelphia, Pennsylvania,
  USA, January 23-24, 2012}. ACM Press, 25--34.

\bibitem[\protect\citeauthoryear{Albert, G\'{o}mez-Zamalloa, Hubert, and
  Puebla}{Albert et~al\mbox{.}}{2007}]{jvm-pe-padl07}
{\sc Albert, E.}, {\sc G\'{o}mez-Zamalloa, M.}, {\sc Hubert, L.}, {\sc and}
  {\sc Puebla, G.} 2007.
\newblock {V}erification of {J}ava {B}ytecode using {A}nalysis and
  {T}ransformation of {L}ogic {P}rograms.
\newblock In {\em Ninth International Symposium on Practical Aspects of
  Declarative Languages (PADL 2007)}. Number 4354 in LNCS. Springer-Verlag,
  124--139.

\bibitem[\protect\citeauthoryear{Apt}{Apt}{1990}]{Apt90}
{\sc Apt, K.~R.} 1990.
\newblock Introduction to logic programming.
\newblock In {\em Handbook of Theoretical Computer Science}, {J.~van Leeuwen},
  Ed. Elsevier, 493--576.

\bibitem[\protect\citeauthoryear{Arzt and Bodden}{Arzt and
  Bodden}{2014}]{DBLP:conf/icse/ArztB14}
{\sc Arzt, S.} {\sc and} {\sc Bodden, E.} 2014.
\newblock {R}eviser: {E}fficiently {U}pdating {IDE}-/{IFDS}-based {D}ata-flow
  {A}nalyses in {R}esponse to {I}ncremental {P}rogram {C}hanges.
\newblock In {\em 36th International Conference on Software Engineering, {ICSE}
  '14, Hyderabad, India - May 31 - June 07, 2014}, {P.~Jalote}, {L.~C. Briand},
  {and} {A.~van~der Hoek}, Eds. {ACM}, 288--298.

\bibitem[\protect\citeauthoryear{Banda and Gallagher}{Banda and
  Gallagher}{2009}]{BandaG08}
{\sc Banda, G.} {\sc and} {\sc Gallagher, J.~P.} 2009.
\newblock Analysis of {L}inear {H}ybrid {S}ystems in {CLP}.
\newblock In {\em Logic-Based Program Synthesis and Transformation, 18th
  International Symposium, LOPSTR 2008, Valencia, Spain, July 17-18, 2008},
  {M.~Hanus}, Ed. Lecture Notes in Computer Science, vol. 5438. Springer,
  55--70.

\bibitem[\protect\citeauthoryear{Bj{\o}rner, Gurfinkel, McMillan, and
  Rybalchenko}{Bj{\o}rner
  et~al\mbox{.}}{2015}]{DBLP:conf/birthday/BjornerGMR15}
{\sc Bj{\o}rner, N.}, {\sc Gurfinkel, A.}, {\sc McMillan, K.~L.}, {\sc and}
  {\sc Rybalchenko, A.} 2015.
\newblock {H}orn {C}lause {S}olvers for {P}rogram {V}erification.
\newblock In {\em Fields of Logic and Computation {II} - Essays Dedicated to
  Yuri Gurevich on the Occasion of His 75th Birthday}, {L.~D. Beklemishev},
  {A.~Blass}, {N.~Dershowitz}, {B.~Finkbeiner}, {and} {W.~Schulte}, Eds.
  Lecture Notes in Computer Science, vol. 9300. Springer, 24--51.

\bibitem[\protect\citeauthoryear{Bj{\o}rner, McMillan, and
  Rybalchenko}{Bj{\o}rner et~al\mbox{.}}{2013}]{DBLP:conf/sas/BjornerMR13}
{\sc Bj{\o}rner, N.}, {\sc McMillan, K.~L.}, {\sc and} {\sc Rybalchenko, A.}
  2013.
\newblock On solving universally quantified {H}orn clauses.
\newblock In {\em SAS}, {F.~Logozzo} {and} {M.~F{\"a}hndrich}, Eds. LNCS, vol.
  7935. Springer, 105--125.

\bibitem[\protect\citeauthoryear{Bossi, Gabbrieli, Levi, and Meo}{Bossi
  et~al\mbox{.}}{1994}]{BGLM94}
{\sc Bossi, A.}, {\sc Gabbrieli, M.}, {\sc Levi, G.}, {\sc and} {\sc Meo, M.}
  1994.
\newblock A compositional semantics for logic programs.
\newblock {\em Theoretical Computer Science\/}~{\em 122,\/}~1,2, 3--47.

\bibitem[\protect\citeauthoryear{Braem, Charlier, Modart, and Hentenryck}{Braem
  et~al\mbox{.}}{1994}]{cardinality-ilps94}
{\sc Braem, C.}, {\sc Charlier, B.~L.}, {\sc Modart, S.}, {\sc and} {\sc
  Hentenryck, P.~V.} 1994.
\newblock Cardinality analysis of {P}rolog.
\newblock In {\em Proc. International Symposium on Logic Programming}. MIT
  Press, Ithaca, NY, 457--471.

\bibitem[\protect\citeauthoryear{Bruynooghe}{Bruynooghe}{1991}]{bruy91}
{\sc Bruynooghe, M.} 1991.
\newblock {A} {P}ractical {F}ramework for the {A}bstract {I}nterpretation of
  {L}ogic {P}rograms.
\newblock {\em Journal of Logic Programming\/}~{\em 10}, 91--124.

\bibitem[\protect\citeauthoryear{Bueno, de~la Banda, Hermenegildo, Marriott,
  Puebla, and Stuckey}{Bueno et~al\mbox{.}}{2001}]{modular-anal-lopstr}
{\sc Bueno, F.}, {\sc de~la Banda, M.~G.}, {\sc Hermenegildo, M.~V.}, {\sc
  Marriott, K.}, {\sc Puebla, G.}, {\sc and} {\sc Stuckey, P.} 2001.
\newblock {A} {M}odel for {I}nter-module {A}nalysis and {O}ptimizing
  {C}ompilation.
\newblock In {\em Logic-based Program Synthesis and Transformation}. Number
  2042 in LNCS. Springer-Verlag, 86--102.

\bibitem[\protect\citeauthoryear{Burke}{Burke}{1990}]{Bur-90}
{\sc Burke, M.} 1990.
\newblock An interval-based approach to exhaustive and incremental
  interprocedural data-flow analysis.
\newblock {\em {ACM} Transactions on Programming Languages and Systems\/}~{\em
  12,\/}~3, 341--395.

\bibitem[\protect\citeauthoryear{Calcagno and Distefano}{Calcagno and
  Distefano}{2011}]{infer-nfm-2011}
{\sc Calcagno, C.} {\sc and} {\sc Distefano, D.} 2011.
\newblock Infer: An automatic program verifier for memory safety of {C}
  programs.
\newblock In {\em {NASA} Formal Methods - Third International Symposium, {NFM}
  2011, Pasadena, CA, USA, April 18-20, 2011. Proceedings}, {M.~G. Bobaru},
  {K.~Havelund}, {G.~J. Holzmann}, {and} {R.~Joshi}, Eds. Lecture Notes in
  Computer Science, vol. 6617. Springer, 459--465.

\bibitem[\protect\citeauthoryear{Carroll and Ryder}{Carroll and
  Ryder}{1988}]{Car-Ryd-88}
{\sc Carroll, M.} {\sc and} {\sc Ryder, B.} 1988.
\newblock Incremental data flow analysis via dominator and attribute updates.
\newblock In {\em 15th {ACM} Symposium on Principles of Programming Languages
  (POPL)}. {ACM} Press, 274--284.

\bibitem[\protect\citeauthoryear{Codish, Debray, and Giacobazzi}{Codish
  et~al\mbox{.}}{1993}]{Co-De-Gi}
{\sc Codish, M.}, {\sc Debray, S.}, {\sc and} {\sc Giacobazzi, R.} 1993.
\newblock {C}ompositional {A}nalysis of {M}odular {L}ogic {P}rograms.
\newblock In {\em ACM SIGPLAN-SIGACT Symposium on Principles of Programming
  Languages POPL'93}. ACM, Charleston, South Carolina, 451--464.

\bibitem[\protect\citeauthoryear{Conway, Namjoshi, Dams, and Edwards}{Conway
  et~al\mbox{.}}{2005}]{DBLP:conf/cav/ConwayNDE05}
{\sc Conway, C.~L.}, {\sc Namjoshi, K.~S.}, {\sc Dams, D.}, {\sc and} {\sc
  Edwards, S.~A.} 2005.
\newblock Incremental algorithms for inter-procedural analysis of safety
  properties.
\newblock In {\em Computer Aided Verification, 17th International Conference,
  {CAV} 2005, Edinburgh, Scotland, UK, July 6-10, 2005}, {K.~Etessami} {and}
  {S.~K. Rajamani}, Eds. Lecture Notes in Computer Science, vol. 3576.
  Springer, 449--461.

\bibitem[\protect\citeauthoryear{Cooper and Kennedy}{Cooper and
  Kennedy}{1984}]{Coo-Ken-84}
{\sc Cooper, K.} {\sc and} {\sc Kennedy, K.} 1984.
\newblock Efficient computation of flow insensitive interprocedural summary
  information.
\newblock In {\em {ACM SIGPLAN} Symposium on Compiler Construction (SIGPLAN
  Notices {\bf 19}(6))}. {ACM} Press, 247--258.

\bibitem[\protect\citeauthoryear{Correas, Puebla, Hermenegildo, and
  Bueno}{Correas et~al\mbox{.}}{2006}]{modbenchmarks-lopstr05}
{\sc Correas, J.}, {\sc Puebla, G.}, {\sc Hermenegildo, M.~V.}, {\sc and} {\sc
  Bueno, F.} 2006.
\newblock {E}xperiments in {C}ontext-{S}ensitive {A}nalysis of {M}odular
  {P}rograms.
\newblock In {\em 15th International Symposium on Logic-based Program Synthesis
  and Transformation (LOPSTR'05)}. Number 3901 in LNCS. Springer-Verlag,
  163--178.

\bibitem[\protect\citeauthoryear{Cousot and Cousot}{Cousot and
  Cousot}{1977}]{Cousot77}
{\sc Cousot, P.} {\sc and} {\sc Cousot, R.} 1977.
\newblock {A}bstract {I}nterpretation: a {U}nified {L}attice {M}odel for
  {S}tatic {A}nalysis of {P}rograms by {C}onstruction or {A}pproximation of
  {F}ixpoints.
\newblock In {\em {ACM} {S}ymposium on {P}rinciples of {P}rogramming
  {L}anguages (POPL'77)}. ACM Press, 238--252.

\bibitem[\protect\citeauthoryear{Cousot and Cousot}{Cousot and
  Cousot}{2002}]{CousotModular02}
{\sc Cousot, P.} {\sc and} {\sc Cousot, R.} 2002.
\newblock {M}odular {S}tatic {P}rogram {A}nalysis, invited paper.
\newblock In {\em Eleventh International Conference on Compiler Construction,
  CC 2002}. Number 2304 in LNCS. Springer, 159--178.

\bibitem[\protect\citeauthoryear{Cousot, Cousot, Feret, Min{\'e}, Mauborgne,
  and Rival}{Cousot et~al\mbox{.}}{2009}]{ccfmmr09}
{\sc Cousot, P.}, {\sc Cousot, R.}, {\sc Feret, J.}, {\sc Min{\'e}, A.}, {\sc
  Mauborgne, L.}, {\sc and} {\sc Rival, X.} 2009.
\newblock Why does {A}strée scale up?
\newblock {\em Formal Methods in System Design (FMSD)\/}~{\em 35,\/}~3
  (December), 229--264.

\bibitem[\protect\citeauthoryear{{De Angelis}, Fioravanti, Pettorossi, and
  Proietti}{{De Angelis} et~al\mbox{.}}{2014}]{DBLP:conf/tacas/AngelisFPP14}
{\sc {De Angelis}, E.}, {\sc Fioravanti, F.}, {\sc Pettorossi, A.}, {\sc and}
  {\sc Proietti, M.} 2014.
\newblock {V}eri{MAP}: {A} {T}ool for {V}erifying {P}rograms through
  {T}ransformations.
\newblock In {\em Tools and Algorithms for the Construction and Analysis of
  Systems - 20th International Conference, {TACAS} 2014, Held as Part of the
  European Joint Conferences on Theory and Practice of Software, {ETAPS} 2014,
  Grenoble, France, April 5-13, 2014. Proceedings}, {E.~{\'{A}}brah{\'{a}}m}
  {and} {K.~Havelund}, Eds. Lecture Notes in Computer Science, vol. 8413.
  Springer, 568--574.

\bibitem[\protect\citeauthoryear{de~Moura and Bj{\o}rner}{de~Moura and
  Bj{\o}rner}{2008}]{z3}
{\sc de~Moura, L.~M.} {\sc and} {\sc Bj{\o}rner, N.} 2008.
\newblock {Z3}: {A}n {E}fficient {SMT} {S}olver.
\newblock In {\em Tools and Algorithms for the Construction and Analysis of
  Systems, 14th International Conference, TACAS 2008}, {C.~R. Ramakrishnan}
  {and} {J.~Rehof}, Eds. Lecture Notes in Computer Science, vol. 4963.
  Springer, 337--340.

\bibitem[\protect\citeauthoryear{Debray, Lopez-Garcia, and Hermenegildo}{Debray
  et~al\mbox{.}}{1997}]{non-failure-iclp97}
{\sc Debray, S.}, {\sc Lopez-Garcia, P.}, {\sc and} {\sc Hermenegildo, M.~V.}
  1997.
\newblock {N}on-{F}ailure {A}nalysis for {L}ogic {P}rograms.
\newblock In {\em 1997 International Conference on Logic Programming}. MIT
  Press, Cambridge, MA, Cambridge, MA, 48--62.

\bibitem[\protect\citeauthoryear{Dumortier, Janssens, Simoens, and {Garc\'{\i}a
  de la Banda}}{Dumortier et~al\mbox{.}}{1993}]{free-def-comb}
{\sc Dumortier, V.}, {\sc Janssens, G.}, {\sc Simoens, W.}, {\sc and} {\sc
  {Garc\'{\i}a de la Banda}, M.} 1993.
\newblock {C}ombining a {D}efiniteness and a {F}reeness {A}bstraction for {CLP}
  {L}anguages.
\newblock In {\em Workshop on Logic Program Synthesis and Transformation}.

\bibitem[\protect\citeauthoryear{Eichberg, Kahl, Saha, Mezini, and
  Ostermann}{Eichberg et~al\mbox{.}}{2007}]{Eichberg2007}
{\sc Eichberg, M.}, {\sc Kahl, M.}, {\sc Saha, D.}, {\sc Mezini, M.}, {\sc and}
  {\sc Ostermann, K.} 2007.
\newblock {\em Automatic Incrementalization of {P}rolog Based Static Analyses}.
\newblock Springer Berlin Heidelberg, Berlin, Heidelberg, 109--123.

\bibitem[\protect\citeauthoryear{F\"{a}hndrich and Logozzo}{F\"{a}hndrich and
  Logozzo}{2011}]{clousot-2010}
{\sc F\"{a}hndrich, M.} {\sc and} {\sc Logozzo, F.} 2011.
\newblock {Static Contract Checking with Abstract Interpretation}.
\newblock In {\em Proceedings of the 2010 International Conference on Formal
  Verification of Object-oriented Software, FoVeOOS'10}. Lecture Notes in
  Computer Science, vol. 6528. Springer-Verlag, Berlin, Heidelberg, 10--30.

\bibitem[\protect\citeauthoryear{Fedyukovich, Gurfinkel, and
  Sharygina}{Fedyukovich et~al\mbox{.}}{2016}]{fedyukovich2016property}
{\sc Fedyukovich, G.}, {\sc Gurfinkel, A.}, {\sc and} {\sc Sharygina, N.} 2016.
\newblock Property directed equivalence via abstract simulation.
\newblock In {\em International Conference on Computer Aided Verification}.
  Springer, 433--453.

\bibitem[\protect\citeauthoryear{Gallagher, Hermenegildo, Kafle, Klemen,
  Lopez-Garcia, and Morales}{Gallagher
  et~al\mbox{.}}{2020}]{big-small-step-vpt2020}
{\sc Gallagher, J.}, {\sc Hermenegildo, M.~V.}, {\sc Kafle, B.}, {\sc Klemen,
  M.}, {\sc Lopez-Garcia, P.}, {\sc and} {\sc Morales, J.} 2020.
\newblock From big-step to small-step semantics and back with interpreter
  specialization (invited paper).
\newblock In {\em Proceedings of the Eighth International Workshop on
  Verification and Program Transformation (VPT 2020)}. Electronic Proceedings
  in Theoretical Computer Science (EPTCS). Open Publishing Association (OPA),
  50--65.
\newblock Co-located with ETAPS 2020.

\bibitem[\protect\citeauthoryear{Grebenshchikov, Gupta, Lopes, Popeea, and
  Rybalchenko}{Grebenshchikov
  et~al\mbox{.}}{2012}]{DBLP:conf/tacas/GrebenshchikovGLPR12}
{\sc Grebenshchikov, S.}, {\sc Gupta, A.}, {\sc Lopes, N.~P.}, {\sc Popeea,
  C.}, {\sc and} {\sc Rybalchenko, A.} 2012.
\newblock {HSF(C)}: {A} {S}oftware {V}erifier {B}ased on {H}orn {C}lauses -
  ({C}ompetition {C}ontribution).
\newblock In {\em TACAS}, {C.~Flanagan} {and} {B.~K{\"o}nig}, Eds. LNCS, vol.
  7214. Springer, 549--551.

\bibitem[\protect\citeauthoryear{Gurfinkel, Kahsai, Komuravelli, and
  Navas}{Gurfinkel et~al\mbox{.}}{2015}]{DBLP:conf/cav/GurfinkelKKN15}
{\sc Gurfinkel, A.}, {\sc Kahsai, T.}, {\sc Komuravelli, A.}, {\sc and} {\sc
  Navas, J.~A.} 2015.
\newblock {T}he {S}ea{H}orn {V}erification {F}ramework.
\newblock In {\em Computer Aided Verification - 27th International Conference,
  {CAV} 2015, San Francisco, CA, USA, July 18-24, 2015, Proceedings, Part {I}}.
  Number 9206 in LNCS. Springer, 343--361.

\bibitem[\protect\citeauthoryear{Henriksen and Gallagher}{Henriksen and
  Gallagher}{2006}]{HGScam06}
{\sc Henriksen, K.~S.} {\sc and} {\sc Gallagher, J.~P.} 2006.
\newblock {A}bstract {I}nterpretation of {PIC} {P}rograms through {L}ogic
  {P}rogramming.
\newblock In {\em {SCAM~'06}, Proceedings of the Sixth IEEE International
  Workshop on Source Code Analysis and Manipulation}. IEEE Computer Society,
  184--196.

\bibitem[\protect\citeauthoryear{Hermenegildo, Bueno, Carro, Lopez-Garcia,
  Mera, Morales, and Puebla}{Hermenegildo
  et~al\mbox{.}}{2012}]{hermenegildo11:ciao-design-tplp}
{\sc Hermenegildo, M.~V.}, {\sc Bueno, F.}, {\sc Carro, M.}, {\sc Lopez-Garcia,
  P.}, {\sc Mera, E.}, {\sc Morales, J.}, {\sc and} {\sc Puebla, G.} 2012.
\newblock {A}n {O}verview of {C}iao and its {D}esign {P}hilosophy.
\newblock {\em Theory and Practice of Logic Programming\/}~{\em 12,\/}~1--2
  (January), 219--252.

\bibitem[\protect\citeauthoryear{Hermenegildo, Puebla, Bueno, and
  Lopez-Garcia}{Hermenegildo et~al\mbox{.}}{2005}]{ciaopp-sas03-journal-scp}
{\sc Hermenegildo, M.~V.}, {\sc Puebla, G.}, {\sc Bueno, F.}, {\sc and} {\sc
  Lopez-Garcia, P.} 2005.
\newblock {I}ntegrated {P}rogram {D}ebugging, {V}erification, and
  {O}ptimization {U}sing {A}bstract {I}nterpretation (and {T}he {C}iao {S}ystem
  {P}reprocessor).
\newblock {\em Science of Computer Programming\/}~{\em 58,\/}~1--2 (October),
  115--140.

\bibitem[\protect\citeauthoryear{Hermenegildo, Puebla, Marriott, and
  Stuckey}{Hermenegildo et~al\mbox{.}}{1995}]{incanal-iclp95}
{\sc Hermenegildo, M.~V.}, {\sc Puebla, G.}, {\sc Marriott, K.}, {\sc and} {\sc
  Stuckey, P.} 1995.
\newblock {I}ncremental {A}nalysis of {L}ogic {P}rograms.
\newblock In {\em International Conference on Logic Programming}. MIT Press,
  797--811.

\bibitem[\protect\citeauthoryear{Hermenegildo, Puebla, Marriott, and
  Stuckey}{Hermenegildo et~al\mbox{.}}{2000}]{incanal-toplas}
{\sc Hermenegildo, M.~V.}, {\sc Puebla, G.}, {\sc Marriott, K.}, {\sc and} {\sc
  Stuckey, P.} 2000.
\newblock {I}ncremental {A}nalysis of {C}onstraint {L}ogic {P}rograms.
\newblock {\em ACM Transactions on Programming Languages and Systems\/}~{\em
  22,\/}~2 (March), 187--223.

\bibitem[\protect\citeauthoryear{Jaffar and Lassez}{Jaffar and
  Lassez}{1987}]{jaff87}
{\sc Jaffar, J.} {\sc and} {\sc Lassez, J.-L.} 1987.
\newblock {C}onstraint {L}ogic {P}rogramming.
\newblock In {\em {ACM} Symposium on Principles of Programming Languages}.
  {ACM}, 111--119.

\bibitem[\protect\citeauthoryear{Jaffar, Murali, Navas, and Santosa}{Jaffar
  et~al\mbox{.}}{2012}]{DBLP:conf/cav/JaffarMNS12}
{\sc Jaffar, J.}, {\sc Murali, V.}, {\sc Navas, J.~A.}, {\sc and} {\sc Santosa,
  A.~E.} 2012.
\newblock {TRACER:} {A} symbolic execution tool for verification.
\newblock In {\em Computer Aided Verification - 24th International Conference,
  {CAV} 2012, Berkeley, CA, USA, July 7-13, 2012 Proceedings}, {P.~Madhusudan}
  {and} {S.~A. Seshia}, Eds. Lecture Notes in Computer Science, vol. 7358.
  Springer, 758--766.

\bibitem[\protect\citeauthoryear{Kafle, Gallagher, and Morales}{Kafle
  et~al\mbox{.}}{2016}]{kafle-cav2016}
{\sc Kafle, B.}, {\sc Gallagher, J.~P.}, {\sc and} {\sc Morales, J.~F.} 2016.
\newblock {RAHFT}: {A} {T}ool for {V}erifying {H}orn {C}lauses {U}sing
  {A}bstract {I}nterpretation and {F}inite {T}ree {A}utomata.
\newblock In {\em Computer Aided Verification - 28th International Conference,
  {CAV} 2016, Toronto, ON, Canada, July 17-23, 2016, Proceedings, Part {I}},
  {S.~Chaudhuri} {and} {A.~Farzan}, Eds. Lecture Notes in Computer Science,
  vol. 9779. Springer, 261--268.

\bibitem[\protect\citeauthoryear{Kahn}{Kahn}{1987}]{Kahn87}
{\sc Kahn, G.} 1987.
\newblock Natural semantics.
\newblock {F.~Brandenburg}, {G.~Vidal-Naque}, {and} {M.~Wirsing}, Eds. Lecture
  Notes in Computer Science, vol. 247. Springer, 22--39.

\bibitem[\protect\citeauthoryear{Kelly, Marriott, S\o{}ndergaard, and
  Stuckey}{Kelly et~al\mbox{.}}{1997}]{clpr-anal}
{\sc Kelly, A.}, {\sc Marriott, K.}, {\sc S\o{}ndergaard, H.}, {\sc and} {\sc
  Stuckey, P.} 1997.
\newblock A generic object oriented incremental analyser for constraint logic
  programs.
\newblock In {\em Proceedings of the 20th Australasian Computer Science
  Conference}. 92--101.

\bibitem[\protect\citeauthoryear{Khedker and Karkare}{Khedker and
  Karkare}{2008}]{DBLP:conf/cc/KhedkerK08}
{\sc Khedker, U.~P.} {\sc and} {\sc Karkare, B.} 2008.
\newblock Efficiency, precision, simplicity, and generality in interprocedural
  data flow analysis: {R}esurrecting the classical call strings method.
\newblock In {\em Compiler Construction, 17th International Conference, {CC}
  2008, Budapest, Hungary, March 29 - April 6, 2008}, {L.~J. Hendren}, Ed.
  Lecture Notes in Computer Science, vol. 4959. Springer, 213--228.

\bibitem[\protect\citeauthoryear{King, Lu, and Genaim}{King
  et~al\mbox{.}}{2006}]{KLG06:ICLP}
{\sc King, A.}, {\sc Lu, L.}, {\sc and} {\sc Genaim, S.} 2006.
\newblock {D}etecting {D}eterminacy in {P}rolog {P}rograms.
\newblock In {\em Logic Programming, 22nd International Conference, ICLP 2006,
  Seattle, WA, USA, August 17-20, 2006, Proceedings}, {S.~Etalle} {and}
  {M.~Truszczynski}, Eds. Lecture Notes in Computer Science, vol. 4079.
  Springer, 132--147.

\bibitem[\protect\citeauthoryear{Krall and Berger}{Krall and
  Berger}{1995a}]{Vienna-abs-mach}
{\sc Krall, A.} {\sc and} {\sc Berger, T.} 1995a.
\newblock Incremental global compilation of {P}rolog with the vienna abstract
  machine.
\newblock In {\em International Conference on Logic Programming}. MIT Press.

\bibitem[\protect\citeauthoryear{Krall and Berger}{Krall and
  Berger}{1995b}]{VAM}
{\sc Krall, A.} {\sc and} {\sc Berger, T.} 1995b.
\newblock The {VAM$_{\rm AI}$} - an abstract machine for incremental global
  dataflow analysis of {Prolog}.
\newblock In {\em ICLP'95 Post-Conference Workshop on Abstract Interpretation
  of Logic Languages}, {M.~G. de~la Banda}, {G.~Janssens}, {and} {P.~Stuckey},
  Eds. Science University of Tokyo, Tokyo, 80--91.

\bibitem[\protect\citeauthoryear{Liqat, Georgiou, Kerrison, Lopez-Garcia,
  Hermenegildo, Gallagher, and Eder}{Liqat
  et~al\mbox{.}}{2016}]{isa-vs-llvm-fopara}
{\sc Liqat, U.}, {\sc Georgiou, K.}, {\sc Kerrison, S.}, {\sc Lopez-Garcia,
  P.}, {\sc Hermenegildo, M.~V.}, {\sc Gallagher, J.~P.}, {\sc and} {\sc Eder,
  K.} 2016.
\newblock {I}nferring {P}arametric {E}nergy {C}onsumption {F}unctions at
  {D}ifferent {S}oftware {L}evels: {ISA} vs. {LLVM IR}.
\newblock In {\em Foundational and Practical Aspects of Resource Analysis: 4th
  International Workshop, FOPARA 2015, London, UK, April 11, 2015. Revised
  Selected Papers}, {M.~V. Eekelen} {and} {U.~D. Lago}, Eds. Lecture Notes in
  Computer Science, vol. 9964. Springer, 81--100.

\bibitem[\protect\citeauthoryear{Liqat, Kerrison, Serrano, Georgiou,
  Lopez-Garcia, Grech, Hermenegildo, and Eder}{Liqat
  et~al\mbox{.}}{2014}]{isa-energy-lopstr13-final}
{\sc Liqat, U.}, {\sc Kerrison, S.}, {\sc Serrano, A.}, {\sc Georgiou, K.},
  {\sc Lopez-Garcia, P.}, {\sc Grech, N.}, {\sc Hermenegildo, M.~V.}, {\sc and}
  {\sc Eder, K.} 2014.
\newblock {E}nergy {C}onsumption {A}nalysis of {P}rograms based on {XMOS}
  {ISA}-level {M}odels.
\newblock In {\em Logic-Based Program Synthesis and Transformation, 23rd
  International Symposium, {LOPSTR} 2013, Revised Selected Papers}, {G.~Gupta}
  {and} {R.~Pe{\~n}a}, Eds. Lecture Notes in Computer Science, vol. 8901.
  Springer, 72--90.

\bibitem[\protect\citeauthoryear{Lloyd}{Lloyd}{1987}]{Lloyd87}
{\sc Lloyd, J.} 1987.
\newblock {\em Foundations of Logic Programming}.
\newblock Springer, second, extended edition.

\bibitem[\protect\citeauthoryear{Lopez-Garcia, Bueno, and
  Hermenegildo}{Lopez-Garcia et~al\mbox{.}}{2010}]{determinacy-ngc09}
{\sc Lopez-Garcia, P.}, {\sc Bueno, F.}, {\sc and} {\sc Hermenegildo, M.~V.}
  2010.
\newblock {A}utomatic {I}nference of {D}eterminacy and {M}utual {E}xclusion for
  {L}ogic {P}rograms {U}sing {M}ode and {T}ype {A}nalyses.
\newblock {\em New Generation Computing\/}~{\em 28,\/}~2, 117--206.

\bibitem[\protect\citeauthoryear{Lopez-Garcia, Darmawan, Klemen, Liqat, Bueno,
  and Hermenegildo}{Lopez-Garcia
  et~al\mbox{.}}{2018}]{resource-verification-tplp18}
{\sc Lopez-Garcia, P.}, {\sc Darmawan, L.}, {\sc Klemen, M.}, {\sc Liqat, U.},
  {\sc Bueno, F.}, {\sc and} {\sc Hermenegildo, M.~V.} 2018.
\newblock {I}nterval-based {R}esource {U}sage {V}erification by {T}ranslation
  into {H}orn {C}lauses and an {A}pplication to {E}nergy {C}onsumption.
\newblock {\em Theory and Practice of Logic Programming, Special Issue on
  Computational Logic for Verification\/}~{\em 18,\/}~2 (March), 167--223.
\newblock arXiv:1803.04451.

\bibitem[\protect\citeauthoryear{Madsen, Yee, and Lhot{\'{a}}k}{Madsen
  et~al\mbox{.}}{2016}]{DBLP:conf/pldi/MadsenYL16}
{\sc Madsen, M.}, {\sc Yee, M.}, {\sc and} {\sc Lhot{\'{a}}k, O.} 2016.
\newblock {From {D}atalog to {FLIX}: a Declarative Language for Fixed Points on
  Lattices}.
\newblock In {\em Proceedings of the 37th {ACM} {SIGPLAN} Conference on
  Programming Language Design and Implementation, {PLDI} 2016, Santa Barbara,
  CA, USA, June 13-17, 2016}, {C.~Krintz} {and} {E.~Berger}, Eds. {ACM},
  194--208.

\bibitem[\protect\citeauthoryear{Marlowe and Ryder}{Marlowe and
  Ryder}{1990}]{Mar-Ryd-90}
{\sc Marlowe, T.} {\sc and} {\sc Ryder, B.} 1990.
\newblock An efficient hybrid algorithm for incremental data flow analysis.
\newblock In {\em 17th {ACM} Symposium on Principles of Programming Languages
  (POPL)}. {ACM} Press, 184--196.

\bibitem[\protect\citeauthoryear{Marriott and Stuckey}{Marriott and
  Stuckey}{1998}]{intro_constraints_stuckey}
{\sc Marriott, K.} {\sc and} {\sc Stuckey, P.~J.} 1998.
\newblock {\em {P}rogramming with {C}onstraints: an {I}ntroduction}.
\newblock MIT Press.

\bibitem[\protect\citeauthoryear{M\'{e}ndez-Lojo, Navas, and
  Hermenegildo}{M\'{e}ndez-Lojo
  et~al\mbox{.}}{2007}]{decomp-oo-prolog-lopstr07}
{\sc M\'{e}ndez-Lojo, M.}, {\sc Navas, J.}, {\sc and} {\sc Hermenegildo, M.}
  2007.
\newblock {A} {F}lexible ({C}){LP}-{B}ased {A}pproach to the {A}nalysis of
  {O}bject-{O}riented {P}rograms.
\newblock In {\em 17th International Symposium on Logic-based Program Synthesis
  and Transformation (LOPSTR 2007)}. Number 4915 in Lecture Notes in Computer
  Science. Springer-Verlag, 154--168.

\bibitem[\protect\citeauthoryear{Muthukumar and Hermenegildo}{Muthukumar and
  Hermenegildo}{1990}]{mcctr-fixpt}
{\sc Muthukumar, K.} {\sc and} {\sc Hermenegildo, M.} 1990.
\newblock {D}eriving {A} {F}ixpoint {C}omputation {A}lgorithm for {T}op-down
  {A}bstract {I}nterpretation of {L}ogic {P}rograms.
\newblock Technical Report ACT-DC-153-90, Microelectronics and Computer
  Technology Corporation (MCC), Austin, TX 78759. April.

\bibitem[\protect\citeauthoryear{Muthukumar and Hermenegildo}{Muthukumar and
  Hermenegildo}{1991}]{freeness-iclp91}
{\sc Muthukumar, K.} {\sc and} {\sc Hermenegildo, M.} 1991.
\newblock {C}ombined {D}etermination of {S}haring and {F}reeness of {P}rogram
  {V}ariables {T}hrough {A}bstract {I}nterpretation.
\newblock In {\em International Conference on Logic Programming (ICLP 1991)}.
  {MIT} {P}ress, 49--63.

\bibitem[\protect\citeauthoryear{Muthukumar and Hermenegildo}{Muthukumar and
  Hermenegildo}{1992}]{ai-jlp}
{\sc Muthukumar, K.} {\sc and} {\sc Hermenegildo, M.} 1992.
\newblock {C}ompile-time {D}erivation of {V}ariable {D}ependency {U}sing
  {A}bstract {I}nterpretation.
\newblock {\em Journal of Logic Programming\/}~{\em 13,\/}~2/3 (July),
  315--347.

\bibitem[\protect\citeauthoryear{Navas, M\'{e}ndez-Lojo, and
  Hermenegildo}{Navas et~al\mbox{.}}{2008}]{NMHLFM08}
{\sc Navas, J.}, {\sc M\'{e}ndez-Lojo, M.}, {\sc and} {\sc Hermenegildo, M.}
  2008.
\newblock {S}afe {U}pper-bounds {I}nference of {E}nergy {C}onsumption for
  {J}ava {B}ytecode {A}pplications.
\newblock In {\em The Sixth NASA Langley Formal Methods Workshop (LFM 08)}.
  29--32.
\newblock {E}xtended {A}bstract.

\bibitem[\protect\citeauthoryear{Navas, M\'{e}ndez-Lojo, and
  Hermenegildo}{Navas et~al\mbox{.}}{2007}]{fixpt-javabytecode-FTfJP07}
{\sc Navas, J.}, {\sc M\'{e}ndez-Lojo, M.}, {\sc and} {\sc Hermenegildo, M.~V.}
  2007.
\newblock {A}n {E}fficient, {C}ontext and {P}ath {S}ensitive {A}nalysis
  {F}ramework for {J}ava {P}rograms.
\newblock In {\em 9th Workshop on Formal Techniques for Java-like Programs
  FTfJP 2007}.

\bibitem[\protect\citeauthoryear{Navas, M\'{e}ndez-Lojo, and
  Hermenegildo}{Navas et~al\mbox{.}}{2009}]{resources-bytecode09}
{\sc Navas, J.}, {\sc M\'{e}ndez-Lojo, M.}, {\sc and} {\sc Hermenegildo, M.~V.}
  2009.
\newblock {U}ser-{D}efinable {R}esource {U}sage {B}ounds {A}nalysis for {J}ava
  {B}ytecode.
\newblock In {\em Proceedings of the Workshop on Bytecode Semantics,
  Verification, Analysis and Transformation (BYTECODE'09)}. Electronic Notes in
  Theoretical Computer Science, vol. 253. {E}lsevier - {N}orth {H}olland,
  65--82.

\bibitem[\protect\citeauthoryear{Perez-Carrasco, Klemen, Lopez-Garcia, Morales,
  and Hermenegildo}{Perez-Carrasco
  et~al\mbox{.}}{2020}]{resources-blockchain-sas20}
{\sc Perez-Carrasco, V.}, {\sc Klemen, M.}, {\sc Lopez-Garcia, P.}, {\sc
  Morales, J.}, {\sc and} {\sc Hermenegildo, M.~V.} 2020.
\newblock {C}ost {A}nalysis of {S}mart {C}ontracts via {P}arametric {R}esource
  {A}nalysis.
\newblock In {\em Proceedings of the 27th Static Analysis Symposium (SAS
  2020)}. LNCS. Springer-Verlag.

\bibitem[\protect\citeauthoryear{Plotkin}{Plotkin}{1981}]{Plotkin1981}
{\sc Plotkin, G.} 1981.
\newblock A structural approach to operational semantics.
\newblock Technical report {DAIMI FN-19}, Computer Science Department, Aarhus
  University, Denmark.

\bibitem[\protect\citeauthoryear{Plotkin}{Plotkin}{2004}]{Plotkin04a}
{\sc Plotkin, G.~D.} 2004.
\newblock A structural approach to operational semantics.
\newblock {\em J. Log. Algebr. Program.\/}~{\em 60-61}, 17--139.

\bibitem[\protect\citeauthoryear{Pollock and Soffa}{Pollock and
  Soffa}{1989}]{Pol-Sof}
{\sc Pollock, L.} {\sc and} {\sc Soffa, M.} 1989.
\newblock An incremental version of iterative data flow analysis.
\newblock {\em {IEEE} Transactions on Software Engineering\/}~{\em 15,\/}~12,
  1537--1549.

\bibitem[\protect\citeauthoryear{Puebla, Correas, Hermenegildo, Bueno,
  {Garc\'{\i}a de la Banda}, Marriott, and Stuckey}{Puebla
  et~al\mbox{.}}{2004}]{mod-an-lopstrbook}
{\sc Puebla, G.}, {\sc Correas, J.}, {\sc Hermenegildo, M.~V.}, {\sc Bueno,
  F.}, {\sc {Garc\'{\i}a de la Banda}, M.}, {\sc Marriott, K.}, {\sc and} {\sc
  Stuckey, P.~J.} 2004.
\newblock {A} {G}eneric {F}ramework for {C}ontext-{S}ensitive {A}nalysis of
  {M}odular {P}rograms.
\newblock In {\em {P}rogram {D}evelopment in {C}omputational {L}ogic, {A}
  {D}ecade of {R}esearch {A}dvances in {L}ogic-{B}ased {P}rogram
  {D}evelopment}, {M.~Bruynooghe} {and} {K.~Lau}, Eds. Number 3049 in LNCS.
  Springer-Verlag, Heidelberg, Germany, 234--261.

\bibitem[\protect\citeauthoryear{Puebla and Hermenegildo}{Puebla and
  Hermenegildo}{1996}]{inc-fixp-sas}
{\sc Puebla, G.} {\sc and} {\sc Hermenegildo, M.~V.} 1996.
\newblock {O}ptimized {A}lgorithms for the {I}ncremental {A}nalysis of {L}ogic
  {P}rograms.
\newblock In {\em International Static Analysis Symposium (SAS 1996)}. Number
  1145 in Lecture Notes in Computer Science. Springer-Verlag, 270--284.

\bibitem[\protect\citeauthoryear{Ramalingam and Reps}{Ramalingam and
  Reps}{1993}]{Ra-Re}
{\sc Ramalingam, G.} {\sc and} {\sc Reps, T.} 1993.
\newblock {A} {C}ategorized {B}ibliography on {I}ncremental {C}omputation.
\newblock In {\em ACM SIGPLAN-SIGACT Symposium on Principles of Programming
  Languages POPL'93}. ACM, Charleston, South Carolina.

\bibitem[\protect\citeauthoryear{Reps, Horwitz, and Sagiv}{Reps
  et~al\mbox{.}}{1995}]{DBLP:conf/popl/RepsHS95}
{\sc Reps, T.~W.}, {\sc Horwitz, S.}, {\sc and} {\sc Sagiv, S.} 1995.
\newblock Precise interprocedural dataflow analysis via graph reachability.
\newblock In {\em POPL}. 49--61.

\bibitem[\protect\citeauthoryear{Robinson}{Robinson}{1965}]{Robinson65}
{\sc Robinson, J.~A.} 1965.
\newblock {A} {M}achine {O}riented {L}ogic {B}ased on the {R}esolution
  {P}rinciple.
\newblock {\em Journal of the {ACM}\/}~{\em 12,\/}~23 (January), 23--41.

\bibitem[\protect\citeauthoryear{Rosen}{Rosen}{1981}]{Ros-81}
{\sc Rosen, B.} 1981.
\newblock Linear cost is sometimes quadratic.
\newblock In {\em Eighth {ACM} Symposium on Principles of Programming Languages
  (POPL)}. {ACM} Press, 117--124.

\bibitem[\protect\citeauthoryear{Rothenberg, Dietsch, and Heizmann}{Rothenberg
  et~al\mbox{.}}{2018}]{DBLP:conf/sas/RothenbergDH18}
{\sc Rothenberg, B.}, {\sc Dietsch, D.}, {\sc and} {\sc Heizmann, M.} 2018.
\newblock Incremental verification using trace abstraction.
\newblock In {\em Static Analysis - 25th International Symposium, {SAS} 2018,
  Freiburg, Germany, August 29-31, 2018}, {A.~Podelski}, Ed. Lecture Notes in
  Computer Science, vol. 11002. Springer, 364--382.

\bibitem[\protect\citeauthoryear{Ryder}{Ryder}{1988}]{Ryd-88}
{\sc Ryder, B.} 1988.
\newblock Incremental data-flow analysis algorithms.
\newblock {\em {ACM} Transactions on Programming Languages and Systems\/}~{\em
  10,\/}~1, 1--50.

\bibitem[\protect\citeauthoryear{Ryder, Marlowe, and Paull}{Ryder
  et~al\mbox{.}}{1988}]{Ryd-Mar-Pau-88}
{\sc Ryder, B.}, {\sc Marlowe, T.}, {\sc and} {\sc Paull, M.} 1988.
\newblock Conditions for incremental iteration: Examples and counterexamples.
\newblock {\em Science of Computer Programming\/}~{\em 11,\/}~1, 1--15.

\bibitem[\protect\citeauthoryear{Sery, Fedyukovich, and Sharygina}{Sery
  et~al\mbox{.}}{2012}]{DBLP:conf/fmcad/SeryFS12}
{\sc Sery, O.}, {\sc Fedyukovich, G.}, {\sc and} {\sc Sharygina, N.} 2012.
\newblock Incremental upgrade checking by means of interpolation-based function
  summaries.
\newblock In {\em Formal Methods in Computer-Aided Design, {FMCAD} 2012,
  Cambridge, UK, October 22-25, 2012}, {G.~Cabodi} {and} {S.~Singh}, Eds.
  {IEEE}, 114--121.

\bibitem[\protect\citeauthoryear{Sharir and Pnueli}{Sharir and
  Pnueli}{1978}]{sharir1978two}
{\sc Sharir, M.} {\sc and} {\sc Pnueli, A.} 1978.
\newblock {\em Two approaches to interprocedural data flow analysis}.
\newblock New York University. Courant Institute of Mathematical Sciences.

\bibitem[\protect\citeauthoryear{Swift}{Swift}{2014}]{swift2014incremental}
{\sc Swift, T.} 2014.
\newblock {Incremental Tabling in Support of Knowledge Representation and
  Reasoning}.
\newblock {\em Theory and Practice of Logic Programming\/}~{\em 14,\/}~4-5,
  553--567.

\bibitem[\protect\citeauthoryear{Szab{\'{o}}, Erdweg, and Voelter}{Szab{\'{o}}
  et~al\mbox{.}}{2016}]{DBLP:conf/kbse/SzaboEV16}
{\sc Szab{\'{o}}, T.}, {\sc Erdweg, S.}, {\sc and} {\sc Voelter, M.} 2016.
\newblock Inca: a {DSL} for the definition of incremental program analyses.
\newblock In {\em Proceedings of the 31st {IEEE/ACM} International Conference
  on Automated Software Engineering, {ASE} 2016, Singapore, September 3-7,
  2016}, {D.~Lo}, {S.~Apel}, {and} {S.~Khurshid}, Eds. {ACM}, 320--331.

\bibitem[\protect\citeauthoryear{Thakur and Nandivada}{Thakur and
  Nandivada}{2020}]{Thakur2020}
{\sc Thakur, M.} {\sc and} {\sc Nandivada, V.~K.} 2020.
\newblock Mix your contexts well: Opportunities unleashed by recent advances in
  scaling context-sensitivity.
\newblock In {\em Proceedings of the 29th International Conference on Compiler
  Construction}. CC 2020. Association for Computing Machinery, New York, NY,
  USA, 27--38.

\end{thebibliography}
\end{small}

\appendix
\clearpage

\section{Additional experimental results}\label{app:benchmarks}

\subsection{Detailed analysis times per step for analysis with
  \texttt{def}}\label{app:abs-graphs}

\vspace*{---8mm}

\newcommand{\includedetgraphsmall}[3]
{
  \begin{minipage}{0.45\textwidth}
    \includegraphics[width=1\textwidth]{#1_#2_#3_details}
  \end{minipage}
}

\newcommand{\detrowsmall}[2]
{
  \hspace{-5mm}
  \includedetgraphsmall{#1}{def}{#2}
}

\begin{figure*}[!ht]
  \centering
  \begin{tabular}{l l}
    \detrowsmall{hanoi}{add} & \detrowsmall{hanoi}{del} \\
    \detrowsmall{aiakl}{add} & \detrowsmall{aiakl}{del} \\
    \detrowsmall{qsort}{add} & \detrowsmall{qsort}{del} \\
    \detrowsmall{progeom}{add} & \detrowsmall{progeom}{del} \\
    \detrowsmall{bid}{add} & \detrowsmall{bid}{del} \\
  \end{tabular}
  \caption{Analysis times (ms) for both experiments with \texttt{def} for \textit{smaller} benchmarks.}
  \label{fig:app-small-details}
\end{figure*}

\begin{figure*}[p]
  \centering
  \begin{tabular}{l l}
    \detrowsmall{rdtok}{add} & \detrowsmall{rdtok}{del} \\
    \detrowsmall{cleandirs}{add}  & \detrowsmall{cleandirs}{del}  \\
    \detrowsmall{read}{add}  & \detrowsmall{read}{del}  \\
    \detrowsmall{warplan}{add}  & \detrowsmall{warplan}{del}  \\
  \end{tabular}
  \caption{Analysis times (ms) for both experiments with \texttt{def}
    for \textit{larger} benchmarks (1).}
  \label{fig:app-details-a}
\end{figure*}

\begin{figure*}[p]
  \centering
  \begin{tabular}{l l}
    \detrowsmall{boyer}{add}  & \detrowsmall{boyer}{del}  \\
    \detrowsmall{peephole}{add}  & \detrowsmall{peephole}{del}  \\
    \detrowsmall{witt}{add}  & \detrowsmall{witt}{del}  \\
    \detrowsmall{ann}{add}  & \detrowsmall{ann}{del}  \\
  \end{tabular}
  \caption{Analysis times (ms) for both experiments with \texttt{def}
    for \textit{larger} benchmarks (2).}
  \label{fig:app-details-b}
\end{figure*}

\subsection{Average analysis times split by domain}
\label{app:abs-times}

\newcommand{\avgtable}[2]
{%
  \begin{table}[!ht]
\captionsetup{margin=-10pt}
\centering
\scriptsize
    \centering
    \ifthenelse{\equal{#1}{del}}{\hspace*{-14mm}}{}
    \ifthenelse{\equal{#1}{pdb}}{\hspace*{-14mm}}{}
    \ifthenelse{\equal{#1}{gr}}{\hspace*{-14mm}}{}
    \ifthenelse{\equal{#1}{shfr}\and\equal{#2}{del}}{\hspace*{-24mm}}{}
    \input{#1_#2_it-time_auto}
    \caption{Analysis times (ms) per benchmark for the \emph{#2} experiment with the
      \texttt{#1} domain.}
  \end{table}%
  \vspace*{-3mm}
}

\begin{tabular}{|>{\tt}l|rrr|rrr|rrr|rrr|}
\toprule
  {} & \multicolumn{3}{l|}{\texttt{mon}} & \multicolumn{3}{l|}{\texttt{mon-inc}} & \multicolumn{3}{l|}{\texttt{mod}} & \multicolumn{3}{l|}{\texttt{mod-inc}} \\
  \midrule
  bench & \colm{ean} &   \colm{ax} & \colm{in} &    \colm{ean} &   \colm{ax} & \colm{in} & \colm{ean} &   \colm{ax} & \colm{in} &  \colm{ean} &   \colm{ax} & \colm{in} \\
\midrule
aiakl       &  1.8 &  4.1 & 1.1 &     1.6 &  4.0 & 1.1 &  1.9 &  16.6 & 0.5 &     1.3 &  10.7 & 0.5 \\
ann         &  4.0 & 28.0 & 1.4 &     3.1 & 15.2 & 1.9 &  1.3 &  32.4 & 0.3 &     1.3 &  27.4 & 0.5 \\
bid         &  2.5 & 10.0 & 1.4 &     1.9 &  5.5 & 1.5 &  1.8 &  22.5 & 0.5 &     1.6 &  19.6 & 0.5 \\
boyer       &  7.6 & 13.2 & 1.1 &     2.6 &  4.1 & 1.4 &  2.2 &   6.4 & 0.3 &     1.5 &   3.9 & 0.6 \\
check\_links &  7.2 & 98.1 & 2.2 &    13.2 & 96.2 & 5.7 &  3.5 & 221.6 & 0.4 &     5.7 & 212.2 & 0.8 \\
cleandirs   &  7.2 & 22.6 & 1.1 &     4.4 & 16.1 & 1.7 &  5.4 &  86.4 & 0.7 &     3.0 &  54.1 & 0.6 \\
hanoi       &  1.2 &  1.8 & 0.8 &     1.2 &  2.1 & 0.8 &  1.3 &   4.7 & 0.4 &     1.4 &   4.5 & 0.5 \\
manag\_proj  & 11.4 & 39.0 & 4.4 &     7.1 & 20.6 & 4.5 &  4.6 &  56.4 & 0.4 &     3.2 &  42.6 & 0.5 \\
peephole    &  3.7 & 20.1 & 1.5 &     2.3 &  7.0 & 1.6 &  1.1 &  25.5 & 0.4 &     1.0 &  20.4 & 0.4 \\
progeom     &  1.9 &  3.9 & 0.9 &     1.5 &  2.7 & 0.9 &  2.2 &  10.0 & 0.7 &     2.2 &  10.1 & 0.9 \\
read        &  9.6 & 31.8 & 1.3 &     4.7 & 20.5 & 1.3 &  7.1 &  32.9 & 0.8 &     4.9 &  28.0 & 0.9 \\
qsort       &  2.0 &  4.0 & 1.2 &     1.5 &  2.6 & 1.2 &  1.5 &   7.3 & 0.5 &     1.5 &   7.2 & 0.6 \\
rdtok       &  5.1 & 13.9 & 1.5 &     5.1 & 21.9 & 1.3 &  3.8 &  12.8 & 0.5 &     4.8 &  23.8 & 0.4 \\
warplan     &  6.0 & 13.5 & 0.9 &     2.3 &  4.4 & 1.2 &  3.4 &  10.6 & 0.6 &     2.1 &   6.3 & 1.0 \\
witt        &  5.9 & 27.1 & 1.4 &     3.6 & 16.5 & 2.3 &  3.0 &  59.8 & 0.3 &     2.2 &  47.1 & 0.4 \\
\bottomrule \end{tabular}

\begin{tabular}{|>{\tt}l|rrr|rrr|rrr|rrr|rrr|rrr|}
\toprule
  {} & \multicolumn{3}{l|}{\texttt{mon}} & \multicolumn{3}{l|}{\texttt{mon-inc}} & \multicolumn{3}{l|}{\texttt{mon-scc}} & \multicolumn{3}{l|}{\texttt{mod}} & \multicolumn{3}{l|}{\texttt{mod-inc}} & \multicolumn{3}{l|}{\texttt{mod-scc}} \\
  \midrule
bench & \colm{ean} &   \colm{ax} & \colm{in} &    \colm{ean} &   \colm{ax} & \colm{in} &    \colm{ean} &   \colm{ax} & \colm{in} & \colm{ean} &     \colm{ax} & \colm{in} &    \colm{ean} &   \colm{ax} & \colm{in} &    \colm{ean} &   \colm{ax} & \colm{in} \\
\midrule
aiakl       &  2.1 &  4.3 & 1.3 &     2.0 &   5.7 & 1.3 &     1.4 &   5.7 & 1.0 &  1.8 &  13.6 & 0.4 &     1.9 &  13.2 & 0.5 &     1.6 &  13.1 & 0.4 \\
ann         &  4.0 & 29.4 & 1.2 &     3.6 &  32.7 & 1.8 &     2.8 &  33.3 & 1.5 &  2.4 &  48.4 & 0.3 &     1.9 &  57.2 & 0.6 &     1.9 &  57.8 & 0.5 \\
bid         &  2.6 &  8.7 & 1.8 &     2.4 &  12.8 & 1.8 &     1.7 &  10.1 & 1.3 &  1.8 &  35.7 & 0.5 &     1.9 &  33.8 & 0.5 &     1.3 &  28.4 & 0.3 \\
boyer       &  7.7 & 14.7 & 1.0 &     6.9 &  15.1 & 1.1 &     1.8 &  14.5 & 1.0 &  2.1 &  12.5 & 0.3 &     2.5 &  15.9 & 0.3 &     1.2 &  15.3 & 0.3 \\
check\_links &  7.4 & 99.3 & 1.8 &    20.0 & 232.9 & 5.7 &    17.9 & 109.9 & 5.5 &  3.7 & 277.0 & 0.3 &    10.0 & 255.1 & 0.8 &     9.2 & 242.6 & 0.6 \\
cleandirs   &  7.5 & 28.9 & 1.4 &     5.3 &  28.5 & 1.5 &     3.1 &  25.4 & 1.3 &  4.8 &  98.4 & 0.5 &     3.6 &  78.1 & 0.6 &     3.1 &  76.8 & 0.5 \\
hanoi       &  1.5 &  3.1 & 0.7 &     1.2 &   2.5 & 0.8 &     1.0 &   2.5 & 0.7 &  1.4 &   5.0 & 0.4 &     1.7 &   6.3 & 0.6 &     1.4 &   6.0 & 0.4 \\
manag\_proj  & 12.0 & 40.4 & 4.3 &     7.7 &  46.3 & 4.1 &     7.1 &  41.7 & 4.3 &  4.4 & 104.8 & 0.3 &     3.5 &  94.4 & 0.4 &     3.3 &  94.3 & 0.4 \\
peephole    &  3.3 & 17.2 & 1.4 &     2.6 &  22.5 & 1.6 &     1.9 &  22.6 & 1.2 &  2.3 &  35.4 & 0.7 &     1.9 &  35.5 & 0.8 &     1.6 &  33.1 & 0.7 \\
progeom     &  2.1 &  4.1 & 1.2 &     1.8 &   5.1 & 1.1 &     1.1 &   4.9 & 0.8 &  2.6 &  20.9 & 0.9 &     2.6 &  18.7 & 0.9 &     2.6 &  23.9 & 0.6 \\
read        &  9.1 & 28.5 & 1.1 &    10.5 &  35.7 & 1.1 &     2.0 &  30.7 & 1.0 &  6.2 &  38.9 & 0.5 &     8.4 &  45.6 & 0.6 &     2.9 &  42.6 & 0.5 \\
qsort       &  1.9 &  3.7 & 1.2 &     1.8 &   4.6 & 1.1 &     1.2 &   4.4 & 0.9 &  1.9 &  11.3 & 0.5 &     1.8 &  10.9 & 0.5 &     1.6 &  11.4 & 0.5 \\
rdtok       &  4.9 & 14.2 & 1.5 &     5.3 &  17.5 & 1.5 &     2.0 &  15.6 & 1.3 &  4.0 &  36.1 & 0.4 &     4.9 &  32.6 & 0.4 &     3.0 &  35.8 & 0.3 \\
warplan     &  5.8 & 16.2 & 1.0 &     3.5 &  14.4 & 1.3 &     1.6 &  15.1 & 0.8 &  3.7 &  29.4 & 0.8 &     2.8 &  23.6 & 0.9 &     1.8 &  22.6 & 0.7 \\
witt        &  6.1 & 26.1 & 1.4 &     4.5 &  32.3 & 2.2 &     3.8 &  32.9 & 2.0 &  3.4 &  90.1 & 0.3 &     2.6 &  70.0 & 0.4 &     2.4 &  69.1 & 0.4 \\
\bottomrule \end{tabular}

\begin{tabular}{|>{\tt}l|rrr|rrr|rrr|rrr|}
\toprule
  {} & \multicolumn{3}{l|}{\texttt{mon}} & \multicolumn{3}{l|}{\texttt{mon-inc}} & \multicolumn{3}{l|}{\texttt{mod}} & \multicolumn{3}{l|}{\texttt{mod-inc}} \\
  \midrule
  bench & \colm{ean} &   \colm{ax} & \colm{in} &    \colm{ean} &   \colm{ax} & \colm{in} & \colm{ean} &   \colm{ax} & \colm{in} &  \colm{ean} &   \colm{ax} & \colm{in} \\
\midrule
aiakl       &  2.5 &   6.9 & 1.2 &     1.9 &   6.5 & 1.2 &  2.1 &  18.3 & 0.5 &     2.2 &  19.0 & 0.6 \\
ann         &  5.2 &  50.0 & 1.0 &     3.4 &  36.5 & 1.7 &  2.1 &  98.5 & 0.3 &     1.8 &  83.9 & 0.5 \\
bid         &  3.1 &  14.8 & 1.7 &     2.3 &   8.7 & 1.8 &  2.4 &  38.8 & 0.5 &     2.2 &  26.9 & 0.5 \\
boyer       & 16.0 &  26.1 & 1.4 &     3.1 &  11.2 & 1.5 &  3.3 &  11.9 & 0.5 &     1.7 &  11.2 & 0.6 \\
check\_links &  8.8 & 158.7 & 2.0 &    13.6 & 160.8 & 5.3 &  5.3 & 312.6 & 0.3 &     6.1 & 278.1 & 0.7 \\
cleandirs   &  5.2 &  23.5 & 1.3 &     3.3 &  17.7 & 1.3 &  3.6 &  72.5 & 0.3 &     3.1 &  64.1 & 0.6 \\
hanoi       &  1.5 &   2.3 & 0.9 &     1.4 &   2.9 & 0.9 &  2.0 &   7.0 & 0.5 &     1.4 &   5.6 & 0.4 \\
manag\_proj  &  6.9 &  13.4 & 5.0 &     5.9 &  10.8 & 4.3 &  2.8 &  10.3 & 0.3 &     2.2 &  12.1 & 0.4 \\
peephole    &  6.8 &  42.0 & 1.7 &     2.3 &  20.2 & 1.4 &  1.6 &  62.4 & 0.4 &     1.5 &  43.9 & 0.5 \\
progeom     &  2.1 &   4.8 & 0.9 &     1.4 &   2.6 & 0.8 &  2.1 &   9.9 & 0.7 &     2.0 &   7.7 & 0.7 \\
read        & 23.3 & 102.8 & 1.1 &     8.7 &  65.6 & 1.1 & 11.1 & 100.7 & 0.6 &     7.5 &  81.4 & 0.8 \\
qsort       &  2.7 &   7.2 & 1.1 &     1.7 &   3.5 & 1.0 &  2.9 &  12.2 & 0.5 &     2.0 &  10.0 & 0.4 \\
rdtok       &  7.4 &  26.0 & 1.2 &     5.8 &  37.4 & 1.4 &  5.1 &  23.7 & 0.4 &     5.6 &  39.3 & 0.5 \\
warplan     &  8.6 &  26.4 & 1.0 &     2.9 &   7.9 & 1.2 &  4.7 &  21.3 & 0.6 &     2.7 &  11.8 & 1.2 \\
witt        &  1.6 &   2.7 & 1.2 &     3.0 &   7.0 & 2.0 &  0.7 &   2.4 & 0.3 &     1.1 &   2.5 & 0.4 \\
\bottomrule \end{tabular}

\begin{tabular}{|>{\tt}l|rrr|rrr|rrr|rrr|rrr|rrr|}
\toprule
  {} & \multicolumn{3}{l|}{\texttt{mon}} & \multicolumn{3}{l|}{\texttt{mon-inc}} & \multicolumn{3}{l|}{\texttt{mon-scc}} & \multicolumn{3}{l|}{\texttt{mod}} & \multicolumn{3}{l|}{\texttt{mod-inc}} & \multicolumn{3}{l|}{\texttt{mod-scc}} \\
  \midrule
bench & \colm{ean} &   \colm{ax} & \colm{in} &    \colm{ean} &   \colm{ax} & \colm{in} &    \colm{ean} &   \colm{ax} & \colm{in} & \colm{ean} &     \colm{ax} & \colm{in} &    \colm{ean} &   \colm{ax} & \colm{in} &    \colm{ean} &   \colm{ax} & \colm{in} \\
\midrule
aiakl       &  2.3 &   5.9 & 1.1 &     2.1 &   8.0 & 1.3 &     1.5 &   7.9 & 0.9 &  2.2 &  20.8 & 0.4 &     1.9 &  16.2 & 0.4 &     1.9 &  18.3 & 0.5 \\
ann         &  5.9 &  74.0 & 1.2 &     3.9 &  59.0 & 1.8 &     3.2 &  55.2 & 1.5 &  6.5 & 134.6 & 0.3 &     3.1 & 127.1 & 0.6 &     2.5 & 121.3 & 0.5 \\
bid         &  2.8 &  12.9 & 1.8 &     2.2 &  16.0 & 1.5 &     1.9 &  15.8 & 1.4 &  2.3 &  45.2 & 0.4 &     2.8 &  44.8 & 0.5 &     1.9 &  39.3 & 0.3 \\
boyer       & 15.7 &  30.5 & 1.3 &    10.3 &  26.2 & 1.1 &     2.0 &  23.6 & 1.0 &  3.1 &  33.1 & 0.3 &     3.4 &  36.7 & 0.4 &     1.6 &  35.6 & 0.5 \\
check\_links & 10.0 & 175.5 & 1.6 &    22.9 & 367.2 & 5.9 &    20.0 & 186.0 & 5.6 &  6.4 & 411.7 & 0.5 &    12.9 & 335.0 & 0.8 &    12.9 & 337.4 & 0.6 \\
cleandirs   &  5.7 &  23.0 & 1.5 &     4.3 &  26.5 & 1.8 &     2.9 &  26.8 & 1.3 &  4.1 &  95.6 & 0.4 &     3.3 &  83.3 & 0.5 &     2.9 &  82.1 & 0.4 \\
hanoi       &  1.7 &   2.6 & 1.0 &     1.4 &   2.7 & 1.0 &     1.0 &   2.7 & 0.6 &  1.7 &   6.4 & 0.3 &     1.8 &   6.4 & 0.4 &     1.8 &   7.9 & 0.5 \\
manag\_proj  &  7.1 &  14.0 & 5.1 &     6.7 &  20.4 & 4.7 &     5.9 &  20.1 & 4.0 &  2.6 &  29.5 & 0.3 &     2.3 &  35.1 & 0.3 &     2.2 &  36.9 & 0.3 \\
peephole    &  7.2 &  50.1 & 1.7 &     3.0 &  46.5 & 1.5 &     2.2 &  47.2 & 1.4 &  4.0 &  66.0 & 1.0 &     3.1 &  61.6 & 1.0 &     2.4 &  61.9 & 1.0 \\
progeom     &  2.1 &   4.5 & 0.9 &     1.8 &   5.9 & 0.9 &     1.1 &   5.4 & 0.8 &  2.7 &  21.0 & 0.7 &     2.5 &  19.2 & 0.7 &     2.3 &  19.4 & 0.7 \\
read        & 22.6 & 109.2 & 1.0 &    22.6 & 118.7 & 1.3 &     2.9 & 101.2 & 0.9 & 11.0 & 110.8 & 0.6 &    13.2 & 113.8 & 0.5 &     4.5 & 113.6 & 0.4 \\
qsort       &  2.5 &   7.0 & 0.9 &     2.3 &   8.4 & 1.2 &     1.4 &   8.7 & 0.8 &  3.5 &  21.5 & 0.4 &     2.6 &  17.2 & 0.4 &     2.4 &  16.9 & 0.4 \\
rdtok       &  7.5 &  29.3 & 1.4 &     6.1 &  27.8 & 1.2 &     2.1 &  26.7 & 1.1 &  5.3 &  53.5 & 0.3 &     6.4 &  44.9 & 0.4 &     3.4 &  43.7 & 0.3 \\
warplan     &  9.0 &  23.1 & 1.0 &     4.8 &  25.4 & 1.3 &     1.9 &  29.6 & 0.9 &  4.9 &  48.1 & 0.8 &     3.8 &  39.6 & 1.2 &     2.4 &  37.2 & 0.7 \\
witt        &  2.0 &   3.0 & 1.4 &     3.5 &   8.4 & 2.2 &     3.1 &   8.4 & 1.9 &  0.9 &   8.2 & 0.3 &     1.2 &   5.3 & 0.4 &     1.0 &   5.2 & 0.3 \\
\bottomrule \end{tabular}

\begin{tabular}{|>{\tt}l|rrr|rrr|rrr|rrr|}
\toprule
  {} & \multicolumn{3}{l|}{\texttt{mon}} & \multicolumn{3}{l|}{\texttt{mon-inc}} & \multicolumn{3}{l|}{\texttt{mod}} & \multicolumn{3}{l|}{\texttt{mod-inc}} \\
  \midrule
  bench & \colm{ean} &   \colm{ax} & \colm{in} &    \colm{ean} &   \colm{ax} & \colm{in} & \colm{ean} &   \colm{ax} & \colm{in} &  \colm{ean} &   \colm{ax} & \colm{in} \\
\midrule
aiakl       &     2.6 &      6.8 &  1.1 &     1.8 &      7.1 &  1.2 &   1.7 &     15.2 & 0.4 &     1.6 &    13.8 & 0.4 \\
ann         &    10.3 &    123.2 &  1.3 &     3.9 &     84.5 &  2.1 &   4.2 &    201.9 & 0.4 &     2.8 &   175.9 & 0.6 \\
bid         &     4.2 &     22.6 &  2.1 &     3.1 &     13.9 &  2.2 &   2.9 &     39.0 & 0.4 &     2.1 &    26.3 & 0.4 \\
boyer       &    31.2 &     47.4 &  1.2 &     3.9 &     24.9 &  1.3 &   6.6 &     28.9 & 0.5 &     2.0 &    26.4 & 0.5 \\
check\_links &    34.0 &    709.6 &  2.6 &    17.5 &    704.8 &  6.0 &  28.1 &  2,012.9 & 0.3 &    10.8 & 1,002.6 & 0.8 \\
cleandirs   &    38.0 &    418.1 &  1.1 &     9.0 &    353.6 &  2.3 &  47.3 &  1,108.3 & 0.7 &    12.9 &   740.4 & 0.8 \\
hanoi       &     4.9 &      9.6 &  1.3 &     3.1 &     10.2 &  1.6 &   5.7 &     22.6 & 0.5 &     3.8 &    18.0 & 0.6 \\
manag\_proj  & 1,151.8 & 17,937.2 & 13.4 &   186.2 & 15,469.2 & 12.5 &  43.8 &  1,589.1 & 0.3 &    17.1 & 1,086.8 & 0.4 \\
peephole    &    19.7 &    155.0 &  1.8 &     3.6 &     58.9 &  1.7 &   4.1 &    167.2 & 0.3 &     3.1 &   120.1 & 0.5 \\
progeom     &     2.9 &      6.6 &  1.2 &     1.7 &      3.3 &  1.0 &   2.7 &     12.3 & 0.7 &     2.1 &     9.4 & 0.7 \\
read        &    84.3 &    413.2 &  1.1 &    24.6 &    321.8 &  1.3 &  37.5 &    424.4 & 1.0 &    16.7 &   341.4 & 0.7 \\
qsort       &     2.6 &      9.1 &  0.9 &     1.8 &      4.9 &  1.0 &   3.2 &     12.6 & 0.4 &     2.5 &    10.2 & 0.6 \\
rdtok       &    14.1 &     61.1 &  1.5 &     8.4 &     67.0 &  1.3 &  10.5 &     58.9 & 0.5 &     8.0 &    69.7 & 0.5 \\
warplan     &    15.9 &     61.1 &  0.9 &     5.3 &     39.1 &  1.2 &   9.1 &    111.9 & 0.9 &     3.8 &    61.1 & 1.1 \\
witt        &   140.4 &  2,494.9 &  1.4 &    28.0 &  1,759.9 &  2.1 & 128.3 & 10,786.0 & 0.3 &    20.6 & 2,062.8 & 0.4 \\
\bottomrule \end{tabular}

\begin{tabular}{|>{\tt}l|rrr|rrr|rrr|rrr|rrr|rrr|}
\toprule
  {} & \multicolumn{3}{l|}{\texttt{mon}} & \multicolumn{3}{l|}{\texttt{mon-inc}} & \multicolumn{3}{l|}{\texttt{mon-scc}} & \multicolumn{3}{l|}{\texttt{mod}} & \multicolumn{3}{l|}{\texttt{mod-inc}} & \multicolumn{3}{l|}{\texttt{mod-scc}} \\
  \midrule
bench & \colm{ean} &   \colm{ax} & \colm{in} &    \colm{ean} &   \colm{ax} & \colm{in} &    \colm{ean} &   \colm{ax} & \colm{in} & \colm{ean} &     \colm{ax} & \colm{in} &    \colm{ean} &   \colm{ax} & \colm{in} &    \colm{ean} &   \colm{ax} & \colm{in} \\
\midrule
aiakl       &     2.6 &      7.4 &  1.2 &     2.0 &      8.2 &  1.2 &     1.5 &      8.1 &  1.0 &   1.6 &     15.0 & 0.3 &     1.8 &     16.3 & 0.4 &     1.5 &     15.1 & 0.3 \\
ann         &    10.1 &    123.6 &  1.3 &     4.3 &    117.9 &  2.0 &     3.1 &    117.2 &  1.5 &   4.5 &    260.4 & 0.3 &     2.9 &    232.0 & 0.6 &     2.3 &    232.1 & 0.6 \\
bid         &     3.9 &     23.4 &  2.1 &     2.6 &     23.8 &  1.7 &     2.2 &     23.2 &  1.5 &   2.8 &     62.1 & 0.3 &     2.7 &     54.4 & 0.4 &     2.4 &     52.3 & 0.3 \\
boyer       &    32.8 &     50.6 &  1.6 &    14.0 &     62.3 &  1.3 &     2.4 &     50.8 &  1.1 &   6.7 &     60.7 & 0.4 &     5.5 &     61.6 & 0.5 &     1.8 &     60.8 & 0.3 \\
check\_links &    33.5 &    740.9 &  2.0 &    36.5 &  1,799.2 &  6.0 &    19.3 &    728.9 &  5.5 &  28.9 &  2,091.5 & 0.3 &    22.4 &  1,362.3 & 0.7 &     9.9 &  1,126.0 & 0.6 \\
cleandirs   &    38.0 &    388.9 &  1.1 &    10.8 &    381.9 &  2.2 &     8.2 &    386.7 &  1.5 &  52.1 &  1,169.5 & 0.9 &    18.4 &    782.8 & 0.8 &    18.4 &    813.0 & 0.7 \\
hanoi       &     5.0 &      9.6 &  1.4 &     3.0 &     10.3 &  1.4 &     2.7 &     10.5 &  1.2 &   5.2 &     20.8 & 0.4 &     3.7 &     18.3 & 0.4 &     4.3 &     21.9 & 0.6 \\
manag\_proj  & 1,131.4 & 18,049.1 & 12.8 &   229.0 & 18,121.6 & 12.9 &   145.1 & 18,081.0 & 12.5 & 366.6 & 37,511.5 & 0.3 &   321.2 & 34,731.8 & 0.4 &   322.5 & 34,924.4 & 0.4 \\
peephole    &    19.2 &    158.2 &  1.6 &     4.9 &    212.4 &  1.6 &     3.0 &    156.4 &  1.4 &   5.2 &    386.1 & 0.4 &     3.7 &    273.3 & 0.5 &     3.4 &    272.6 & 0.5 \\
progeom     &     2.7 &      6.0 &  1.0 &     2.3 &      7.7 &  1.1 &     1.4 &      7.7 &  0.9 &   3.5 &     29.0 & 0.8 &     3.0 &     24.4 & 0.8 &     2.5 &     22.8 & 0.6 \\
read        &    85.9 &    415.5 &  1.3 &    71.5 &    423.7 &  1.6 &     6.2 &    390.0 &  1.0 &  37.1 &    418.6 & 0.8 &    38.9 &    435.4 & 0.7 &     8.1 &    408.5 & 0.5 \\
qsort       &     2.9 &      9.7 &  1.3 &     2.5 &     11.1 &  1.1 &     1.4 &     10.6 &  0.8 &   3.4 &     22.4 & 0.3 &     3.9 &     24.8 & 0.6 &     3.5 &     25.0 & 0.4 \\
rdtok       &    13.1 &     60.4 &  1.4 &    12.0 &     63.4 &  1.6 &     2.9 &     67.4 &  1.3 &  11.5 &    117.7 & 0.3 &    12.4 &     88.4 & 0.6 &     7.9 &     91.6 & 0.4 \\
warplan     &    16.3 &     61.6 &  1.1 &     7.7 &     64.2 &  1.1 &     2.5 &     65.7 &  1.0 &   9.6 &    159.7 & 0.9 &     5.6 &    109.0 & 1.2 &     3.9 &    114.6 & 0.9 \\
witt        &   137.9 &  2,447.8 &  1.3 &    38.8 &  2,424.4 &  2.2 &    17.6 &  2,424.8 &  1.8 & 192.5 & 19,997.6 & 0.3 &    49.0 &  3,144.8 & 0.4 &    49.1 &  3,192.7 & 0.4 \\
\bottomrule \end{tabular}

\clearpage
\subsection{Speedup split by domain}
\label{app:speedups}

\newcommand{\speeduptable}[1]
{%
\captionsetup{margin=-10pt}
  \begin{table}[!ht]
    \scriptsize
    \input{#1_add_speedup}
    \input{#1_del_speedup}
    \caption{Speedups of the clause \emph{addition} (left) and \emph{deletion} (right) experiments with \texttt{#1}.}
  \end{table}%
  \vspace*{-3mm}
}
\captionsetup{margin=-10pt}
  \begin{table}[!ht]
    \scriptsize
    \begin{tabular}{|>{\tt}l|r|r|r|}
\toprule
  \multicolumn{1}{|l|}{\multirow{2}{*}{\texttt{bench}}} & \multicolumn{3}{c|}{\texttt{mod-inc}}  \\
                                                        & vs.  & vs.  & vs.  \\
                                                        & \texttt{mon} & \texttt{mon-inc} &  \texttt{mod} \\
\midrule
            aiakl &              1.4 &                  1.2 &              1.4 \\
              ann &              3.0 &                  2.3 &              0.9 \\
              bid &              1.6 &                  1.2 &              1.1 \\
            boyer &              5.1 &                  1.8 &              1.5 \\
      check\_links &              1.3 &                  2.3 &              0.6 \\
        cleandirs &              2.4 &                  1.5 &              1.8 \\
            hanoi &              0.9 &                  0.9 &              1.0 \\
manag\_proj &              3.5 &                  2.2 &              1.5 \\
         peephole &              3.7 &                  2.3 &              1.1 \\
          progeom &              0.9 &                  0.7 &              1.0 \\
      read &              2.0 &                  1.0 &              1.5 \\
            qsort &              1.3 &                  1.0 &              1.0 \\
            rdtok &              1.1 &                  1.1 &              0.8 \\
          warplan &              2.8 &                  1.1 &              1.6 \\
             witt &              2.6 &                  1.6 &              1.3 \\
\bottomrule \end{tabular}
    \begin{tabular}{|>{\tt}l|r|r|r|r|r|}
  \toprule
  \multicolumn{1}{|l|}{\multirow{2}{*}{\texttt{bench}}} & \multicolumn{5}{c|}{\texttt{mod-scc}}  \\
                                                        & vs. & vs. & vs. & vs. & vs. \\
                                                        &  \texttt{mon} &  \texttt{mon-inc} &  \texttt{mon-scc} &  \texttt{mod} &  \texttt{mod-inc} \\
\midrule
            aiakl &              1.3 &                  1.3 &                  0.9 &              1.1 &                  1.2 \\
              ann &              2.1 &                  1.9 &                  1.5 &              1.3 &                  1.0 \\
              bid &              1.9 &                  1.8 &                  1.3 &              1.4 &                  1.5 \\
            boyer &              6.5 &                  5.8 &                  1.5 &              1.8 &                  2.1 \\
      check\_links &              0.8 &                  2.2 &                  1.9 &              0.4 &                  1.1 \\
        cleandirs &              2.4 &                  1.7 &                  1.0 &              1.5 &                  1.2 \\
            hanoi &              1.1 &                  0.9 &                  0.7 &              1.0 &                  1.2 \\
manag\_proj &              3.6 &                  2.3 &                  2.1 &              1.3 &                  1.0 \\
         peephole &              2.1 &                  1.7 &                  1.2 &              1.4 &                  1.2 \\
          progeom &              0.8 &                  0.7 &                  0.4 &              1.0 &                  1.0 \\
      read &              3.1 &                  3.6 &                  0.7 &              2.1 &                  2.9 \\
            qsort &              1.2 &                  1.1 &                  0.7 &              1.2 &                  1.1 \\
            rdtok &              1.6 &                  1.8 &                  0.7 &              1.3 &                  1.6 \\
          warplan &              3.2 &                  1.9 &                  0.9 &              2.0 &                  1.5 \\
             witt &              2.6 &                  1.9 &                  1.6 &              1.4 &                  1.1 \\
\bottomrule \end{tabular}
    \caption{Speedups of the clause \emph{addition} (left) and \emph{deletion} (right) experiments with \texttt{pdb}.}
  \end{table}%
  \vspace*{-3mm}

\captionsetup{margin=-10pt}
  \begin{table}[!ht]
    \scriptsize
    \begin{tabular}{|>{\tt}l|r|r|r|}
\toprule
  \multicolumn{1}{|l|}{\multirow{2}{*}{\texttt{bench}}} & \multicolumn{3}{c|}{\texttt{mod-inc}}  \\
                                                        & vs.  & vs.  & vs.  \\
                                                        & \texttt{mon} & \texttt{mon-inc} &  \texttt{mod} \\
\midrule
            aiakl &              1.1 &                  0.9 &              0.9 \\
              ann &              2.9 &                  1.9 &              1.1 \\
              bid &              1.4 &                  1.1 &              1.1 \\
            boyer &              9.2 &                  1.8 &              1.9 \\
      check\_links &              1.4 &                  2.2 &              0.9 \\
        cleandirs &              1.7 &                  1.1 &              1.2 \\
            hanoi &              1.0 &                  1.0 &              1.4 \\
manag\_proj &              3.1 &                  2.7 &              1.3 \\
         peephole &              4.6 &                  1.5 &              1.1 \\
          progeom &              1.1 &                  0.7 &              1.0 \\
      read &              3.1 &                  1.2 &              1.5 \\
            qsort &              1.3 &                  0.8 &              1.5 \\
            rdtok &              1.3 &                  1.0 &              0.9 \\
          warplan &              3.2 &                  1.1 &              1.7 \\
             witt &              1.5 &                  2.8 &              0.7 \\
\bottomrule \end{tabular}
    \begin{tabular}{|>{\tt}l|r|r|r|r|r|}
\toprule
  \multicolumn{1}{|l|}{\multirow{2}{*}{\texttt{bench}}} & \multicolumn{5}{c|}{\texttt{mod-scc}}  \\
                                                        & vs. & vs. & vs. & vs. & vs. \\
                                                        &  \texttt{mon} &  \texttt{mon-inc} &  \texttt{mon-scc} &  \texttt{mod} &  \texttt{mod-inc} \\
\midrule
            aiakl &              1.2 &                  1.1 &                  0.8 &              1.1 &                  1.0 \\
              ann &              2.4 &                  1.6 &                  1.3 &              2.6 &                  1.2 \\
              bid &              1.5 &                  1.1 &                  1.0 &              1.2 &                  1.4 \\
            boyer &              9.8 &                  6.4 &                  1.2 &              1.9 &                  2.1 \\
      check\_links &              0.8 &                  1.8 &                  1.5 &              0.5 &                  1.0 \\
        cleandirs &              1.9 &                  1.5 &                  1.0 &              1.4 &                  1.1 \\
            hanoi &              1.0 &                  0.8 &                  0.5 &              1.0 &                  1.0 \\
manag\_proj &              3.2 &                  3.0 &                  2.6 &              1.2 &                  1.0 \\
         peephole &              3.0 &                  1.2 &                  0.9 &              1.6 &                  1.3 \\
          progeom &              0.9 &                  0.8 &                  0.5 &              1.2 &                  1.1 \\
      read &              5.0 &                  5.0 &                  0.6 &              2.4 &                  2.9 \\
            qsort &              1.0 &                  1.0 &                  0.6 &              1.5 &                  1.1 \\
            rdtok &              2.2 &                  1.8 &                  0.6 &              1.6 &                  1.9 \\
          warplan &              3.7 &                  2.0 &                  0.8 &              2.0 &                  1.6 \\
             witt &              2.0 &                  3.4 &                  3.0 &              0.9 &                  1.2 \\
\bottomrule \end{tabular}
    \caption{Speedups of the clause \emph{addition} (left) and \emph{deletion} (right) experiments with \texttt{gr}.}
  \end{table}%
  \vspace*{-3mm}

\captionsetup{margin=-10pt}
  \begin{table}[!ht]
    \scriptsize
    \begin{tabular}{|>{\tt}l|r|r|r|}
\toprule
  \multicolumn{1}{|l|}{\multirow{2}{*}{\texttt{bench}}} & \multicolumn{3}{c|}{\texttt{mod-inc}}  \\
                                                        & vs.  & vs.  & vs.  \\
                                                        & \texttt{mon} & \texttt{mon-inc} &  \texttt{mod} \\
\midrule
            aiakl &              1.7 &                  1.2 &              1.1 \\
              ann &              3.7 &                  1.4 &              1.5 \\
              bid &              2.0 &                  1.5 &              1.4 \\
            boyer &             16.0 &                  2.0 &              3.4 \\
      check\_links &              3.1 &                  1.6 &              2.6 \\
        cleandirs &              2.9 &                  0.7 &              3.7 \\
            hanoi &              1.3 &                  0.8 &              1.5 \\
manag\_proj &             67.2 &                 10.9 &              2.6 \\
         peephole &              6.4 &                  1.2 &              1.3 \\
          progeom &              1.4 &                  0.8 &              1.3 \\
      read &              5.0 &                  1.5 &              2.2 \\
            qsort &              1.0 &                  0.7 &              1.3 \\
            rdtok &              1.8 &                  1.0 &              1.3 \\
          warplan &              4.2 &                  1.4 &              2.4 \\
             witt &              6.8 &                  1.4 &              6.2 \\
\bottomrule \end{tabular}
    \begin{tabular}{|>{\tt}l|r|r|r|r|r|}
  \toprule
  \multicolumn{1}{|l|}{\multirow{2}{*}{\texttt{bench}}} & \multicolumn{5}{c|}{\texttt{mod-scc}}  \\
                                                        & vs. & vs. & vs. & vs. & vs. \\
                                                        &  \texttt{mon} &  \texttt{mon-inc} &  \texttt{mon-scc} &  \texttt{mod} &  \texttt{mod-inc} \\
\midrule
            aiakl &              1.8 &                  1.3 &                  1.0 &              1.1 &                  1.2 \\
              ann &              4.4 &                  1.9 &                  1.4 &              2.0 &                  1.3 \\
              bid &              1.6 &                  1.1 &                  0.9 &              1.2 &                  1.1 \\
            boyer &             17.8 &                  7.6 &                  1.3 &              3.6 &                  3.0 \\
      check\_links &              3.4 &                  3.7 &                  2.0 &              2.9 &                  2.3 \\
        cleandirs &              2.1 &                  0.6 &                  0.4 &              2.8 &                  1.0 \\
            hanoi &              1.2 &                  0.7 &                  0.6 &              1.2 &                  0.9 \\
manag\_proj &              3.5 &                  0.7 &                  0.4 &              1.1 &                  1.0 \\
         peephole &              5.7 &                  1.5 &                  0.9 &              1.5 &                  1.1 \\
          progeom &              1.1 &                  0.9 &                  0.6 &              1.4 &                  1.2 \\
      read &             10.6 &                  8.8 &                  0.8 &              4.6 &                  4.8 \\
            qsort &              0.8 &                  0.7 &                  0.4 &              1.0 &                  1.1 \\
            rdtok &              1.7 &                  1.5 &                  0.4 &              1.5 &                  1.6 \\
          warplan &              4.2 &                  2.0 &                  0.6 &              2.5 &                  1.4 \\
             witt &              2.8 &                  0.8 &                  0.4 &              3.9 &                  1.0 \\
\bottomrule \end{tabular}
    \caption{Speedups of the clause \emph{addition} (left) and \emph{deletion} (right) experiments with \texttt{shfr}.}
  \end{table}%
  \vspace*{-3mm}

\clearpage
\subsection{Accummulated analysis times}

\begin{figure*}[!ht]
  \hfill \hspace{6cm}
  \includegraphics[width=0.15\textwidth]{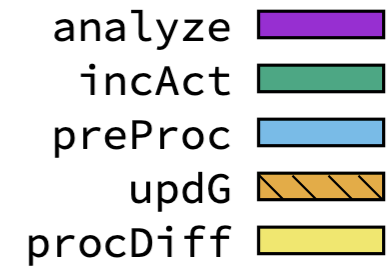}

  \vspace{-1cm}
  \hspace{-7mm}

  \begin{tabular}{l}
    \multicolumn{1}{c}{\texttt{pdb}} \\
    \includeaccgraphbig{pdb}{add} \vspace{-5mm}\\
    \benchnamessmall \\
    \multicolumn{1}{c}{\texttt{shfr}}\\
    \includeaccgraphbig{shfr}{add} \vspace{-5mm} \\
    \benchnamessmall \\
  \end{tabular}
   \caption{Accumulated analysis time (normalized w.r.t \texttt{mon}) adding
    clauses. The order inside each set of bars is:
  |{\tt mon}|{\tt mon\_inc}|{\tt mod}|{\tt mod\_inc}|.}
  \label{fig:app-add-acc}
\end{figure*}

\begin{figure*}[!ht]
  \hfill \hspace{6cm}
  \includegraphics[width=0.15\textwidth]{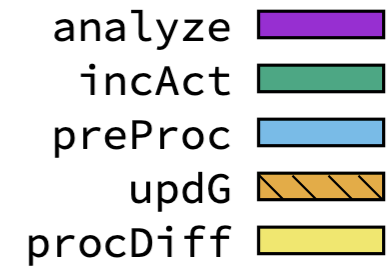}

  \vspace{-1cm}
  \hspace{-7mm}

  \begin{tabular}{l}
    \multicolumn{1}{c}{\texttt{pdb}} \\
    \includeaccgraphbig{pdb}{del}  \vspace{-5mm} \\
    \benchnamessmall \\
    \multicolumn{1}{c}{\texttt{shfr}} \\
    \includeaccgraphbig{shfr}{del}  \vspace{-5mm} \\
    \benchnamessmall \\
  \end{tabular}
  \caption{Accumulated analysis time (normalized w.r.t \texttt{mon}) deleting
    clauses. The order inside each set of bars is:
    |{\tt mon}|{\tt mon\_td}|{\tt mon\_scc}|{\tt mod}|{\tt mod\_td}|{\tt
      mod\_scc}|.}
  \label{fig:app-del-acc}
\end{figure*}

\clearpage
\subsection{Speedup vs. size of the analysis}

\begin{figure}[!ht]
  \centering
  \includegraphics[]{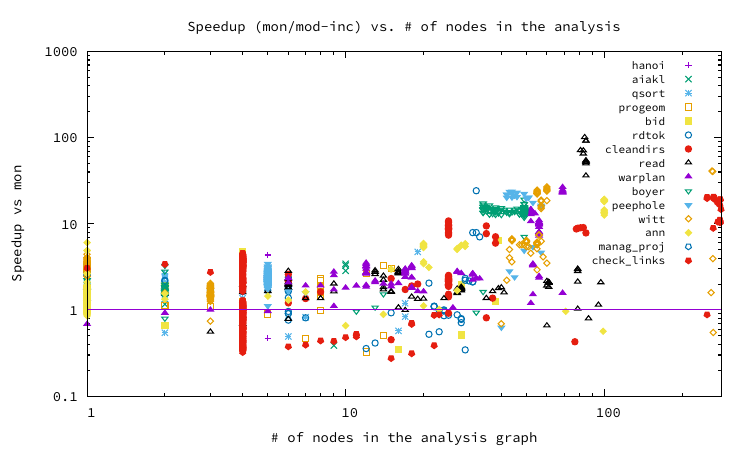}
  \caption{Speedup vs monolithic depending on the number of nodes in the analysis graph.}
  \label{fig:app-speed-def-mon-top}
\end{figure}

\begin{figure}[!ht]
  \centering
  \includegraphics[]{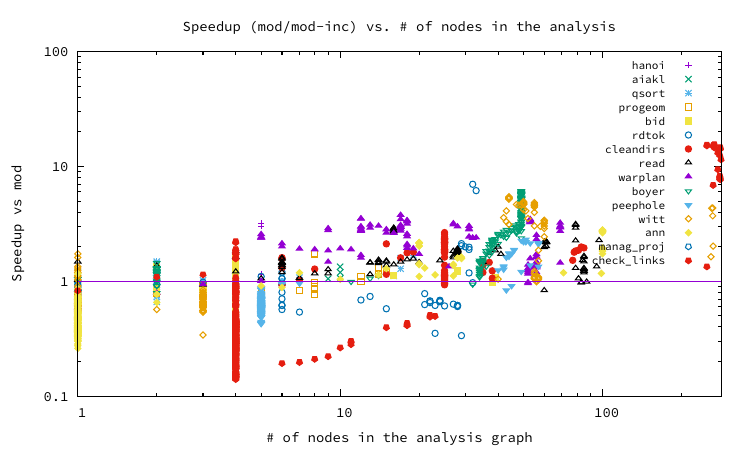}
  \caption{Speedup vs modular depending on the number of nodes in the analysis graph.}
  \label{fig:app-speed-def-mod-top}
\end{figure}


\label{lastpage}
\end{document}